\documentclass[nofootinbib,floats,aps,superscriptaddress,preprint]{revtex4}
\usepackage{graphicx}
\usepackage{amssymb,amsmath,amsbsy}
\usepackage{mathrsfs}

\newlength{\captsize}           \let\captsize=\small
\newlength{\captwidth}          
\newlength{\beforetableskip}    
\newcommand{\capt}[1]{\begin{minipage}{\captwidth}
              \let\normalsize=\captsize
              \caption[0]{#1}
              \end{minipage}\\ \vspace{\beforetableskip}}

\def\iso{\mathchoice{\cong}{\cong}{\isoS}{\cong}}
\def\isoS{\vbox{\baselineskip 0pt  \lineskip 0.5pt
    \ialign{$ \mathsurround=0pt  \scriptstyle \hfil ## \hfil $\crcr
        \sim \crcr = \crcr}}}

\newcommand{\mathbold}[1]{\mbox{\boldmath $\bf #1$}}
\def\ddel{\!\!\mathrel{\raise1.5ex\hbox{$\leftrightarrow$\kern-.85em
\lower1.7ex\hbox{$\partial$}}}}

\def\Tr{{\rm Tr}}
\def\abar{{\bar a}}
\def\bbar{{\bar b}}
\def\cbar{{\bar c}}
\def\dbar{{\bar d}}
\def\ebar{{\bar e}}
\def\fbar{{\bar f}}
\def\gbar{{\bar g}}
\def\hb{{\bar h}}

\def\lam{\lambda}

\def\wtil{\widetilde}
\def\ben{\begin{enumerate}}
\def\een{\end{enumerate}}
\def\beq{\begin{equation}}
\def\eeq{\end{equation}}
\def\beqa{\begin{eqnarray}}
\def\eeqa{\end{eqnarray}}

\def\ifmath#1{\relax\ifmmode #1\else $#1$\fi}
\def\lsim{\mathrel{\raise.3ex\hbox{$<$\kern-.75em\lower1ex\hbox{$\sim$}}}}
\def\gsim{\mathrel{\raise.3ex\hbox{$>$\kern-.75em\lower1ex\hbox{$\sim$}}}}
\def\sect#1{section~\ref{#1}}
\def\sects#1#2{sections~\ref{#1} and \ref{#2}}

\def\eq#1{eq.~(\ref{#1})}
\def\Ref#1{ref.~\cite{#1}}
\def\Refs#1#2{refs.~\cite{#1} and \cite{#2}}

\def\eqs#1#2{eqs.~(\ref{#1}) and (\ref{#2})}

\def\eqst#1#2{eqs.~(\ref{#1})--(\ref{#2})}
\def\eqthree#1#2#3{eqs.~(\ref{#1}), (\ref{#2}) and (\ref{#3})}
\def\Eq#1{Eq.~(\ref{#1})}
\def\Eqst#1#2{Eqs.~(\ref{#1})--(\ref{#2})}

\def\Eqst#1#2{Eqs.~(\ref{#1})--(\ref{#2})}
\def\App#1{Appendix~\ref{#1}}

\def\anti{\overline}

\def\mud{M_U}
\def\mdd{M_D}
\def\vev#1{\langle #1 \rangle}
\def\qlo{Q^0_L}
\def\uro{U^0_R}
\def\dro{D^0_R}
\def\ql{Q_L}
\def\ur{U_R}
\def\dr{D_R}
\def\eiuo{\eta_1^{U,0}}
\def\eiiuo{\eta_2^{U,0}}
\def\eido{\eta_1^{D,0}}
\def\eiido{\eta_2^{D,0}}

\def\eiuoa{\eta_a^{U,0}}
\def\eidoa{\eta_a^{D,0}}

\def\eiua{\eta_a^U}
\def\eida{\eta_a^D}

\def\cbma{c_{\beta-\alpha}}

\def\sbma{s_{\beta-\alpha}}

\def\phm{\phantom{-}}
\def\phaa{\phantom{AA}}

\def\beq{\begin{equation}}
\def\eeq{\end{equation}}
\def\ifmath#1{\relax\ifmmode #1\else $#1$\fi}

\def\call{\mathcal{L}}

\def\sb  {s_{\beta}}
\def\cb  {c_{\beta}}

\def\tanb{\tan\beta}

\def\hl{h^0}
\def\ha{A^0}
\def\hh{H^0}
\def\hpm{{H^\pm}}

\def\go{G^0}

\def\mha{m_{\ha}}
\def\mhl{m_{\hl}}
\def\mhh{m_{\hh}}
\def\mhpm{m_{\hpm}}

\def\ls#1{\ifmath{_{\lower1.5pt\hbox{$\scriptstyle #1$}}}}
\def\lss#1{\ifmath{^{\,\lower2.5pt\hbox{$\scriptstyle #1$}}}}
\def\lsup#1{^{\lower 6pt\hbox{$\scriptstyle#1$}}}
\def\llsup#1{^{\lower 3pt\hbox{$\scriptstyle#1$}}}
\def\lasup#1{^{\lower 2pt\hbox{$\scriptstyle#1$}}}
\def\nicefrac#1#2{\hbox{$\frac{#1}{#2}$}}

\def\half{\ifmath{{\textstyle{\frac{1}{2}}}}}

\def\quarter{\ifmath{{\textstyle{\frac{1}{4}}}}}
\def\eighth{\ifmath{{\textstyle{\frac{1}{8}}}}}

\renewcommand{\Re}{{\rm Re}}
\renewcommand{\Im}{{\rm Im}}

\begin{document}
%
\preprint{
\vbox{\vspace*{2cm}
      \hbox{SCIPP-06/01}
      \hbox{hep-ph/0602242}
      \hbox{February, 2006}
}}
\vspace*{4cm}

\title{Basis-independent methods for the two-Higgs-doublet model \\
II. The significance of $\mathbold{\tan\beta}$}
\author{Howard E. Haber}
\author{Deva O'Neil}
\affiliation{Santa Cruz Institute for Particle Physics  \\
   University of California, Santa Cruz, CA 95064, U.S.A. \\
\vspace{2cm}}

\begin{abstract}
  
In the most general two-Higgs-doublet model (2HDM), there is no
distinction between the two complex hypercharge-one SU(2)$\ls{\rm L}$ doublet
scalar fields, $\Phi_a$ ($a=1,2$).  Thus, any two orthonormal linear
combinations of these two fields can serve as a basis for the
Lagrangian.  All physical observables of the model must be
basis-independent.  For example, $\tan\beta\equiv
\vev{\Phi_2^0}/\vev{\Phi_1^0}$ is basis-dependent and thus cannot be
a physical parameter of the model.  In this paper, we provide a
basis-independent treatment of the Higgs sector with particular
attention to the neutral Higgs boson mass-eigenstates, which generically
are \textit{not} eigenstates of CP.  We then demonstrate that all physical
Higgs couplings are indeed independent of $\tan\beta$.  In
specialized versions of the 2HDM, $\tan\beta$ can be promoted to a
physical parameter of the Higgs-fermion interactions.  In
the most general 2HDM, the Higgs-fermion couplings can be expressed
in terms of a number of physical ``$\tan\beta$--like'' parameters
that are manifestly basis-independent.  The minimal supersymmetric
extension of the Standard Model provides a simple framework for
exhibiting such effects.

\end{abstract}

\maketitle

\section{Introduction}  \label{sec:intro}

The two-Higgs doublet model (2HDM) is one of the most well studied
extensions of the Standard Model.  Various motivations for adding a second
hypercharge-one complex Higgs doublet to the Standard Model
have been advocated in the literature%
~\cite{Lee:1973iz,Peccei:1977hh,susyhiggs,ghhiggs,hhg,Carena:2002es,branco}.
Perhaps the best motivated of these two-Higgs doublet models is the minimal
supersymmetric extension of the Standard Model (MSSM)~\cite{susyreviews},
which requires
a second Higgs doublet (and its supersymmetric fermionic partners) in
order to preserve the cancellation of gauge anomalies.

In most cases, the structure of the 2HDM is constrained in some way.
For example, in many of the early two-Higgs doublet models proposed
in the literature, a discrete symmetry was introduced that restricted
the most general form of the Higgs scalar potential and the
Higgs-fermion interactions~\cite{desh,type1,type2,hallwise,lavoura2}.  
In the MSSM, this discrete symmetry is not
present, but the imposition of supersymmetry on the dimension-four
terms of the Lagrangian yields similar restrictions on the
Higgs-fermion interactions and
even more stringent restrictions on the scalar potential.

It is tempting to relax these constraints and study the most general
2HDM consistent with the electroweak SU(2)$\ls{\rm L}\times$U(1)$_{\rm Y}$
gauge symmetry.  However,
phenomenology dictates that we choose the Higgs-fermion
interactions with some care~\cite{Weinberg,Georgi} to avoid
neutral Higgs-mediated flavor-changing neutral currents (FCNCs)
at tree level.  In the most general 2HDM, these effects are present,
and can only be suppressed (to avoid conflict with observed data)
by a significant fine-tuning of the Higgs-fermion interactions (to
ensure that certain couplings are small enough in magnitude).
Theoretically, it is more natural to introduce a symmetry to
completely remove the tree-level FCNC effects.  Both the discrete
symmetries alluded to above and supersymmetry provide just such a
natural mechanism.

Nevertheless, symmetries are often broken.   Supersymmetry is not an
exact symmetry, and it is possible to imagine small violations of the
discrete symmetries generated from new physics at the TeV-scale.  In
both cases, when the TeV-scale physics is integrated out, the
effective low-energy theory can resemble the most general 2HDM, albeit
with small couplings of the dangerous interactions that can potentially
yield tree-level FCNC effects.

The LHC will soon provide the first comprehensive look at the
TeV-scale.  Experiments from this collider may yield the first hints
of the existence of a non-minimal Higgs sector.   Precision Higgs
studies are one of the primary motivations for the development of the
International Linear Collider (ILC)~\cite{ilcbook}.  Data from the ILC could provide
detailed evidence of a 2HDM structure responsible for electroweak
symmetry breaking.  To make further progress, one must measure the
Higgs sector observables with some precision
in order to reconstruct (as best as one can) the Higgs Lagrangian.
Given that we will not know \textit{a priori} the theoretical
principles that constrain the Higgs sector, it is critical to develop
techniques for identifying experimental observables with the physical
parameters of the model.  In this context, a \textit{physical}
parameter is one that is measurable (in principle) without imposing
any simplifying theoretical assumptions.

Perhaps the simplest example of an unphysical parameter of the 2HDM is
the well known quantity
\beq \label{tanbdef}
\tan\beta\equiv\frac{\vev{\Phi_2^0}}{\vev{\Phi_1^0}}\,,
\eeq
given by the ratio
of the two neutral Higgs field vacuum expectation values.  The problem
with this quantity is that its definition assumes that one can
distinguish between
the two \textit{identical} hypercharge-one Higgs doublet fields.  In
the most general 2HDM, there is no preferred choice of \textit{basis}
of scalar fields $\Phi_1$--$\Phi_2$.  Any two sets of scalar doublets
related by a global $2\times 2$ unitary transformation are equally valid
choices.  Clearly, $\tan\beta$ is a basis-dependent quantity, and hence
is not a physical parameter.  Only basis-independent quantities can
be physical.

In the more specialized 2HDMs, a preferred basis is singled out.
For example, in the models with discrete symmetries and in the MSSM,
the Higgs potential exhibits a special form in
the preferred basis.  Then $\tan\beta$ can be defined with respect to
this basis and thereby is promoted to a physical parameter.  However, if
we allow for symmetry-breaking effects, the effective low-energy theory
is a completely general 2HDM (albeit with certain relations inherited
from the more fundamental theory).  In this case, the identification of
$\tan\beta$ as a physical parameter is more subtle.  Moreover, in
order to perform truly model-independent analyses of the Higgs precision
data, one should refrain from any additional theoretical assumptions, in
which case $\tan\beta$ once again is relegated to the class of
basis-dependent (and hence unphysical) parameters.

In \Ref{davidson}, a basis-independent formalism was advocated in order
to avoid the potential problems associated with unphysical parameters.
In particular, the importance of a basis-independent form for all
Higgs couplings was stressed.  These are the quantities that one wishes
to extract from precision Higgs experiments.  For example, in the case
of the Higgs-fermion interactions, the parameter $\tan\beta$ never
appears.  Instead, one can identify various
basis-independent $\tan\beta$-like parameters that can be identified
with ratios of physical couplings.  One of the main
shortcomings of \Ref{davidson} is that this program was only carried out
in the approximation of a CP-conserving scalar potential.  However, in
the most general 2HDM, the scalar potential possesses complex couplings
that can generate CP-violating effects.  Among the most important of
these effects is the mixing of CP-even and CP-odd scalar eigenstates to
produce neutral Higgs mass-eigenstates of indefinite CP quantum numbers.  
In this paper, we
complete the analysis of \Ref{davidson} by identifying the
basis-independent form for the Higgs couplings, allowing for the most
general CP-violating effects.

In \sect{sec:two}, we review the basis-independent formalism of
\Ref{davidson}.  This formalism was inspired by an elegant presentation
of the 2HDM in \Ref{branco}.
Although any basis choice is as good as any other basis
choice, the Higgs basis (defined to be a basis
in which one of the two neutral scalar fields has zero vacuum
expectation value) possesses some invariant
features.   In \sect{sec:three}, we
review the construction of the Higgs basis and use the basis-independent
formalism to highlight the invariant qualities of this basis choice.
Ultimately, we are interested in the Higgs mass-eigenstates.
In the most general CP-violating 2HDM, three neutral Higgs states
mix to form mass-eigenstates that are not eigenstates of CP.  
In \sect{sec:four}, we demonstrate how
to define basis-independent Higgs mixing parameters that are crucial for
deriving an invariant form for the Higgs couplings.  

In \sect{sec:five}
and \sect{sec:six} we provide the explicit basis-independent forms for
the Higgs couplings to bosons (gauge bosons and Higgs boson self-couplings)
and fermions (quarks and leptons), respectively.  Finally, in
\sect{sec:seven} we return to the question of the significance of
$\tan\beta$.  We demonstrate in a one-generation model how to define
three basis-independent $\tan\beta$-like parameters 
in terms of physical Higgs-fermion couplings.  In special
cases, these three parameters all reduce to the usual $\tan\beta$.
However, in the most general case, these three parameters can
differ.  The detection of such differences would yield important
clues to the fundamental nature of the 2HDM theoretical structure.
Conclusions and an outlook to future work are addressed in
\sect{sec:eight}.  Some additional details are relegated to a set of
four appendices.  In particular, Appendix~\ref{app:four}
provides the link between
the most general 2HDM considered in this paper and the more common
CP-conserving 2HDM that is often treated in the literature.

The basis-independent formalism has attracted some attention
during the past year.  In refs.~\cite{davidson,cpbasis,Branco:2005em,ivanov},
these techniques have been exploited to great
advantage in the study of the CP-violating structure of the 2HDM
(and extend the results originally obtained in \Refs{cpx}{cpx2}.)
In addition, the importance of the global U(2) transformation of the
two Higgs-doublet fields (and the subgroup of U(1)$\times$U(1) rephasing
transformations) has been emphasized, 
and some of their implications for 2HDM phenomenology have
been explored recently in \Ref{Ginzburg:2004vp}.

\section{The basis-independent formalism}
\label{sec:two}

The fields of the two-Higgs-doublet model (2HDM) consist of two
identical complex hypercharge-one, SU(2)$\ls{\rm L}$ doublet scalar fields
$\Phi_a(x)\equiv (\Phi^+_a(x)\,,\,\Phi^0_a(x))$, 
where $a=1,2$ labels the two Higgs doublet fields, and will 
be referred to as the Higgs ``flavor'' index.   
The Higgs doublet fields can always be redefined
by an arbitrary non-singular complex transformation
$\Phi_a\to B_{ab}\Phi_b$, where the matrix $B$ depends on eight real
parameters.  However, four of these parameters can be used
to transform the scalar field kinetic energy terms 
into canonical form.\footnote{That is, starting from
$\mathscr{L}_{\rm KE}=a\,(D_\mu\Phi_1)^\dagger(D_\mu\Phi_1)+
b\,(D_\mu\Phi_2)^\dagger(D_\mu\Phi_2)
+\bigl[c\,(D_\mu\Phi_1)^\dagger(D_\mu\Phi_2)+{\rm h.c.}\bigr]$, 
where $a$ and $b$ are real and $c$ is complex, one
can always find a (non-unitary) transformation
$B$ that removes the four real degrees of freedom
corresponding to $a$, $b$ and $c$ and sets $a=b=1$ and $c=0$.
Mathematically, such a transformation is an element of the coset space
GL$(2,\mathbb{C})/$U(2).}
The most general redefinition of the scalar fields [which leaves invariant
the form of the canonical kinetic 
energy terms 
$\mathscr{L}_{\rm KE}=(D_\mu\Phi)^\dagger_{\abar} (D^\mu\Phi)_a$] 
corresponds to a global U(2) transformation,
$\Phi_a\to U_{a\bbar}\Phi_b$ [and $\Phi_\abar^\dagger\to\Phi_\bbar^\dagger
U^\dagger_{b\abar}$], where the $2\times 2$ unitary matrix $U$ satisfies
$U^\dagger_{b\abar}U_{a\cbar}=\delta_{b\cbar}$.  In our index
conventions, replacing an unbarred index with a barred index is
equivalent to complex conjugation.   We only allow sums over 
barred--unbarred index pairs, which are
performed by employing
the U(2)-invariant tensor $\delta_{a\bbar}$.  
The basis-independent formalism consists of writing all equations
involving the Higgs sector fields in a U(2)-covariant form.
Basis-independent quantities can then be identified as 
U(2)-invariant scalars, which are easily identified as products of
tensor quantities with
all barred--unbarred index pairs summed with no Higgs
flavor indices left over.

We begin with the most general 2HDM scalar potential.  An explicit
form for the scalar potential in a generic basis is given in 
Appendix~\ref{app:one}.
Following \Refs{branco}{davidson}, the scalar potential can be written
in U(2)-covariant form:
\beq \label{genericpot}
\mathcal{V}=Y_{a\bbar}\Phi_\abar^\dagger\Phi_b
+\half Z_{a\bbar c\dbar}(\Phi_\abar^\dagger\Phi_b)
(\Phi_\cbar^\dagger\Phi_d)\,,
\eeq
where the indices $a$, $\bbar$, $c$ and $\dbar$
are labels with respect to the two-dimensional Higgs
flavor space and $Z_{a\bbar c\dbar}=Z_{c\dbar a\bbar}$.
The hermiticity of $\mathcal{V}$ yields
$Y_{a \bbar}= (Y_{b \abar})^\ast$ and
$Z_{a\bbar c\dbar}= (Z_{b\abar d\cbar})^\ast$.
Under a U(2) transformation, the tensors $Y_{a\bbar}$ and
$Z_{a\bbar c\dbar}$ transform covariantly:
$Y_{a\bbar}\to U_{a\cbar}Y_{c\dbar}U^\dagger_{d\bbar}$
and $Z_{a\bbar c\dbar}\to U_{a\ebar}U^\dagger_{f\bbar}U_{c\gbar}
U^\dagger_{h\dbar} Z_{e\fbar g\hb}$.  Thus, the scalar potential
$\mathcal{V}$ is a U(2)-scalar.  The interpretation of these results
is simple.  Global U(2)-flavor
transformations of the two Higgs doublet fields do not change the
functional form of the scalar potential.  However, the coefficients of
each term of the potential depends on the choice of basis.  The
transformation of these coefficients under a U(2) basis change are
precisely the transformation laws of $Y$ and $Z$ given above.

We shall assume that the vacuum of the theory respects the electromagnetic
U(1)$_{\rm EM}$ gauge symmetry.  In this case, the non-zero vacuum
expectation values of $\Phi_a$ must be aligned.  The standard
convention is to make a gauge-SU(2)$\ls{\rm L}$ 
transformation (if necessary) such
that the lower (or second) component of the doublet fields correspond
to electric charge $Q=0$.  In this case, the most general
U(1)$_{\rm EM}$-conserving vacuum expectation values are:
\beq \label{emvev}
\langle \Phi_a
\rangle={\frac{v}{\sqrt{2}}} \left(
\begin{array}{c} 0\\ \widehat v_a \end{array}\right)\,,\qquad
{\rm with}\qquad
\widehat v_a \equiv e^{i\eta}\left(
\begin{array}{c} \cb\,\\ \sb\,e^{i\xi} \end{array}\right)
\,,
\eeq
where $v\equiv 2m_W/g=246$~GeV and $\widehat v_a$ is a vector of unit norm.
The overall phase $\eta$ is arbitrary.  By convention, we take
$0\leq\beta\leq\pi/2$ and $0\leq\xi<2\pi$.
Taking the derivative of \eq{genericpot} with respect to $\Phi_b$, and
setting $\vev{\Phi^0_a}=v_a/\sqrt{2}$, we find the covariant form for
the scalar potential minimum conditions:
\beq \label{potmingeneric}
v\,\widehat v_\abar^\ast\,
[Y_{a\bar b}+\half v^2 Z_{a\bbar c\dbar}\, \widehat v_\cbar^\ast\,
\widehat v_d]=0 \,.
\eeq

Before proceeding, let us consider the most general global-U(2)
transformation~(see p.~5 of \Ref{murnaghan}):
\beq \label{utransform}
U=e^{i\psi}
\left(\begin{array}{cc} e^{i\gamma}\cos\theta &
\quad e^{-i\zeta}\,\sin\theta  \\
 -e^{i\zeta}\,\sin\theta & \quad e^{-i\gamma}\,\cos\theta
\end{array}\right)\,,
\eeq
where $-\pi\leq \theta\,,\,\psi <\pi$ and
$-\pi/2\leq \zeta\,,\,\gamma\leq\pi/2$ defines the closed and bounded
U(2) parameter space.  If we fix $\psi=0$, then the $U$ span an SU(2)
matrix subgroup of U(2), and $\{e^{i\psi}\}$ constitutes a U(1)
subgroup of U(2). More precisely, 
U(2)~$\iso$~SU(2)$\times$U(1)$/\mathbb{Z}_2$.
In the scalar sector, this U(1) coincides with
global hypercharge U(1)$_{\rm Y}$.   However,
the former U(1) is distinguished from hypercharge by the fact
that it has no effect on the other fields of the Standard Model.

Because the scalar potential is \textit{invariant}
under U(1)$_{\rm Y}$ hypercharge 
transformations,\footnote{The SU(2)$\ls{\rm L}\times$U(1)$_{\rm Y}$
gauge transformations act on the fields of the Standard Model, but
do \textit{not} transform the coefficients of 
the terms appearing in the Lagrangian.}
it follows that $Y$ and $Z$ are invariant under U(1)-flavor
transformations.  Thus, from the standpoint of the Lagrangian, only
SU(2)-flavor transformations correspond to a change of basis.
Nevertheless, the vacuum expectation value $\widehat v$ does change by
an overall phase under flavor-U(1) transformations.  Thus, it is
convenient to expand our definition of the basis to include the phase
of $\widehat v$.  In this convention, all
U(2)-flavor transformations correspond to a change of basis.
The reason for this choice is that it permits us to expand our
potential list of basis-independent quantities to include quantities
that depend on $\widehat v$. Since $\Phi_a\to U_{a\bbar}\Phi_b$ it
follows that $\widehat v_a\to U_{a\bbar} \widehat v_b$, and the
covariance properties of quantities that depend on $\widehat v$ are
easily discerned.

The unit vector $\widehat v_a$ can also be regarded as an
eigenvector of unit norm of the Hermitian matrix
$V_{a\bbar}\equiv \widehat v_a \widehat v_{\bbar}^*$.  The overall phase of
$\hat v_a$ is not determined in this definition, but as noted above
different phase choices are related by U(1)-flavor transformations.
Since $V_{a\bbar}$ is hermitian, it possesses a second eigenvector of unit norm
that is orthogonal to $\widehat v_a$.  We denote this eigenvector by
$\widehat w_a$, which satisfies:
\beq \label{vw}
\widehat v_{\bbar}^* \widehat w_b=0\,.
\eeq
The most general solution to \eq{vw}, up to an overall multiplicative
phase factor, is:
\beq \label{wdef}
\widehat w_b \equiv \widehat
v_\abar^\ast\epsilon_{ab}=e^{-i\eta}\left(
\begin{array}{c} -\sb\,e^{-i\xi}\\ \cb\end{array}\right)\,.\
\eeq
That is,
we have chosen a convention in which $\widehat w_b \equiv e^{i\chi}\widehat
v_\abar^\ast\epsilon_{ab}$, where $\chi=0$.
Of course, $\chi$ is not fixed by \eq{vw};
the existence of this
phase choice is reflected in the non-uniqueness of 
the Higgs basis, as discussed in \sect{sec:three}.

The inverse relation to \eq{wdef} is
easily obtained: $\widehat v^\ast_{\abar}=
\epsilon_{\abar\bbar}\,\widehat w_b$.
Above, we have introduced two Levi-Civita tensors
with $\epsilon_{12}=-\epsilon_{21}=1$ and $\epsilon_{11}=\epsilon_{22}=0$.
However, $\epsilon_{ab}$ and $\epsilon_{\abar\bbar}$ are not proper
tensors with respect to the full flavor-U(2) group (although these are
invariant SU(2)-tensors).  Consequently, $\widehat w_a$ does not
transform covariantly with respect to the full flavor-U(2) group. If
we write $U=e^{i\psi}\widehat U$, with $\det\widehat U=1$ (and
$\det U=e^{2i\psi}$), it is simple to check that under a
U(2) transformation
\beq \label{uvtransform}
\widehat v_a\to U_{a\bbar}\widehat v_b\qquad {\rm implies~that}\qquad
\widehat w_a\to ({\rm det}~U)^{-1}\,U_{a\bbar\,} \widehat w_b\,.
\eeq

Henceforth, we shall define a pseudotensor\footnote{In tensor calculus,
analogous quantities are usually referred to as
tensor densities or relative tensors~\cite{synge}.}
as a tensor
that transform covariantly with respect to the flavor-SU(2)
subgroup but whose transformation law
with respect to the full flavor-U(2)
group is only covariant modulo an overall nontrivial phase equal to
some integer power of $\det U$.  Thus, $\widehat w_a$ is a pseudovector.
However, we can use $\widehat w_a$ to construct proper tensors.  For
example, the Hermitian matrix $W_{a\bbar}\equiv\widehat w_a
\widehat w^*_{\bbar}=\delta_{a\bbar}-V_{a\bbar}$ is a proper
second-ranked tensor.

Likewise, a pseudoscalar (henceforth referred to as 
a pseudo-invariant) is defined
as a quantity that transforms
under U(2) by multiplication by
some integer power of $\det U$.  We reiterate that pseudo-invariants
\textit{cannot} be physical observables as the latter must be true
U(2)-invariants.

\section{The Higgs bases}
\label{sec:three}

Once the scalar potential minimum is determined, which defines
$\widehat v_a$, one class of basis choices is uniquely selected.
Suppose we begin in a generic $\Phi_1$--$\Phi_2$ basis.
We define new Higgs doublet fields:
\beq \label{hbasisdef}
H_1=(H_1^+\,,\,H_1^0)\equiv \widehat v_{\abar}^*\Phi_a\,,\qquad\qquad
H_2=(H_2^+\,,\,H_2^0)\equiv \widehat w_{\abar}^*\Phi_a= \epsilon_{\bbar\abar}
\widehat v_b\Phi_a \,.
\eeq
The transformation between
the generic basis and the Higgs basis, $H_a = \widehat U_{a \bar{b}} \Phi_b$, is given by the following
flavor-SU(2) matrix:
\beq \label{ugenhiggs}
\widehat U=\left(\begin{array}{cc}\widehat v_1^* &\quad \widehat v_2^*\\
\widehat w_1^* &\quad \widehat w_2^*\end{array}\right)=
\left(\begin{array}{cc}\phm\widehat v_1^* &\quad \widehat v_2^*\\
-\widehat v_2 &\quad \widehat v_1\end{array}\right)\,.
\eeq
This defines a particular Higgs basis.

Inverting \eq{hbasisdef} yields:
\beq \label{hbasis}
\Phi_a=H_1 \widehat v_a+ H_2 \widehat w_a = H_1 \widehat v_a+ H_2 \widehat v^*_{\bbar}\epsilon_{ba}\,.
\eeq
The definitions of $H_1$ and $H_2$ imply that
\beq \label{higgsvevs}
\vev{H_1^0}=\frac{v}{\sqrt{2}}\,,\qquad\qquad \vev{H_2^0}=0\,,
\eeq
where we have used \eq{vw} and the fact that
$\widehat v^{\,*}_{\abar}\,\widehat v_a=1$.

The Higgs basis is not unique.  Suppose one begins in a generic
$\Phi'_1$--$\Phi'_2$ basis, where $\Phi'_a=V_{a\bbar}\Phi_b$ and
$\det V\equiv e^{i\chi}\neq 1$.  If we now define:
\beq \label{hbasispdef}
H'_1\equiv \widehat v_{\abar}^*\Phi'_a\,,\qquad\qquad
H'_2\equiv \widehat w_{\abar}^*\Phi'_a\,,
\eeq
then
\beq \label{hbasischi}
H'_1=H_1\,,\qquad\qquad H'_2=(\det V)H_2= e^{i\chi} H_2\,.
\eeq
That is, $H_1$ is an invariant field, whereas $H_2$ is pseudo-invariant
with respect to arbitrary U(2) transformations.
In particular, the unitary matrix
\beq \label{ud}
U_D\equiv\left(\begin{array}{cc} 1 &\quad 0\\ 0 &\quad e^{i\chi}\end{array}
\right)
\eeq
transforms from the unprimed Higgs basis to the primed Higgs basis.
The phase angle
$\chi$ parameterizes the class of Higgs bases.  From the definition of
$H_2$ given in \eq{hbasisdef}, this phase freedom can be attributed
to the choice of an overall phase in the definition of $\widehat
w$ as discussed in \sect{sec:two}.  This phase freedom will be
reflected by the appearance of pseudo-invariants in the study of the
Higgs basis.  However, pseudo-invariants are useful in that they can
be combined to create true invariants, which are candidates for
observable quantities.

It is now a simple matter to insert
\eq{hbasis} into \eq{genericpot} to obtain:
\beqa \label{hbasispot}
\mathcal{V}&=& Y_1 H_1^\dagger H_1+ Y_2 H_2^\dagger H_2
+[Y_3 H_1^\dagger H_2+{\rm h.c.}]\nonumber \\[5pt]
&&\quad
+\half Z_1(H_1^\dagger H_1)^2 +\half Z_2(H_2^\dagger H_2)^2
+Z_3(H_1^\dagger H_1)(H_2^\dagger H_2)
+Z_4( H_1^\dagger H_2)(H_2^\dagger H_1) \nonumber \\[5pt]
&&\quad +\left\{\half Z_5 (H_1^\dagger H_2)^2
+\big[Z_6 (H_1^\dagger H_1)
+Z_7 (H_2^\dagger H_2)\big]
H_1^\dagger H_2+{\rm h.c.}\right\}\,,
\eeqa
where $Y_1$, $Y_2$ and $Z_{1,2,3,4}$ are U(2)-invariant quantities and
$Y_3$ and $Z_{5,6,7}$ are pseudo-invariants.
The explicit forms for the Higgs basis coefficients have been given in
\Ref{davidson}.  The invariant coefficients are conveniently expressed
in terms of the second-ranked tensors $V_{a\bbar}$ and $W_{a\bbar}$
introduced in \sect{sec:two}:
\beqa \label{invariants}
Y_1 &\equiv& \Tr(YV)\,,\qquad\qquad\qquad\,\, Y_2 \equiv \Tr(YW)\,,\nonumber\\
Z_1 &\equiv& Z_{a\bbar c\dbar}\,V_{b\abar}V_{d\cbar}\,,\qquad\qquad\,\,\,\,
Z_2 \equiv Z_{a\bbar c\dbar}\,W_{b\abar}W_{d\cbar}\,,\qquad\qquad \nonumber\\
Z_3 &\equiv& Z_{a\bbar c\dbar}\,V_{b\abar}W_{d\cbar}\,,\qquad\qquad\,\,\,
Z_4 \equiv Z_{a\bbar c\dbar}\,V_{b\cbar}W_{d\abar}\,,
\eeqa
whereas the pseudo-invariant coefficients are given by:
\beqa \label{pseudoinvariants}
\hspace{-0.3in} Y_3 &\equiv&
Y_{a\bbar}\,\widehat v_\abar^\ast\, \widehat w_b\,,\qquad\qquad\qquad
Z_5 \equiv
Z_{a\bbar c\dbar}\,\widehat v_\abar^\ast\, \widehat w_b\,
\widehat v_\cbar^\ast\, \widehat w_d\,,\nonumber \\ \hspace{-0.3in}
Z_6 &\equiv&  Z_{a\bbar c\dbar}\,\widehat v_\abar^\ast\,\widehat v_b\,
\widehat v_\cbar^\ast\, \widehat w_d\,,\qquad\quad
Z_7 \equiv
     Z_{a\bbar c\dbar}\,\widehat v_\abar^\ast\, \widehat w_b\,
\widehat w_\cbar^\ast\,\widehat w_d\,.
\eeqa
The invariant coefficients are manifestly real, whereas the
pseudo-invariant coefficients are potentially complex.

Using \eq{uvtransform}, it follows that under a flavor-U(2)
transformation specified by the matrix $U$, the pseudo-invariants
transform as:
\beq \label{tpseudo}
[Y_3, Z_6, Z_7]\to (\det U)^{-1}[Y_3, Z_6, Z_7]\qquad {\rm and} \qquad
Z_5\to  (\det U)^{-2} Z_5 \,.
\eeq
One can also deduce \eq{tpseudo} from \eq{hbasispot} by
noting that $\mathcal{V}$ and $H_1$ are invariant
whereas $H_2$ is pseudo-invariant field that is transforms as:
\beq \label{h2pseudo}
H_2\to (\det U)H_2\,.
\eeq

In the class of Higgs bases defined by \eq{hbasischi}, $\widehat{v}=(1,0)$
and $\widehat{w}=(0,1)$, which are independent of the angle $\chi$
that distinguishes among different Higgs bases.  That is,
under the phase transformation specified by \eq{ud}, both $\widehat v$
and $\widehat w$ are unchanged.
Inserting these values of $\widehat v$ and $\widehat w$ into
\eqs{invariants}{pseudoinvariants} yields the coefficients
of the Higgs basis scalar potential.  For example, the coefficient of
$H_1^\dagger H_2$ is given by $Y_{12}=Y_3$ in the unprimed Higgs basis
and $Y'_{12}=Y'_3$ in the primed Higgs basis.  Using \eq{tpseudo}, it
follows that
$Y'_{12}=Y_{12} e^{-i\chi}$, which is consistent with the matrix
transformation law $Y'=U_D Y U_D^\dagger$.

From the four complex pseudo-invariant coefficients, one can form
four independent real invariants $|Y_3|$, $|Z_{5,6,7}|$ and three
invariant relative phases $\arg(Y_3^2 Z_5^*)$, 
$\arg(Y_3 Z_6^*)$ and $\arg(Y_3 Z_7^*)$.
Including the six invariants of \eq{invariants}, we have 
therefore identified thirteen
independent invariant real degrees of freedom prior to imposing the
scalar potential minimum conditions.  \Eq{potmingeneric} then imposes three
additional conditions on the set of thirteen invariants\footnote{The 
second condition of
\eq{hbasismincond} is a complex equation that can be rewritten in
terms of invariants: $|Y_3|=\half|Z_6|v^2$ and $Y_3
Z_6^*=-\half|Z_6|^2 v^2$.}
\beq \label{hbasismincond}
Y_1=-\half Z_1 v^2\,,\qquad\qquad\qquad
Y_3=-\half Z_6 v^2\,.
\eeq
This leaves eleven independent real degrees of freedom 
(one of which is the vacuum expectation
value~$v=246$~GeV) that specify the 2HDM parameter space.

The doublet of scalar fields in the Higgs basis
can be parameterized as follows:
\beq
\label{hbasisfields}
H_1=\left(\begin{array}{c}
G^+ \\ {\frac{1}{\sqrt{2}}}\left(v+\varphi_1^0+iG^0\right)\end{array}
\right)\,,\qquad
H_2=\left(\begin{array}{c}
H^+ \\ {\frac{1}{\sqrt{2}}}\left(\varphi_2^0+ia^0\right)\end{array}
\right)\,,
\eeq
and the corresponding hermitian conjugated fields are likewise defined.
We identify $G^\pm$ as a charged Goldstone boson pair
and $G^0$ as the CP-odd neutral Goldstone boson.\footnote{The definite CP
property of the neutral Goldstone boson persists even if the Higgs Lagrangian
is CP-violating (either explicitly or spontaneously), 
as shown in \App{app:four}.}
In particular, the identification of $G^0=\sqrt{2}\,\Im\, H_1^0$
follows from the fact that we have
defined the Higgs basis [see \eqs{hbasisdef}{higgsvevs}]
such that $\vev{H_1^0}$ is real and non-negative.  Of the remaining fields,
$\varphi_1^0$ is a CP-even neutral scalar field,  
$\varphi_2^0$ and $a^0$ are states of indefinite CP quantum 
numbers,\footnote{The CP-properties of 
the neutral scalar fields (in the Higgs basis) can be determined
by studying the pattern of gauge boson/scalar boson couplings and the 
scalar self-couplings in the interaction Lagrangian (see \sect{sec:five}).
If the scalar potential is CP-conserving, then two orthogonal
linear combinations of $\varphi_2^0$ and $a^0$ can be found that are
eigenstates of CP.  
By an appropriate rephasing of $H_2$ 
(which corresponds to some particular choice among the possible Higgs
bases) such that all the coefficients of the scalar potential in the
Higgs basis are real, one can then identify $\varphi_2^0$ as a CP-even 
scalar field and $a^0$ as a CP-odd scalar field.  See \App{app:four}
for further details.}
and $H^\pm$ is the physical charged Higgs boson pair.
If the Higgs sector is CP-violating, then
$\varphi_1^0$, $\varphi_2^0$, and $a^0$ all mix to produce three
physical neutral Higgs mass-eigenstates of indefinite CP quantum numbers.  

\section{The physical Higgs mass-eigenstates}
\label{sec:four}

To determine the Higgs mass-eigenstates, one must examine the terms of
the scalar potential that are quadratic in the scalar fields (after
minimizing the scalar potential and defining shifted scalar
fields with zero vacuum expectation values).  
This procedure is carried out in Appendix~\ref{app:two} starting 
from a generic basis.  However, there is an advantage in     
performing the computation in the Higgs basis
since the corresponding scalar potential coefficients 
are
invariant or pseudo-invariant quantities [\eqst{hbasispot}{pseudoinvariants}].
This will allow us to identify U(2)-invariants
in the Higgs mass diagonalization procedure.

Thus, we proceed by inserting \eq{hbasis}
into \eq{genericpot} and examining the terms linear and quadratic in
the scalar fields.  The requirement that the coefficient of the linear term
vanishes corresponds to the scalar potential minimum conditions
[\eq{hbasismincond}].  These conditions are then used in the evaluation
of the coefficients of the terms quadratic in the fields.  One can easily
check that no quadratic terms involving the Goldstone boson fields survive
(as expected, since the Goldstone bosons are massless).  This confirms
our identification of the Goldstone fields in \eq{hbasisfields}.
The charged Higgs boson mass is also easily determined:
\beq \label{hplus}
m_{H^\pm}^2=Y_{2}+\half Z_3 v^2\,.
\eeq
The three remaining neutral fields mix, and
the resulting neutral Higgs
squared-mass matrix in the $\varphi_1^0$--$\varphi_2^0$--$a^0$ basis is:
\beq \label{matrix33}
\mathcal{M}=v^2\left( \begin{array}{ccc}
Z_1&\,\, \Re(Z_6) &\,\, -\Im(Z_6)\\
\Re(Z_6)  &\,\, \half\left[Z_3+Z_4+\Re(Z_5)\right]+Y_2/v^2 & \,\,
- \half \Im(Z_5)\\ -\Im(Z_6) &\,\, - \half \Im(Z_5) &\,\,
 \half\left[Z_3+Z_4-\Re(Z_5)\right]+Y_2/v^2\end{array}\right).
\eeq
Note that $\mathcal{M}$ depends implicitly on the choice of Higgs
basis [\eq{hbasischi}] via the
$\chi$-dependence of the pseudo-invariants $Z_5$ and $Z_6$.
Moreover, the real and imaginary parts of these pseudo-invariants mix
if $\chi$ is changed.  Thus, $\mathcal{M}$ does not
possess simple transformation
properties under arbitrary flavor-U(2) transformations.
Nevertheless, we demonstrate
below that the eigenvalues and normalized eigenvectors are
U(2)-invariant.  First, we compute the characteristic equation:
\beq \label{charpoly}
\det(\mathcal{M}-xI)=-x^3+\Tr(\mathcal{M})\,x^2
-\half\left[(\Tr\mathcal{M})^2-\Tr(\mathcal{M}^2)\right]x+
\det(\mathcal{M})\,,
\eeq
where $I$ is the $3\times 3$ identity matrix. [The coefficient of
$x$ in \eq{charpoly} is particular to $3\times 3$
matrices (see Fact 4.9.3 of \Ref{matrixref}).]
Explicitly,
\beqa
\Tr(\mathcal{M}) &=& 2Y_2+(Z_1+Z_3+Z_4)v^2\,,\nonumber\\
\Tr(\mathcal{M}^2) &=& Z_1^2 v^4
+\half v^4\left[(Z_3+Z_4)^2+|Z_5|^2+4|Z_6|^2\right]
+2Y_2[Y_2+(Z_3+Z_4)v^2]\,,\nonumber \\
\det(\mathcal{M}) &=& \quarter\left\{Z_1 v^6[(Z_3+Z_4)^2-|Z_5|^2]
-2v^4[2Y_2+(Z_3+Z_4)v^2]|Z_6|^2\right. \nonumber \\
&&\qquad\qquad\qquad \left. +4Y_2 Z_1 v^2[Y_2+(Z_3+Z_4)v^2]
+2v^6\Re(Z_5^* Z_6^2)\right\}\,.
\eeqa
Clearly, all the coefficients of the characteristic polynomial are
U(2)-invariant.  Since the roots of this polynomial are the
squared-masses of the physical Higgs bosons, it follows that the
physical Higgs masses are basis-independent as required.
Since $\mathcal{M}$ is a real symmetric matrix, the eigenvalues of
$\mathcal{M}$ are real.  However, if any of these eigenvalues are
negative, then the extremal solution of \eq{potmingeneric} with $v\neq 0$ 
is \textit{not} a minimum of the scalar potential.  
The requirements that
$\mhpm^2>0$ [\eq{hplus}] and the positivity of the squared-mass
eigenvalues of $\mathcal{M}$ provide basis-independent conditions for
the desired spontaneous symmetry breaking pattern specified by \eq{emvev}.

The real symmetric squared-mass matrix $\mathcal{M}$ can be diagonalized by
an orthogonal transformation
\beq \label{rmrt}
R\mathcal{M} R^T=\mathcal{M}_D\equiv {\rm diag}~(m_1^2\,,\,m_2^2\,,\,m_3^2)\,,
\eeq
where $RR^T=I$ and the $m_k^2$ are the eigenvalues of $\mathcal{M}$
[\textit{i.e.}, the roots of \eq{charpoly}].
A~convenient form for $R$ is:
\beqa \label{rmatrix}
R=R_{12}R_{13}R_{23} &=&\left( \begin{array}{ccc}
c_{12}\,\, &-s_{12}\quad &0\\
s_{12}\,\, &\phm c_{12}\quad &0\\
0\,\, &\phm 0\quad &1\end{array}\right)\left( \begin{array}{ccc}
c_{13}\quad &0\,\, &-s_{13}\\
0\quad & 1\,\,&\phm 0\\
s_{13}\quad &0\,\, &\phm c_{13}\end{array}\right) \left( \begin{array}{ccc}
1\quad &0\,\, &\phm 0\\
0\quad &c_{23}\,\, &-s_{23}\\
0\quad &s_{23}\,\, &\phm c_{23}\end{array}\right) \nonumber \\[10pt]
&=&
\left( \begin{array}{ccc}
c_{13}c_{12}\quad &-c_{23}s_{12}-c_{12}s_{13}s_{23}\quad &-c_{12}c_{23}s_{13}
+s_{12}s_{23}\\[6pt]
c_{13}s_{12}\quad &c_{12}c_{23}-s_{12}s_{13}s_{23}\quad
& -c_{23}s_{12}s_{13}-c_{12}s_{23}\\
s_{13}\quad &c_{13}s_{23}\quad &c_{13}c_{23}\end{array}\right)\,,
\eeqa
where $c_{ij}\equiv \cos\theta_{ij}$ and $s_{ij}\equiv\sin\theta_{ij}$.
Note that $\det R=1$, although we could have chosen an orthogonal
matrix with determinant equal to $-1$ by choosing $-R$ in place
of $R$.  In addition, if we take the range of the angles to be
$-\pi\leq\theta_{12}$, $\theta_{23}<\pi$ and $|\theta_{13}|\leq\pi/2$,
then we cover the complete parameter space of SO(3) matrices
(see p.~11 of \Ref{murnaghan}).  That is, we work in a convention
where $c_{13}\geq 0$.
However, this parameter space includes points that simply correspond
to the redefinition of two of the Higgs mass-eigenstate fields by
their negatives.  Thus, we may reduce the parameter
space further and define all Higgs mixing angles modulo~$\pi$.
We shall verify this assertion at the end of this section.

The neutral Higgs mass-eigenstates are denoted by $h_1$, $h_2$ and
$h_3$:
\beq \label{rotated}
\left( \begin{array}{c}
h_1\\ h_2\\h_3 \end{array}\right)=R \left(\begin{array}{c} \varphi_1^0\\
\varphi_2^0\\ a^0\end{array}\right)\,.
\eeq
It is often convenient to choose a convention for the mass
ordering of the $h_k$ such that $m_1\leq m_2\leq m_3$.

Since the mass-eigenstates $h_k$ do not depend on
the initial basis choice, they must be U(2)-invariant fields.
In order to present a formal proof of this assertion, we need to
determine the transformation properties of the elements of $R$ under
an arbitrary U(2) transformation.  In principle, these can be
determined from \eq{rmrt}, using the fact that the $m_k^2$ are
invariant quantities.  However, the form of
$\mathcal{M}$ is not especially convenient for this purpose as noted
below \eq{matrix33}.  This can be ameliorated by introducing 
the unitary matrix:
\beq
W=\left(\begin{array}{ccc} 1&\qquad 0&\qquad 0\\
0 &\qquad 1/\sqrt{2} &\qquad 1/\sqrt{2} \\
0  &\qquad -i/\sqrt{2} &\qquad i/\sqrt{2}\end{array}\right)\,,
\eeq
and rewriting \eq{rmrt} as
\beq \label{rwmrw}
(RW)(W^\dagger \mathcal{M}W)(RW)^\dagger=\mathcal{M}_D
={\rm diag}~(m_1^2\,,\,m_2^2\,,\,m_3^2)\,.
\eeq
A straightforward calculation yields:
\beqa
W^\dagger \mathcal{M}W &=& v^2\left(\begin{array}{ccc} Z_1 &\quad
\nicefrac{1}{\sqrt{2}}Z_6  &\quad \nicefrac{1}{\sqrt{2}}Z_6^* \\
 \nicefrac{1}{\sqrt{2}}Z_6^*  &\quad \half(Z_3+Z_4)+Y_2/v^2 &\quad
\half Z_5^*\\  \nicefrac{1}{\sqrt{2}}Z_6  &\quad \half Z_5 &\quad
 \half(Z_3+Z_4)+Y_2/v^2\end{array}\right)\,,\label{WMW}\\[12pt]
RW&=&\left(\begin{array}{ccc}q_{11} 
&\qquad  \nicefrac{1}{\sqrt{2}}q^*_{12}\,e^{i\theta_{23}}
&\qquad \nicefrac{1}{\sqrt{2}} q_{12}\,e^{-i\theta_{23}} \\[4pt]
q_{21} &\qquad  \nicefrac{1}{\sqrt{2}}q^*_{22}\,e^{i\theta_{23}}
&\qquad \nicefrac{1}{\sqrt{2}}q_{22}\,e^{-i\theta_{23}} \\[4pt]
q_{31} &\qquad  \nicefrac{1}{\sqrt{2}}q^*_{32}\,e^{i\theta_{23}}
&\qquad \nicefrac{1}{\sqrt{2}} q_{32}\,e^{-i\theta_{23}}
\end{array}\right)\,,\label{RW}
\eeqa
where
\beqa \label{qkldef}
\!\!\!\!\! q_{11}&=&c_{13}c_{12}\,,\qquad\qquad\qquad\,\,\,
q_{21}=c_{13}s_{12}\,,\qquad\qquad\qquad
q_{31}=s_{13}\,,\nonumber \\
\!\!\!\!\! q_{12}&=&-s_{12}-ic_{12}s_{13}\,,\qquad\quad\!
q_{22}= c_{12}-is_{12}s_{13}\,,\qquad\quad\,
q_{32}= ic_{13}\,.
\eeqa

The matrix $RW$ defined in \eq{RW} is unitary
and satisfies $\det RW=i$. Evaluating this determinant yields:
\beq \label{detrw}
\half\sum_{j,k,\ell=1}^3\,\epsilon_{jk\ell} q_{j1}\Im(q^*_{k2} q_{\ell 2})=1\,,
\eeq
while unitarity implies:
\beqa 
&&\qquad\qquad\qquad\quad
\Re\left(q_{k1} q_{\ell 1}^* + q_{k2}q_{\ell 2}^* \right)=\delta_{k\ell}\,,
\label{unitarity1}\\[6pt]
&&\sum_{k=1}^3\,|q_{k1}|^2=\half\sum_{k=1}^3\,|q_{k2}|^2=1\,,\qquad\qquad
\sum_{k=1}^3\,q_{k2}^{\,2}=\sum_{k=1}^3\,q_{k1}q_{k2}=0\,.\label{unitarity2}
\eeqa
These results can be used to prove the identity~\cite{cpcarlos}:
\beq \label{epsid}
q_{j1}=\half\sum_{k,\ell=1}^3\,\epsilon\ls{jk\ell}\,\Im(q^*_{k2}q_{\ell 2})\,.
\eeq

Since the matrix elements of
$W^\dagger\mathcal{M}W$ only involve invariants and pseudo-invariants,
we may use \eq{rwmrw} to determine the flavor-U(2)
transformation properties of $q_{k\ell}$ and $e^{i\theta_{23}}$.  
The resulting transformation laws are:
\beq \label{qtrans}
q_{k\ell}\to q_{k\ell}\,,\qquad {\rm and}\qquad
e^{i\theta_{23}}\to (\det U)^{-1}e^{i\theta_{23}}\,,
\eeq
under a U(2) transformation $U$.  That is, the $q_{k\ell}$ are
invariants, or equivalently
$\theta_{12}$ and $\theta_{13}$ (modulo $\pi$)
are U(2)-invariant angles, whereas
$e^{i\theta_{23}}$ is a pseudo-invariant.
\Eq{qtrans} is critical for the rest of the paper.  Finally, to show
that the Higgs mass-eigenstates are invariant fields, we rewrite
\eq{rotated} as
\beq \label{rotated2}
\left( \begin{array}{c}
h_1\\ h_2\\h_3 \end{array}\right)=RW \left(\begin{array}{c}
\sqrt{2}\,\Re H_1^0-v\\ H_2^0\\H_2^{0\,\dagger} \end{array}\right)\,.
\eeq
Since the $q_{k\ell}$, $H_1$ and 
the product $e^{i\theta_{23}}H_2$ are U(2)-invariant quantities, 
it follows that the $h_k$ are invariant fields.

The transformation laws given in 
\eqs{tpseudo}{qtrans} imply that the quantities 
$Z_5\,e^{-2i\theta_{23}}$, 
$Z_6\, e^{-i\theta_{23}}$ and $Z_7\, e^{-i\theta_{23}}$ 
are U(2)-invariant.  These combinations will appear
in the physical Higgs boson self-couplings of \sect{sec:five} and in the
expressions for the invariant mixing angles 
given in Appendix~\ref{app:three}.  With this in
mind, it is useful to rewrite the neutral Higgs mass diagonalization equation
[\eq{rmrt}] as follows.  With $R\equiv R_{12}R_{13}R_{23}$ given by
\eq{rmatrix}, 
\beq \label{mtilmatrix}
\widetilde{\mathcal{M}}\equiv R_{23}\mathcal{M}R_{23}^T=
v^2\left( \begin{array}{ccc}
Z_1&\,\, \Re(Z_6 \, e^{-i\theta_{23}}) &\,\, -\Im(Z_6 \, e^{-i\theta_{23}})\\
\Re(Z_6 e^{-i\theta_{23}}) &\,\,\Re(Z_5 \,e^{-2i\theta_{23}})+ A^2/v^2 & \,\,
- \half \Im(Z_5 \,e^{-2i\theta_{23}})\\ -\Im(Z_6 \,e^{-i\theta_{23}})
 &\,\, - \half \Im(Z_5\, e^{-2i\theta_{23}}) &\,\, A^2/v^2\end{array}\right).
\eeq
where 
${A}^2$ is defined by:
\beq \label{madef}
{A}^2\equiv Y_2+\half[Z_3+Z_4-\Re(Z_5 e^{-2i\theta_{23}})]v^2\,.
\eeq
The diagonal neutral Higgs squared-mass matrix is then given by:
\beq \label{diagtil}
\widetilde{R}\,\widetilde{\mathcal{M}}\,\widetilde{R}^T=\mathcal{M}_D={\rm
 diag}(m_1^2\,,\,m_2^2\,,\,m_3^2)\,, 
\eeq
where the diagonalizing matrix $\widetilde{R}\equiv R_{12}R_{13}$ 
depends only on $\theta_{12}$ and $\theta_{13}$:
\beq \label{rtil}
\widetilde R=\left(\begin{array}{ccc}c_{12}c_{13} & \quad -s_{12} &
\quad -c_{12}s_{13} \\ c_{13}s_{12} & \quad c_{12} & \quad-s_{12}s_{13}\\
s_{13} & \quad 0 & \quad c_{13}\end{array}\right)\,.
\eeq
\Eqst{mtilmatrix}{rtil} provide a manifestly U(2)-invariant 
squared-mass matrix
diagonalization, since the elements of  $\widetilde{R}$
and $\widetilde{\mathcal{M}}$ are invariant quantities.

In this section, all computations were carried out by first
transforming to the Higgs basis.  
The advantage of this procedure is that one can readily
identify the relevant invariant and pseudo-invariant quantities
involved in the determination of the Higgs mass-eigenstates.  We may now
combine \eqs{hbasisdef}{rotated2} to obtain explicit expressions for the 
Higgs mass-eigenstate fields $h_k$ in terms of the scalar fields in
the generic basis $\Phi_a$.  
Since these expressions do not depend
on the Higgs basis, one could have obtained the 
results for the Higgs mass-eigenstates directly without reference
to Higgs basis quantities.  In Appendix~\ref{app:two}, we present a
derivation starting from the generic basis, which produces the following
expressions for the Higgs mass-eigenstates (and the
Goldstone boson) in terms of the generic basis fields:
\beq \label{hmassinv}
h_k=\frac{1}{\sqrt{2}}\left[\overline\Phi_{\abar}\lsup{0\,\dagger}
(q_{k1} \widehat v_a+q_{k2}\widehat w_a e^{-i\theta_{23}})
+(q^*_{k1}\widehat v^*_{\abar}+q^*_{k2}\widehat w^*_{\abar}e^{i\theta_{23}})
\overline\Phi_a\lsup{0}\right]\,,
\eeq
for $k=1,\ldots,4$, where $h_4=G^0$.  The shifted neutral fields are defined
by $\overline\Phi_a\lsup{0}\equiv \Phi_a^0-v\widehat v_a/\sqrt{2}$ and
the $q_{k\ell}$ are defined for $k=1,2,3$ and $\ell=1,2$ by \eq{qkldef}.
To account for the Goldstone boson ($k=4$) we have also introduced:
$q_{41}=i$ and $q_{42}=0$.
For the reader's convenience, the explicit forms for the $q_{k\ell}$
are displayed in Table~\ref{tab1}.
\begin{table}[h!]
\centering
\caption{The U(2)-invariant quantities $q_{k\ell}$ are functions of the
the neutral Higgs mixing angles $\theta_{12}$ and $\theta_{13}$, where
$c_{ij}\equiv\cos\theta_{ij}$ and $s_{ij}\equiv\sin\theta_{ij}$.\\
\label{tab1}}
\begin{tabular}{|c||c|c|}\hline
$\phaa k\phaa $ &\phaa $q_{k1}\phaa $ & \phaa $q_{k2} \phaa $ \\ \hline
$1$ & $c_{12} c_{13}$ & $-s_{12}-ic_{12}s_{13}$ \\
$2$ & $s_{12} c_{13}$ & $c_{12}-is_{12}s_{13}$ \\
$3$ & $s_{13}$ & $ic_{13}$ \\
$4$ & $i$ & $0$ \\ \hline
\end{tabular}
\end{table}

\noindent
Since the $q_{k\ell}$ are U(2)-invariant and
$\widehat w_a e^{-i\theta_{23}}$ is a \textit{proper} vector under
U(2) transformations, it follows that
\eq{hmassinv} provides a
U(2)-invariant expression for the Higgs mass-eigenstates.
It is now a simple matter to invert \eq{hmassinv} to obtain
\beq \label{master}
\Phi_a=\left(\begin{array}{c}G^+\widehat v_a+H^+ \widehat w_a\\[6pt]
\displaystyle
\frac{v}{\sqrt{2}}\widehat v_a+\frac{1}{\sqrt{2}}\sum_{k=1}^4
\left(q_{k1}\widehat v_a+q_{k2}e^{-i\theta_{23}}\widehat w_a\right)h_k
\end{array}\right)\,,
\eeq
where $h_4\equiv G^0$.  The form of the charged upper component of $\Phi_a$
is a consequence of \eq{hbasis}.
The U(2)-covariant expression for $\Phi_a$ in terms of the 
Higgs mass-eigenstate scalar fields given by \eq{master}
is one of the central results of
this paper.
In \sects{sec:five}{sec:six}, we shall employ 
this result for $\Phi_a$ in the computation of
the Higgs couplings of the 2HDM.

Finally, we return to the question of the domains of the angles
$\theta_{ij}$.  We assume that $Z_6\equiv |Z_6|e^{i\theta_6}\neq 0$
(the special case of $Z_6=0$ is treated at the end of \App{app:three}).
Since $e^{-i\theta_{23}}$ is a pseudo-invariant, we prefer to deal with
the invariant angle $\phi$:
\beq \label{invang}
\phi\equiv\theta_6-\theta_{23}\,,\qquad {\rm where}
\qquad \theta_6\equiv\arg Z_6\,.
\eeq
As shown in Appendix~\ref{app:three}, the invariant angles
$\theta_{12}$, $\theta_{13}$ and
$\phi$ are determined modulo $\pi$ in terms of invariant combinations of
the scalar potential parameters.
This domain is smaller than the one defined by
$-\pi\leq\theta_{12}$, $\theta_{23}<\pi$ and $|\theta_{13}|\leq\pi/2$,
which covers the parameter space of SO(3) matrices.
Since the U(2)-invariant mass-eigenstate fields $h_k$ are real, one
can always choose to redefine any one of the $h_k$ by its negative.
Redefining two of the three Higgs fields $h_1$, $h_2$ and $h_3$
by their negatives\footnote{In order
to have an odd number of Higgs mass-eigenstates
redefined by their negatives, one would have to employ an orthogonal
Higgs mixing matrix with $\det~R=-1$.} is
equivalent to multiplying two of the rows of $R$ by $-1$.  
In particular,
\beqa
\hspace{-0.25in} \theta_{12}\to\theta_{12}\pm\pi &\Longrightarrow&
h_1\to -h_1~{\rm and}~ h_2\to -h_2  
\,, \label{signflip12} \\
\hspace{-0.25in} \phi\to\phi\pm\pi\,,
\quad\!\! \theta_{13}\to -\theta_{13}\,,\quad\!\!  
\theta_{12}\to \pm\pi-\theta_{12}
&\Longrightarrow&
 h_1\to -h_1~{\rm and}~h_3\to -h_3
\,,\label{signflip13}\\
\hspace{-0.25in} \theta_{13}\to \theta_{13}\pm\pi\,,\quad\!\!  
\theta_{12}\to -\theta_{12}
&\Longrightarrow&
 h_1\to -h_1~{\rm and}~h_3\to -h_3
\,,\label{signflip13p}\\
\hspace{-0.25in}\phi\to\phi\pm\pi\,,
\quad\!\! \theta_{13}\to -\theta_{13}\,,
\quad\!\! \theta_{12}\to -\theta_{12} &\Longrightarrow&
 h_2\to -h_2~{\rm and}~h_3\to -h_3
\,,\label{signflip23}\\
\hspace{-0.25in} \theta_{13}\to \theta_{13}\pm\pi\,,
\quad\!\! \theta_{12}\to \pm\pi-\theta_{12} &\Longrightarrow&
 h_2\to -h_2~{\rm and}~h_3\to -h_3
\,.\label{signflip23p}
\eeqa
This means that if we adopt a convention in which $c_{12}$,
$c_{13}$ and $\sin\phi$ are 
non-negative, with the angles defined modulo $\pi$,
then the sign of the Higgs mass-eigenstate fields
will be fixed.  

Given a choice of the overall sign
conventions of the neutral Higgs fields, the number of solutions for
the invariant angles $\theta_{12}$, $\theta_{13}$ and $\phi$ 
modulo $\pi$ are in one-to-one correspondence with the 
possible mass orderings of the $m_k$ (except at 
certain singular points of the parameter space\footnote{At singular
points of the parameter space corresponding to two (or three) mass-degenerate 
neutral Higgs bosons,
some (or all) of the invariant Higgs mixing angles are indeterminate.
An indeterminate invariant angle also arises in the case of $Z_6=0$
and $c_{13}=0$ as explained at the end of \App{app:three}.\label{fnmass}}).
For example,  note that
\beq \label{signfliph1h2} 
\hspace{-0.25in} 
\theta_{12}\to\theta_{12}\pm\pi/2 \Longrightarrow
 h_1\to \mp h_2~{\rm and}~ h_2\to \pm h_1\,.
\eeq
That is, two solutions for $\theta_{12}$ exist modulo $\pi$.
If $m_1< m_2$, then \eq{m2m1} implies that 
the solutions for $\theta_{12}$ and $\phi$
are correlated such that $s_{12}\cos\phi\geq 0$, and (for fixed $\phi$) 
only one $\theta_{12}$ solution
modulo~$\pi$ survives. 
The corresponding effects on the invariant angles that result from swapping
other pairs of neutral Higgs fields are highly
non-linear and cannot be simply exhibited in closed form.
Nevertheless, we can use the results of Appendix~\ref{app:three}
to conclude that for $m_{1,2}<m_3$
(in a convention where $\sin\phi\geq 0$),  \eq{phieq} yields
$s_{13}\leq 0$, and for $m_1<m_2<m_3$,
\eq{imz56f} implies that $\sin 2\theta_{56}\cos\phi\geq 0$, where
$\theta_{56}\equiv -\half\arg(Z_5^* Z_6^2)$.

The sign of the neutral Goldstone field is conventional, but is not
affected by the choice of Higgs mixing angles.  Finally, we note that
the charged fields $G^\pm$ and $H^\pm$ are complex.  \Eq{master}
implies that $G^\pm$ is an invariant field and $H^\pm$
is a pseudo-invariant field that transforms as:
\beq \label{hplustrans}
H^\pm\to (\det U)^{\pm 1}\,H^\pm
\eeq
with respect to U(2) transformations.
That is, once the Higgs Lagrangian is written in terms
the Higgs mass-eigenstates and the Goldstone bosons, one is still free
to rephase the charged fields.  By convention, we shall fix this phase
according to \eq{master}.

\section{Higgs couplings to bosons}
\label{sec:five}

We begin by computing the Higgs self-couplings in terms of
U(2)-invariant quantities.   First, we use \eq{master} to obtain:
\beqa \label{phiphi}
\Phi^\dagger_{\abar}\Phi_b &=& \half v^2 V_{b\abar}+
vh_k\left[V_{b\abar}\,\Re~q_{k1}+\half\left(\widehat v_b\widehat w^*_{\abar}
q^*_{k2}e^{i\theta_{23}}+\widehat v^*_{\abar}\widehat w_b
q_{k2}e^{-i\theta_{23}}\right)\right]\nonumber \\
&&+\half h_j h_k\left[V_{b\abar}\Re(q_{j1}^* q_{k1})+
W_{b\abar}\Re(q_{j2}^* q_{k2})
+\widehat v_b\widehat w^*_{\abar} q^*_{j_2}q_{k1}
e^{i\theta_{23}}+\widehat v^*_{\abar}\widehat w_b q^*_{j1}
q_{k2}e^{-i\theta_{23}}\right] \nonumber \\
&& +G^+G^- V_{b\abar}+H^+H^- W_{b\abar}+G^- H^+ \widehat
v^*_{\abar}\widehat w_b+ G^+ H^- \widehat w^*_{\abar}\widehat v_b\,,
\eeqa
where repeated indices are summed over and $j,k=1,\ldots,4$.
We then insert \eq{phiphi} into \eq{genericpot}, and expand out the
resulting expression.  We shall write:
\beq
\mathcal{V}=\mathcal{V}_0+\mathcal{V}_2+\mathcal{V}_3+\mathcal{V}_4\,,
\eeq
where the subscript indicates the overall degree of the fields that appears in
the polynomial expression.  $\mathcal{V}_0$ is a constant of no
significance and $\mathcal{V}_1=0$ by the scalar potential minimum condition.
$\mathcal{V}_2$ is obtained in Appendix~\ref{app:three}.  In this
section, we focus on the cubic Higgs self-couplings that reside in
$\mathcal{V}_3$ and the quartic Higgs self-couplings that reside in
$\mathcal{V}_4$.

Using \eqs{invariants}{pseudoinvariants}, one can
express $\mathcal{V}_3$ and $\mathcal{V}_4$ in terms of the invariants
($Y_1$, $Y_2$ and $Z_{1,2,3,4}$) and pseudo-invariants ($Y_3$, $Z_{5,6,7}$).
In the resulting expressions, we have eliminated $Y_1$ and $Y_3$ by the
scalar potential minimum conditions [\eq{hbasismincond}].
The cubic Higgs couplings are governed by the following terms of the
scalar potential:
\beqa \label{hcubic}
\mathcal{V}_3&=&\half v\, h_j h_k h_\ell
\biggl[q_{j1}q^*_{k1}\Re(q_{\ell 1}) Z_1
+q_{j2}q^*_{k2}\,\Re(q_{\ell 1})(Z_3+Z_4) +
\Re(q^*_{j1} q_{k2}q_{\ell 2}Z_5\,
e^{-2i\theta_{23}}) \nonumber \\
&&\qquad\qquad\qquad\quad
+\Re\left([2q_{j1}+q^*_{j1}]q^*_{k1}q_{\ell 2}Z_6\,e^{-i\theta_{23}}\right)
+\Re (q_{j2}^*q_{k2}q_{\ell 2}Z_7\,e^{-i\theta_{23}})
\biggr]\nonumber \\
&& \hspace{-0.2in} +v\,h_k
G^+G^-\biggl[\Re(q_{k1})Z_1+\Re(q_{k2}\,e^{-i\theta_{23}}Z_6)\biggr]
+v\,h_k H^+H^-\biggl[\Re(q_{k1})Z_3+\Re(q_{k2}\,e^{-i\theta_{23}}Z_7)\biggr]
\nonumber \\
&& \hspace{-0.2in} +\half v \,h_k\biggl\{G^-H^+\,e^{i\theta_{23}}
\left[q^*_{k2} Z_4
+q_{k2}\,e^{-2i\theta_{23}}Z_5+2\Re(q_{k1})Z_6
\,e^{-i\theta_{23}}\right]+{\rm h.c.}\biggr\}\,,
\eeqa
where there is an implicit sum over the repeated indices\footnote{Note that the
sum over repeated indices can be rewritten by appropriately symmetrizing
the relevant coefficients.  For example, $\sum_{jk\ell} g_{jk\ell}\,h_j h_k
h_\ell= \sum_{j\leq k\leq\ell} h_j h_k h_\ell\,[g_{jk\ell}+{\rm perm}]$,
where ``perm'' is an instruction to add additional terms (as needed) 
such that the indices $j$, $k$ and $\ell$ appear in all possible 
\textit{distinct} permutations.\label{fn1}}
$j$, $k$, $\ell=1,2,3,4$.  Since the neutral Goldstone boson field is
denoted by $h_4\equiv G^0$, we can extract the cubic couplings of
$G^0$ by using $q_{41}=i$ and $q_{42}=0$.  The only cubic
Higgs--$G^0$ couplings
that survive are:
\beqa \label{G3}
\mathcal{V}_{3G}&=&
\half v\,\sum_{k=1}^3\sum_{\ell =1}^3\,G^0
h_k h_\ell\biggl[\Im( q_{k2} q_{\ell 2}Z_5\,e^{-2i\theta_{23}})+
2q_{k1}\,\Im\left(q_{\ell 2} Z_6\,e^{-i\theta_{23}}\right)\biggr]
\nonumber \\
&& +\half v\,\sum_{\ell=1}^3\,G^0 G^0 h_\ell\biggl[q_{\ell 1}
Z_1+\Re(q_{\ell 2}Z_6e^{-i\theta_{23}})\biggr]\,,
\eeqa
where we have used the fact that $q_{j1}$ is real for $j=1,2,3$.

At the end of the \sect{sec:four}, we noted that $H^+$ is a
pseudo-invariant field.  However $e^{i\theta_{23}}H^+$ is a U(2)-invariant
field [see \eqs{qtrans}{hplustrans}], and it is precisely this
combination that shows up in \eq{hcubic}.  Moreover, 
as shown in \sect{sec:four}, the $q_{k\ell}$
and the quantities $Z_5\,e^{-2i\theta_{23}}$, $Z_6\,e^{-i\theta_{23}}$
and $Z_7\,e^{-i\theta_{23}}$ are also invariant with respect
to flavor-U(2) transformations.  Thus, we conclude that
\eq{hcubic} is U(2)-invariant as required.

The quartic Higgs couplings are governed by the following terms of the
scalar potential:
\beqa
&& \hspace{-0.18in}
\mathcal{V}_4=\eighth h_j h_k h_l h_m
\biggl[q_{j1}q_{k1}q^*_{\ell 1}q^*_{m1}Z_1
+q_{j2}q_{k2}q^*_{\ell 2}q^*_{m2}Z_2
+2q_{j1}q^*_{k1}q_{\ell 2}q^*_{m2}(Z_3+Z_4)\nonumber \\[5pt]
&&\quad\,\,
+2\Re(q^*_{j1}q^*_{k1}q_{\ell 2}q_{m2}Z_5\,e^{-2i\theta_{23}})
+4\Re(q_{j1}q^*_{k1}q^*_{\ell 1}q_{m2}Z_6\,e^{-i\theta_{23}})
+4\Re(q^*_{j1}q_{k2}q_{\ell
  2}q^*_{m2}Z_7\,e^{-i\theta_{23}})\biggr]\nonumber \\[5pt]
&&  +\half h_j h_k G^+ G^-\biggl[q_{j1}q^*_{k1} Z_1 + q_{j2}q^*_{k2}Z_3
+2\Re(q_{j1}q_{k2}Z_6\,e^{-i\theta_{23}})\biggr]
 \nonumber \\[5pt]
&&  +\half h_j h_k H^+ H^-\biggl[q_{j2}q^*_{k2} Z_2 + q_{j1}q^*_{k1}Z_3
+2\Re(q_{j1}q_{k2}Z_7\,e^{-i\theta_{23}})\biggr] \nonumber \\
&& +\half h_j h_k\biggl\{G^- H^+\,e^{i\theta_{23}} \left[q_{j1}q^*_{k2}Z_4
+ q^*_{j1}q_{k2}Z_5\,e^{-2i\theta_{23}}+q_{j1}q^*_{k1}Z_6
\,e^{-i\theta_{23}}+q_{j2}q^*_{k2}Z_7\,e^{-i\theta_{23}}\right]+{\rm h.c.}
\biggr\} \nonumber \\[5pt]
&&
 +\half Z_1 G^+ G^- G^+ G^- +\half Z_2 H^+H^- H^+ H^-
+ (Z_3+Z_4)G^+ G^- H^+ H^- +\half Z_5 H^+H^+G^-G^-
\nonumber \\[5pt]
&&
+\half Z_5^* H^-H^- G^+G^+
+ G^+G^-(Z_6 H^+ G^-\! + Z_6^* H^- G^+) + H^+H^-(Z_7 H^+ G^-\! + Z_7^* H^- G^+)
\,,\eeqa
where there is an implicit sum over the repeated indices
$j$, $k$, $\ell$, $m=1,2,3,4$.
One can check the U(2)-invariance of
$\mathcal{V}_4$ by noting that $Z_5 H^+ H^+$, $Z_6 H^+$ and $Z_7 H^+$ are
U(2)-invariant combinations.\footnote{It is instructive to
write, \textit{e.g.}, $Z_6 H^+ = (Z_6\,e^{-i\theta_{23}})
(H^+\,e^{i\theta_{23}})$, \textit{etc.} to exhibit the well-known
U(2)-invariant combinations.}
It is again straightforward to isolate the
quartic couplings of the neutral Goldstone boson ($h_4\equiv G^0$):
\beqa
&&\mathcal{V}_{4G}=\eighth q_{11}^4 Z_1 G^0 G^0 G^0 G^0
+\half \Im(q_{m2}Z_6\,e^{-i\theta_{23}})\,G^0 G^0 G^0 h_m
 \nonumber\\[5pt]
&&\,\,\,
+\quarter G^0 G^0 h_\ell h_m\biggl[q_{\ell 1}q_{m1}Z_1
+q_{\ell 2}q^*_{m2}(Z_3+Z_4)-\Re(q_{\ell 2} q_{m2} Z_5 e^{-2i\theta_{23}})
+2q_{\ell 1}\Re(q_{m2} Z_6 e^{-i\theta_{23}})\biggr] \nonumber \\[5pt]
&&\,\,\,
+\half G^0 h_k h_\ell h_m
\biggl[q_{k1}\Re(q_{\ell 2} q_{m2} Z_5 e^{-2i\theta_{23}})
+q_{k1} q_{\ell 1}\Re(q_{m2} Z_6 e^{-i\theta_{23}})
+\Re(q_{k2} q_{\ell 2}q^*_{m2} Z_7 e^{-i\theta_{23}})\biggr] \nonumber \\[5pt]
&&\,\,\,
- \Im(q_{m2} Z_6 \,e^{-i\theta_{23}})\,G^+G^-G^0 h_m
- \Im(q_{m2} Z_7 \,e^{-i\theta_{23}})\,H^+H^-G^0 h_m
 \nonumber \\[5pt]
&&\,\,\,
+\half i\,G^0 h_m \biggl\{G^-H^+e^{i\theta_{23}}\left[q^*_{m2}Z_4-
q_{m2}Z_5 e^{-2i\theta_{23}}\right]+{\rm h.c.}\biggr\}
\nonumber \\[5pt]
&&\,\,\,
+\half Z_1 G^0G^0G^+G^- +\half Z_3 G^0G^0H^+H^- +\half Z_6 G^0 G^0G^-H^+
+\half Z_6^* G^0 G^0G^+H^-\,,
\eeqa
where the repeated indices $k$, $\ell$, $m=1,2,3$ are summed over.

The Feynman rules are obtained by multiplying the relevant terms of
the scalar potential by $-iS$, where the symmetry factor 
$S=\prod_i n_i!$ for an the interaction term that possesses $n_i$ 
identical particles of type $i$.
Explicit forms for the $q_{k\ell}$ in terms of the invariant mixing angles
$\theta_{12}$ and $\theta_{13}$ are displayed in Table~\ref{tab1}.
For example, the Feynman rule for the cubic self-coupling of the lightest
neutral Higgs boson is given by $ig(h_1 h_1 h_1)$ where
\beqa
g(h_1 h_1 h_1)&=& -3v\biggl[Z_1 c_{12}^3 c_{13}^3
+(Z_3+Z_4)c_{12} c_{13} |s_{123}|^2
+c_{12} c_{13}\,\Re(s_{123}^2 Z_5\,e^{-2i\theta_{23}})
\nonumber \\ && \qquad
-3 c_{12}^2 c_{13}^2\,\Re(s_{123} Z_6\,e^{-i\theta_{23}})
-|s_{123}|^2\,\Re(s_{123} Z_7\, e^{-i\theta_{23}})
\biggr]\,,
\eeqa
where $s_{123}\equiv s_{12}+ic_{12}s_{13}$.
Similarly, the Feynman rule for the quartic self-coupling of the lightest
neutral Higgs boson is given by $ig(h_1 h_1 h_1 h_1)$ where
\beqa
g(h_1 h_1 h_1 h_1)&=& -3\biggl[Z_1 c_{12}^4 c_{13}^4 + Z_2|s_{123}|^4
+2(Z_3+Z_4)c_{12}^2 c_{13}^2 |s_{123}|^2
+ 2c_{12}^2 c_{13}^2\,\Re(s_{123}^2 Z_5\, e^{-2i\theta_{23}}) \nonumber \\
&& \qquad
- 4 c_{12}^3 c_{13}^3\,\Re(s_{123} Z_6\, e^{-i\theta_{23}})
-4 c_{12}c_{13}|s_{123}|^2\,\Re(s_{123}Z_7\, e^{-i\theta_{23}})
\biggr]\,.
\eeqa

We turn next to the coupling of the Higgs bosons to the gauge bosons.
These arise from the Higgs boson kinetic energy terms when the
partial derivatives are replaced by the gauge covariant derivatives:
$\mathscr{L}_{\rm KE}=D^\mu\Phi_{\abar}^\dagger D_\mu\Phi_a$.
In the SU(2)$\ls{\rm L}\times$U(1) electroweak gauge theory,
\beq \label{covder}
D_\mu\Phi_a=\left(\begin{array}{c} \displaystyle
\partial_\mu\Phi^+_a+\left[\frac{ig}{c_W}\left(\half-s_W^2\right)Z_\mu
+ieA_\mu\right]\Phi^+_a+\frac{ig}{\sqrt{2}}W_\mu^+\Phi^0_a \\[8pt]
\displaystyle \partial_\mu\Phi^0_a-\frac{ig}{2c_W}Z_\mu\Phi_a^0+
\frac{ig}{\sqrt{2}}W_\mu^-\Phi^+_a\end{array}\right)\,,
\eeq
where $s_W\equiv\sin\theta_W$ and $c_W\equiv\cos\theta_W$.  Inserting
\eq{covder} into $\mathscr{L}_{\rm KE}$ yields the Higgs boson--gauge boson
interactions in the generic basis.  Finally, we use \eq{master} to
obtain the interaction Lagrangian of the gauge bosons with the
physical Higgs boson mass-eigenstates.  The resulting interaction
terms are:
\beqa
\mathscr{L}_{VVH}&=&\left(gm_W W_\mu^+W^{\mu\,-}+\frac{g}{2c_W}
m_Z Z_\mu Z^\mu\right)\Re(q_{k1}) h_k \nonumber \\[5pt]
&&
+em_WA^\mu(W_\mu^+G^-+W_\mu^-G^+)
-gm_Zs_W^2 Z^\mu(W_\mu^+G^-+W_\mu^-G^+)
\,, \\[10pt]
\mathscr{L}_{VVHH}&=&\left[\quarter g^2  W_\mu^+W^{\mu\,-}
+\frac{g^2}{8c_W^2}Z_\mu Z^\mu\right]
\Re(q_{j1}^* q_{k1}+q_{j2}^* q_{k2})\,h_j h_k 
+\biggl[\half g^2 W_\mu^+ W^{\mu\,-}
\nonumber \\[5pt]
&& \qquad
+e^2A_\mu A^\mu+\frac{g^2}{c_W^2}\left(\half -s_W^2\right)^2Z_\mu Z^\mu
+\frac{2ge}{c_W}\left(\half -s_W^2\right)A_\mu Z^\mu\biggr](G^+G^-+H^+H^-)
\nonumber \\[5pt]
&&
+\biggl\{
\left(\half eg A^\mu W_\mu^+ -\frac{g^2s_W^2}{2c_W}Z^\mu W_\mu^+\right)
(q_{k1}G^-+q_{k2}\,e^{-i\theta_{23}}H^-)h_k +{\rm h.c.}\biggr\}
\,,
\eeqa
and
\beqa
\hspace{-1.3in}
\mathscr{L}_{VHH}=\frac{g}{4c_W}\,\Im(q_{j1}q^*_{k1}+q_{j2}q^*_{k2})
Z^\mu h_j\ddel_\mu h_k &&\nonumber \\[5pt]
&& \hspace{-2.5in}
-\half g\biggl\{iW_\mu^+\left[q_{k1} G^-\ddel\lsup{\,\mu} h_k+
q_{k2}e^{-i\theta_{23}}H^-\ddel\lsup{\,\mu} h_k\right]
+{\rm h.c.}\biggr\}\nonumber \\[5pt]
&& \hspace{-2.5in}
+\left[ieA^\mu+\frac{ig}{c_W}\left(\half -s_W^2\right)
Z^\mu\right](G^+\ddel_\mu G^-+H^+\ddel_\mu H^-)\,,\label{VHH}
\eeqa
where the repeated indices $j,k=1,\ldots,4$ are summed over.  The
neutral Goldstone boson interaction terms can be ascertained by taking
$h_4\equiv G^0$:
\beqa
\!\!\mathscr{L}_{VG}&\!=\!&\left[\quarter g^2  W_\mu^+W^{\mu\,-}
+\frac{g^2}{8c_W^2}Z_\mu Z^\mu\right]G^0 G^0
+\biggl\{
\half ieg A^\mu W_\mu^+ G^- G^0
-\frac{ig^2s_W^2}{2c_W}Z^\mu W_\mu^+
G^-G^0 +{\rm h.c.}\!\biggr\}
\nonumber \\[6pt]
&& 
+\frac{g}{2c_W} \Re(q_{k1}) Z^\mu G^0\ddel_\mu h_k
+\half g\left(W_\mu^+G^-\ddel\lsup{\,\mu}G^0+W_\mu^-G^+\ddel\lsup{\,\mu}G^0
\right)\,.
\eeqa

Once again, we can verify by inspection that the Higgs boson--vector
boson interactions are U(2)-invariant.  Moreover, one can derive
numerous relations among these couplings using the properties of the
$q_{k\ell}$.  In particular,  
\eqst{unitarity1}{epsid} imply the following relations among the 
Higgs boson--vector boson 
couplings~\cite{cpcarlos,Gunion:1997aq,Grzadkowski:1999ye}:
\beqa
&& g(ZZh_j)= m_Z\sum_{k,\ell=1}^3
\,\epsilon\ls{jk\ell}\, g(Zh_k h_\ell)\,,\qquad (j=1,2,3)\,,
\label{id1}\\[6pt]
&&\sum_{k=1}^3\,[g(VVh_k)]^2 = \frac{g^2 m_V^4}{m^2_W}\,,
\qquad\qquad\quad V=W^\pm~{\rm or}~Z\,,\label{id2}
\\[6pt]
&&\!\!\!\!\sum_{1\leq j<k\leq 3}
\,[g(Zh_j h_k)]^2 = \frac{g^2}{4c_W^2}\,,\label{id3}
\eeqa
\beq
g(ZZh_j)g(ZZh_k)+4m_Z^2\sum_{\ell=1}^3\,g(Zh_j h_\ell)g(Zh_k h_\ell)=
\frac{g^2 m_Z^2}{c_W^2}\,\delta_{jk}\,,\label{id4}
\eeq
where the Feynman rules for the $VVh_k$ and $Zh_j h_k$ vertices
are given by $ig^{\mu\nu}\,g(VVh_k)$ and $(p_k-p_j)^\mu\,g(Zh_j h_k)$,
respectively,
and the four-momenta $p_j$, $p_k$ of the neutral 
Higgs bosons $h_j$, $h_k$ point into the
vertex.\footnote{The Feynman rule for the $ZZh_k$ vertex 
includes a factor of two relative to the coefficient of the
corresponding term in $i\mathscr{L}_{VVH}$ due to the identical
$Z$ bosons. The Feynman rule for the
$Zh_j h_k$ vertex is given by $\half(g/c_W)\Im[q_{j1}q_{k1}^*+
q_{j2}q_{k2}^*](p_k-p_j)^\mu$.  Here, the factor of two relative to
the corresponding term in \eq{VHH} arises from the implicit double sum over
$j$ and $k$ in the Lagrangian.  Note that the rule for the $Zh_j h_k$
vertex does not depend on the ordering of $j$ and $k$.}
Note that \eq{id4} holds for $j,k=1,2,3,4$.

\section{Higgs couplings to fermions}
\label{sec:six}

The most general Yukawa
couplings of Higgs bosons to fermions yield neutral Higgs-mediated
flavor-changing neutral currents at
tree-level~\cite{Weinberg,Georgi}.  Typically, these
couplings are in conflict with the experimental bounds on FCNC
processes.  Thus, most model builders impose restrictions on the
structure of the Higgs fermion couplings to avoid the potential for
phenomenological disaster.  However, even in the case of the most general
Higgs-fermion couplings, parameter regimes exist where FCNC effects
are sufficiently under control.  In the absence of new physics beyond
the 2HDM, such parameter regimes are unnatural (but can be arranged
with fine-tuning).  In models such as the minimal supersymmetric
extension of the Standard Model (MSSM), supersymmetry-breaking effects
generate all possible  Higgs-fermion Yukawa couplings allowed by
electroweak gauge invariance.  Nevertheless, the FCNC effects are
one-loop suppressed and hence phenomenologically acceptable.

In this section, we will study the basis-independent description of
the Higgs-fermion interaction.  In a generic basis,
the so-called type-III model~\cite{typeiii,davidson} of Higgs fermion
interactions is governed by the following interaction Lagrangian:
\beq \label{ymodeliii0}
-\mathscr{L}_{\rm Y}
=\anti \qlo\, \wtil\Phi_1\eiuo\,  \uro +\anti Q_L^0\,\Phi_1(\eido)^\dagger
\,\dro + \anti \qlo\, \wtil\Phi_2\eiiuo\, \uro 
+\anti \qlo\, \Phi_2(\eiido)^\dagger \,\dro +{\rm h.c.}\,,
\eeq
where $\Phi_{1,2}$ are the Higgs doublets, $\wtil\Phi_i\equiv
i\sigma_2 \Phi^*_i$,
$\qlo $ is the weak isospin quark doublet,
and $\uro$, $\dro$ are weak isospin quark singlets.
[The right and left-handed fermion fields are defined as usual:
$\psi_{R,L}\equiv P_{R,L}\psi$, where $P_{R,L}\equiv \half(1\pm\gamma_5)$.]
Here, $\qlo $, $\uro $,
$\dro $ denote the interaction basis quark fields, which
are vectors in the quark
flavor space, and $\eta_1^{Q,0}$ and $\eta_2^{Q,0}$ ($Q=U\,,\,D$)
are four $3\times 3$ matrices in quark flavor space.
We have omitted the leptonic couplings in \eq{ymodeliii0};
these are obtained from \eq{ymodeliii0}
with the obvious substitutions $Q_L^0\to L_L^0$ and $D_R^0\to E_R^0$.  
(In the absence of right-handed neutrinos, there is no analog of $U_R^0$.)

The derivation of the couplings of the physical Higgs bosons with the
quark mass-eigenstates was given in \Ref{davidson} in the case of a
CP-conserving Higgs sector.  Here, we generalize that discussion to
the more general case of a CP-violating Higgs sector.  The first step
is to identify the quark mass-eigenstates.  This is accomplished by
setting the scalar fields to their vacuum expectation values and
performing unitary transformations of the left and right-handed up and
down quark multiplets such that the resulting quark mass matrices
are diagonal with non-negative entries.  In more detail, we define
left-handed and right-handed quark mass-eigenstate fields:
\beqa \label{biunitary}
&& P_L U=V_L^U P_L U^0\,,\qquad P_R U=V_R^U P_R U^0\,,\nonumber \\
&& P_L D=V_L^D P_L D^0\,,\qquad P_R D=V_R^D P_R D^0\,,
\eeqa
and the Cabibbo-Kobayashi-Maskawa (CKM) matrix is defined by $K\equiv V_L^U
V_L^{D\,\dagger}$.  In addition, we introduce ``rotated'' Yukawa coupling
matrices:
\beq
\eiua\equiv V_L^U \,\eiuoa\, V_R^{U\,\dagger}\,,\qquad\qquad
{\eida}\equiv V_R^D\, {\eidoa}\, V_L^{D\,\dagger}\,,
\eeq
(note the different ordering of $V_L^Q$ and $V_R^Q$ 
in the definitions of $\eta_a^Q$ for $Q=U$, $D$). We then
rewrite \eq{ymodeliii0} in terms of the quark mass-eigenstate
fields and the transformed couplings:
\beq \label{ymodeliii}
-\mathscr{L}_{\rm Y}=\anti Q_L \wtil\Phi_{\abar}\eta^U_a \ur
+\anti Q_L\Phi_a \eta^{D\,\dagger}_{\abar}\dr +{\rm h.c.}\,,
\eeq
where $\eta^Q_a\equiv (\eta^Q_1\,,\,\eta^Q_2)$ is a vector
with respect to the Higgs flavor-U(2) space.  Under a
U(2)-transformation of the scalar fields, $\eta^Q_a\to
U_{a\bbar}\eta^Q_b$ and 
$\eta^{Q\,\dagger}_\abar\to\eta^{Q\,\dagger}_\bbar U^\dagger_{b\abar}$. 
Hence, the Higgs--quark Lagrangian is
U(2)-invariant.  We can construct basis-independent
couplings following the strategy of \sect{sec:three}
by transforming to the Higgs basis.  Using \eq{hbasis}, we can rewrite
\eq{ymodeliii} in terms of Higgs basis scalar fields:
\beq \label{hbasisymodeliii}
-\mathscr{L}_{\rm Y}=\anti \ql (\wtil H_1\kappa^U+\wtil H_2\rho^U) \ur
+\anti Q_L (H_1\kappa^{D\,\dagger}+ H_2\rho^{D\,\dagger}) \dr +{\rm h.c.}\,,
\eeq
where
\beq \label{kapparho}
\kappa^{Q}\equiv \widehat v^*_{\abar}\,\eta^{Q}_a\,,\qquad\qquad
\rho^{Q}\equiv \widehat w^*_{\abar}\,\eta^{Q}_a\,.
\eeq
Inverting \eq{kapparho} yields:
\beq \label{kapparhoinv}
\eta^Q_a=\kappa^Q\widehat v_a+\rho^Q\widehat w_a\,.
\eeq
Under a U(2) transformation, $\kappa^Q$ is invariant, whereas $\rho^Q$
is a pseudo-invariant that transforms as:
\beq \label{rhotrans}
\rho^Q\to (\det U)\rho^Q\,.
\eeq
By construction, $\kappa^U$ and $\kappa^D$ are proportional to the
(real non-negative) diagonal quark mass matrices $M_U$ and $M_D$,
respectively.  In particular, the $M_Q$ 
are obtained by inserting \eq{higgsvevs} into
\eq{hbasisymodeliii}, which yields:
\beqa
M_U=\frac{v}{\sqrt{2}}\kappa^U&=&{\rm diag}(m_u\,,\,m_c\,,\,m_t)
= V_L^U M_U^{0} V_R^{U\,\dagger}\,, \label{diagumass}\\[6pt] 
M_D=\frac{v}{\sqrt{2}}\kappa^{D\,\dagger}&=&{\rm diag}(m_d\,,\,m_s\,,\,m_b)
= V_L^D M_D^{0} V_R^{D\,\dagger}\,, \label{diagdmass}
\eeqa
where $M_U^0\equiv (v/\sqrt{2}) \widehat v^*_{\abar}\,\eta^{U,0}_a$
and $M_D^0\equiv (v/\sqrt{2}) \widehat v_a\,\eta^{D,0\,\dagger}_{\abar}$.
That is, we have chosen the unitary matrices $V^U_L$, $V^U_R$,
$V^D_L$ and $V^D_R$ 
such that $M_D$ and $M_U$ are diagonal matrices with
real non-negative entries.\footnote{This can be
accomplished by the singular-value decompositions of the
complex matrices $M_U^0$ and $M_D^0$~\cite{horn}.}
In contrast, the $\rho^Q$ are independent complex $3\times 3$
matrices.

In order to obtain the interactions of the physical Higgs bosons with
the quark mass-eigenstates, we do not require the intermediate step
involving the Higgs
basis.  Instead, we insert \eq{master} into \eq{ymodeliii} and obtain:
\beqa \label{hffu2}
&& \hspace{-0.5in}
-\mathscr{L}_Y = \frac{1}{v}\overline D
\biggl\{M_D (q_{k1} P_R + q^*_{k1} P_L)+\frac{v}{\sqrt{2}}
\left[q_{k2}\,[e^{i\theta_{23}}\rho^D]^\dagger P_R+
q^*_{k2}\,e^{i\theta_{23}}\rho^D P_L\right]\biggr\}Dh_k \nonumber \\[5pt]
&&\quad  \hspace{-0.2in}
+\frac{1}{v}\overline U
\biggl\{M_U (q_{k1} P_L + q^*_{k1} P_R)+\frac{v}{\sqrt{2}}
\left[q^*_{k2}\,e^{i\theta_{23}}\rho^U P_R+
q_{k2}\,[e^{i\theta_{23}}\rho^U]^\dagger P_L\right]\biggr\}U h_k
\nonumber \\[5pt]
&&\quad \hspace{-0.3in}
+\biggl\{\overline U\left[K[\rho^D]^\dagger 
P_R-[\rho^U]^\dagger KP_L\right] DH^+
+\frac{\sqrt{2}}{v}\,\overline U\left[K\mdd P_R-\mud KP_L\right] DG^+
+{\rm h.c.}\biggr\}\,,
\eeqa
where $k=1,\ldots\,4$.  
Since $e^{i\theta_{23}}\rho^Q$ and $[\rho^Q]^\dagger H^+$ are U(2)-invariant,
it follows that
\eq{hffu2} is a basis-independent representation of the Higgs--quark
interactions.

The neutral Goldstone boson interactions ($h_4\equiv G^0$)
are easily isolated:
\beq \label{YG}
-\mathscr{L}_{YG}=\frac{i}{v}\left[\overline DM_D\gamma\ls{5}D
-\overline UM_U\gamma\ls{5}U\right]G^0\,.
\eeq
In addition, since the $q_{k1}$ are real for $k=1,2,3$, it 
follows that the piece of 
the neutral Higgs--quark couplings proportional to the quark mass
matrix is of the form $v^{-1}\overline Q \,M_Q \,q_{k1}\, Q\,h_k$.

The couplings of the neutral Higgs bosons to quark pairs
are generically CP-violating as a result
of the complexity of the $q_{k2}$ and the fact that the matrices
$e^{i\theta_{23}}\rho^Q$ are not generally hermitian or
anti-hermitian.  (Invariant conditions for the CP-invariance of these couplings
are given in \App{app:four}).  
\Eq{hffu2} also exhibits Higgs-mediated
FCNCs at tree-level by virtue of the fact that the $\rho^Q$ are not
flavor-diagonal.  Thus, for a phenomenologically
acceptable theory, the off-diagonal elements of $\rho^Q$ must be
small.

\section{The significance of $\mathbold{\tan\beta}$}
\label{sec:seven}

In \sects{sec:five}{sec:six}, we have written out the entire
interaction Lagrangian for the Higgs bosons of the 2HDM.  Yet, the
famous parameter $\tan\beta$, given by $\tan\beta\equiv v_2/v_1$
in a generic basis [see \eq{emvev}], does not appear in any
physical Higgs (or Goldstone) boson 
coupling.  This is rather surprising given the
large literature of 2HDM phenomenology in which the parameter
$\tan\beta$ is ubiquitous.  For example, 
numerous methods have been proposed for
measuring $\tan\beta$ at future 
colliders~\cite{ilctanb1,tanbprecision1,ilctanb2,ilctanb3,boosetal,plctanb1,plctanb2,lhctanb}.
In a generic basis, one can also define
the relative phase of the two vacuum expectation values,
$\xi={\rm arg}~(v_2 v_1^*)$.  However, neither $\tanb$ nor $\xi$
are basis-independent.  One can remove $\xi$ by rephasing one
of the two Higgs doublet fields, and both $\xi$ and
$\tan\beta$ can be removed
entirely by transforming to the Higgs basis.  Thus, in a
general 2HDM, $\tan\beta$ is an unphysical parameter with no 
significance \textit{a priori}.

The true significance of $\tanb$ emerges only in specialized
versions of the 2HDM, where
$\tan\beta$ is promoted to a
physical parameter.  As noted in
\sect{sec:six}, the general 2HDM generally
predicts FCNCs in conflict with experimental data.  One way to avoid
this phenomenological problem is to constrain the theoretical
structure of the 2HDM.  Such constraints often pick out a preferred
basis.  Relative to that basis, $\tan\beta$ is then a meaningful
parameter.

The most common 2HDM constraint is the requirement that
some of the Higgs-fermion Yukawa couplings vanish in a
``preferred'' basis.
This leads to the well known type-I and type-II 2HDMs~\cite{hallwise}
(henceforth
called 2HDM-I and 2HDM-II).  In the 2HDM-I, there exists a 
preferred basis where
$\eta^U_2=\eta^D_2=0$~\cite{type1,hallwise}.
In the 2HDM-II, there exists a preferred basis where
$\eta^U_1=\eta^D_2=0$~\cite{type2,hallwise}.
These conditions can be enforced by a suitable
symmetry.  For example, the MSSM possesses a type-II Higgs-fermion
interaction, in which case the supersymmetry guarantees
that $\eta^U_1=\eta^D_2=0$.  In non-supersymmetric models, appropriate
discrete symmetries can be found to enforce the type-I or type-II
Higgs-fermion couplings.\footnote{These discrete symmetries also
imply that some of the coefficients of the scalar potential must also
vanish in the same preferred basis~\cite{type1,type2,hallwise,lavoura2}.}

The conditions for type-I and type-II Higgs-fermion interactions given
above are basis-dependent.  But, there is also a basis-independent
criterion that was first given in \Ref{davidson}:\footnote{In this
paper, we have slightly modified our definition of the Yukawa
coupling.  What is called $\eta^D$ in \Ref{davidson} is called
$\eta^{D\,\dagger}$ here.}
\beqa
\epsilon_{\abar\bbar}\eta^D_a\eta^U_b=
\epsilon_{ab}\eta^{D\,\dagger}_{\abar}\eta^{U\,\dagger}_{\bbar}&=&0\,,
\qquad \hbox{\rm type-I}
\,,\label{typecond1} \\
\delta_{a\bbar}\,\eta^{D\,\dagger}_{\abar}\eta^U_b &=& 0
\,,\qquad \hbox{\rm type-II}\,.
\label{typecond2}
\eeqa

We can now prove that $\tan\beta$ is a physical parameter in the 2HDM-II
(we leave the corresponding analysis for the 2HDM-I to the 
reader\footnote{\Eq{typecond1} involves pseudo-invariant quantities.
Nevertheless, setting these quantities to zero yields a U(2)-invariant
condition.}).
In the preferred basis where $\eta^U_1=\eta^D_2=0$, we shall denote:
$\widehat v=e^{i\eta}(\cos\beta\,,\,
\sin\beta\,e^{i\xi})$ and $\widehat w=e^{-i\eta}(-\sin\beta
e^{-i\xi}\,,\,\cos\beta)$.  Evaluating $\kappa^Q=\widehat v^*\cdot\eta^Q$
and $\rho^Q=\widehat w^*\cdot\eta^Q$ in the
preferred basis, and recalling that the $\kappa^Q$ are diagonal 
real matrices, it follows that:
\beq  \label{modeliitanbdefs}
\mathbold{\it I}e^{-i(\xi+2\eta)}\tanb=-\rho^{D\,\dagger}(\kappa^D)^{-1}
=(\rho^U)^{-1}\kappa^U\,,
\eeq
where $\mathbold{\it I}$ is the identity matrix in quark flavor space
and $\kappa^Q=\sqrt{2}M_Q/v$ [see \eqs{diagumass}{diagdmass}].
These two definitions are consistent if
$\kappa^D\kappa^U+\rho^{D\,\dagger}\rho^U=0$ is satisfied. 
But the latter is equivalent to the type-II condition [which can be verified
by inserting \eq{kapparhoinv} into \eq{typecond2}].

To understand the phase factor that appears in \eq{modeliitanbdefs},
we note that only unitary matrices of the form
$U={\rm diag}(e^{i\chi_1}\,,\,e^{i\chi_2})$ that span a
U(1)$\times$U(1) subgroup of the flavor-U(2) group preserve the
the type-II conditions $\eta^U_1=\eta^D_2=0$ in the preferred basis.
Under transformations of this type, $\eta\to \eta+\chi_1$ and
$\xi\to\xi+\chi_2-\chi_1$.   Using \eq{rhotrans}, it follows that
$\rho^Q\to e^{i(\chi_1+\chi_2)}\rho^Q$ .  Hence $\rho^Q e^{-i(\xi+2\eta)}$
is invariant with respect to such U(1)$\times$U(1) transformations.
We conclude that \eq{modeliitanbdefs} is covariant with respect to
transformations that preserve the type-II condition.

The conditions specified in \eq{modeliitanbdefs} are quite
restrictive.  In particular, they determine the matrices
$\rho^Q$:
\beq \label{phfacts}
\rho^D e^{-i(\xi+2\eta)}=\frac{-\sqrt{2}M_D \tan\beta}{v}\,,\qquad
\rho^U e^{-i(\xi+2\eta)}=\frac{\sqrt{2}M_U \cot\beta}{v}\,.
\eeq
Up to an overall phase, $\rho^U$ and $\rho^D$ are real diagonal matrices
with non-negative entries.  There is also some interesting
information in the phase factors of \eq{phfacts}.  Although  the $\rho^Q$ are
pseudo-invariants, we have noted below
\eq{hffu2} that $e^{i\theta_{23}}\rho^Q$ is U(2)-invariant.
This means that the phase factor $e^{-i(\theta_{23}+\xi+2\eta)}$ is a physical
parameter.  
Moreover, we can now define $\tan\beta$ as a physical parameter of the 2HDM-II
as follows:
\beq
\tan\beta=\frac{v}{3\sqrt{2}}\,\bigl|\Tr\left(\rho^D M_D^{-1}\right)\bigr|\,,
\eeq
where $0\leq\beta\leq\pi/2$.  This is a manifestly
basis-independent definition, so $\tan\beta$ is indeed physical.

In Higgs studies at future colliders, suppose one encounters phenomena
that appear consistent with a 2HDM.  It may not be readily apparent
that there is any particular structure in the Higgs-fermion
interactions.  In particular, it could be that \eq{modeliitanbdefs} is
simply false.  A safe strategy is to always measure physical
quantities, which must be U(2)-invariant.  Here is a modest proposal,
assuming that the Yukawa couplings of the Higgs bosons to the third
generation fermions
dominate, in which case we can ignore the effects of the first
two generations.\footnote{This is probably not a bad assumption,
since $\kappa^Q$ is proportional to the quark mass 
matrix $M_Q$.}  In a one-generation model, one can introduce
three $\tanb$-like parameters
\beq \label{tanblike}
\tanb_b\equiv \frac{v}{\sqrt{2}}\frac{|\rho^D|}{m_b}\,,\qquad
\tanb_t\equiv  \frac{\sqrt{2}}{v}\frac{m_t}{|\rho^U|}\,,\qquad
\tanb_\tau\equiv \frac{v}{\sqrt{2}}\frac{|\rho^E|}{m_\tau}\,,
\eeq
where $\tan\beta_\tau$ is analogous to $\tan\beta_d$ and depends on
the third generation Higgs-lepton interaction.  In a
type-II model, one indeed has $\tanb_b=\tanb_t=\tanb_\tau=\tanb$.
In the more general
(type-III) 2HDM, there is no reason for the three parameters
above to coincide.  However, these three parameters are indeed
U(2)-invariant quantities, and thus correspond to physical observables
that can be measured in the laboratory.
The interpretation of these parameters is straightforward.  In the
Higgs basis, up and down-type quarks interact with both Higgs doublets.
But, clearly there exists some basis (\textit{i.e.}, a rotation by an angle
$\beta_t$ from the Higgs basis) for which only one of the two
up-type quark Yukawa couplings is non-vanishing.
This defines the physical angle $\beta_t$.  The interpretation of the
other two angles is similar.

Since the phase of $e^{i\theta_{23}}\rho^Q$ is a physical parameter,
one can generalize \eq{tanblike} by defining
\beq \label{tanblike2}
e^{i(\theta_{23}-\chi_b)}\tanb_b\equiv \frac{v}{\sqrt{2}}
\frac{\rho^{D\,*}}{m_b}\,,\qquad
e^{i(\theta_{23}-\chi_t)}\tanb_t\equiv \frac{\sqrt{2}}{v}
\frac{m_t}{\rho^U}\,,
\eeq
and similarly for $\tan\beta_\tau$.  Thus, in addition to three
$\tan\beta$-like parameters, there are three independent
physical phases $\chi_b$, $\chi_t$ and $\chi_\tau$
that could in principle be deduced from experiment.  Of course,
in the 2HDM-II,
one must have $\beta_b=\beta_t=\beta_\tau$ and
$\chi_b=\chi_t=\chi_\tau$.

A similar analysis can be presented for the case of the 2HDM-I.  In
this case, one is led to define slightly different $\tan\beta$-like
physical parameters.  But, these would be related to those defined in
\eq{tanblike} in a simple way.  A particular choice could be motivated
if one has evidence that that either the type-I or type-II conditions
are approximately satisfied.

We conclude this section by illustrating the utility of this approach
in the case of the MSSM.  This example has already been presented in
\Ref{davidson} in the case of a CP-conserving Higgs sector.  We
briefly explain how that analysis is generalized in the case of a
CP-violating Higgs sector.
The MSSM Higgs sector is a CP-conserving type-II 2HDM in
the limit of exact supersymmetry.  However, when supersymmetry breaking
effects are taken into account, loop corrections to the Higgs
potential and the Higgs-fermion interactions can lead to both
CP-violating effects in the Higgs sector, and the (radiative) generation of
the Higgs-fermion Yukawa couplings that are absent in the type-II
limit.  In particular, in the approximation that
supersymmetric masses are significantly
larger than $m_Z$, the effective Lagrangian that describes the
coupling of the Higgs bosons to the third generation quarks is given
(in the notation of \cite{Carena:2002es}) by:
\beq \label{leff}
-\call_{\rm eff}=
(h_b+\delta h_b)(\anti q_L\Phi_1) b_R
+ (h_t+\delta h_t)(\anti q_L \wtil\Phi_2) t_R 
+\Delta h_b\,(\anti q_L \Phi_2) b_R +\Delta h_t\,(\anti q_L \wtil\Phi_1) t_R
+{\rm h.c.}\,,
\eeq
where $\anti q_L\equiv (\anti u_L\,,\,\anti d_L)$.
Note that the terms proportional to $\Delta h_{b}$ and
$\Delta h_{t}$, which are absent in the tree-level MSSM, 
are generated at one-loop due to supersymmetry-breaking effects
Thus, we identify $\eta^{D}=((h_b+\delta h_b)^*\,,\,\Delta h_b^*)$
and $\eta^{U}=(\Delta h_t\,,\,h_t+\delta h_t)$.
The tree-level MSSM is CP-conserving, 
and $\xi=0$ in the supersymmetric basis.
At one-loop, CP-violating effects can shift $\xi$ away from
zero, and we shall denote this quantity by $\Delta\xi$.\footnote{In
practice, one would rephase the fields after computing the radiative
corrections.  But, since we are advocating basis-independent methods
in this paper, there is no need for us to do this.}
Evaluating $\kappa^Q=\widehat v^*\cdot\eta^Q$
and $\rho^Q=\widehat w^*\cdot\eta^Q$ as we did above \eq{modeliitanbdefs},
\beqa
\hspace{-0.3in}
e^{i\eta}\kappa^D
&=& c_\beta (h_b+\delta h_b)^*+e^{-i\Delta\xi}s_\beta(\Delta h_b)^*
\,,\quad\,\,
e^{-i\eta}\rho^D = -e^{i\Delta\xi} s_\beta (h_b+\delta h_b)^*
+c_\beta(\Delta h_b)^*,\\
\hspace{-0.3in}
e^{i\eta}\kappa^U
&=& c_\beta \Delta h_t+e^{-i\Delta\xi}s_\beta(h_t+\delta h_t)\,,\qquad\quad
\,
e^{-i\eta}\rho^U = -e^{i\Delta\xi} s_\beta \Delta h_t
+c_\beta(h_t+\delta h_t)\,.
\eeqa
By definition, the $\kappa^Q$ are real and non-negative, and related
to the top and bottom quark masses via \eqs{diagumass}{diagdmass}.
Thus, the tree-level relations between $m_b$, $m_t$ and $h_b$, $h_t$
respectively are modified~\cite{deltab}:\footnote{If 
one of the Higgs fields is rephased in order to
remove the phase $\Delta\xi$, then one simultaneously rephases
$\Delta h_{b,t}$ such that the quantities $\Delta h_{b,t} e^{i\Delta\xi}$ 
are invariant
with respect to the rephasing.  In particular,
$h_b$ and $h_t$ are not rephased, since these tree-level
quantities are always real and positive
and proportional to the tree-level values of $m_b$
and $m_t$, respectively.}
\beqa 
m_b=\frac{v\kappa^D}{\sqrt{2}}=
\frac{v \cb h_b}{\sqrt{2}}\left[1+\Re\left(\frac{\delta h_b}{h_b}+
\frac{\Delta h_b}{h_b}e^{i\Delta\xi}\tan\beta\right)\right]
\equiv \frac{v \cb h_b}{\sqrt{2}}
\left[1+\Re(\Delta_b)\right]\,,\label{modmassesb}\\[6pt]
m_t=\frac{v\kappa^U}{\sqrt{2}}=
\frac{v \sb h_t}{\sqrt{2}}\left[1+\Re\left(\frac{\delta h_t}{h_t}+
\frac{\Delta h_t}{h_t}e^{i\Delta\xi}\cot\beta\right)\right]
\equiv \frac{v \sb h_t}{\sqrt{2}}
\left[1+\Re(\Delta_t)\right]\,,\label{modmassest}
\eeqa
which define the complex quantities $\Delta_b$ and 
$\Delta_t$.\footnote{In deriving 
\eqs{modmassesb}{modmassest}, we computed $\kappa^Q=|\kappa^Q|$ by
expanding up to linear order
in the one-loop quantities $\Delta h_{b,t}$ and
$\delta h_{b,t}$.  Explicit
expressions for $\Delta_b$ and $\Delta_t$ in terms of supersymmetric masses and
parameters, and references to the original literature
can be found in \Ref{Carena:2002es}.}

\Eq{tanblike2} then yields:
\beqa
\tan\beta_b&=&\left|\frac{-e^{-i\Delta\xi} s_\beta (h_b+\delta h_b)
+c_\beta\Delta h_b}
{c_\beta (h_b+\delta h_b)+e^{i\Delta\xi}s_\beta\Delta h_b}\right|\,,\qquad
\,\chi_b=\theta_{23}+\psi_b+\eta\,,\\[6pt]
\tan\beta_t&=&\left|\frac{\cb\Delta h_t
+e^{-i\Delta\xi} s_\beta (h_t+\delta h_t)}
{-\Delta h_t\sb e^{i\Delta\xi}+c_\beta(h_t+\delta h_t)}\right|\,,\qquad
\quad\chi_t=\theta_{23}+\psi_t+\eta\,,
\eeqa
where $\psi_{t,b}\equiv\arg(e^{-i\eta}\rho^{U,D})$.
Expanding the numerators and denominators 
above and dropping terms of 
quadratic order in the one-loop quantities, we end up with
\beqa 
\tan\beta_b&=&\frac{\tan\beta}{1+\Re\,\Delta_b}
\left[1+\frac{1}{\sb^2}\,\Re\left(\frac{\delta h_b}{h_b}
-\cb^2\Delta_b\right)\right]\,,\label{tanbb} \\[8pt]
\cot\beta_t&=&\frac{\cot\beta}{1+\Re\,\Delta_t}
\left[1+\Re\left(\Delta_t-\frac{1}{\cb\sb}\frac{\Delta h_t}{h_t}
e^{i\Delta\xi}\right)\right]\,.\label{tanbt} 
\eeqa

We have chosen to write $\tan\beta_b/\tan\beta$ in terms of $\Delta_b$
and $\delta h_b/h_b$, and $\cot\beta_t/\cot\beta$ in terms of $\Delta_t$
and $\Delta h_t/h_t$ in order to emphasize 
the large $\tanb$ behavior of the deviations of these quantities from one.
In particular,
keeping only the leading $\tan\beta$-enhanced corrections, 
\eqs{modmassesb}{modmassest} imply that\footnote{Because the one-loop
corrections $\delta h_b$, $\Delta h_b$, $\delta h_t$ and $\Delta h_t$
depend only on Yukawa and gauge couplings and the supersymmetric
particle masses, they
contain no hidden $\tan\beta$ enhancements 
or suppressions~\cite{Carena:2001uj}.} 
\beq
\Delta_b\simeq e^{i\Delta\xi}\,\frac{\Delta h_b}{h_b}\,\tanb\,,\qquad\qquad
\Delta_t\simeq  \frac{\delta h_t}{h_t}\,.
\eeq
That is, the complex quantity $\Delta_b$ is $\tanb$-enhanced.
In typical models at large $\tanb$, the quantity $|\Delta_b|$ can be
of order 0.1 or larger and of either sign.  
Thus, keeping only the one-loop corrections that are
$\tan\beta$-enhanced,\footnote{In \Ref{davidson} the one-loop
$\tan\beta$-enhanced correction to $\cot\beta_t$ was incorrectly omitted.}
\beq
\tan\beta_b\,\simeq\,\frac{\tanb}{1+\Re\,\Delta_b}
\,,\qquad\quad
\cot\beta_t\simeq \cot\beta\left[
1-\tan\beta\,\Re\left(\frac{\Delta h_t}{h_t}e^{i\Delta\xi}\right)\right]\,.
\eeq
Thus, we have expressed the basis-independent quantities $\tan\beta_b$
and $\tan\beta_t$ in
terms of parameters that appear in the natural basis of the MSSM Higgs
sector.  Indeed, we find that  $\tan\beta_b\neq\tan\beta_t$ as a consequence of
supersymmetry-breaking loop-effects.

\section{Discussion and Conclusions}
\label{sec:eight}

In this paper, we have completed the theoretical development of the
basis-independent treatment of the two-Higgs doublet model (2HDM)
that was initiated in \Ref{davidson}.  In particular, we
focused on the construction of quantities that are invariant with
respect to global U(2) transformations of the form $\Phi_a\to
U_{a\bbar}\Phi_b$.  Such invariant quantities are basis-independent and
can therefore be associated with the physical parameters of the model.
We have also emphasized the utility of pseudo-invariant quantities that
are modified by a phase factor (equal to some integer power of $\det U$)
under U(2)-transformations.  Although such quantities are not
observables, any two of them can be combined to form an invariant
quantity.

The main accomplishment of this
paper was the treatment of the Higgs mass-eigenstates that allows for
the most general set of CP-violating Higgs couplings.  In this
most general case, three neutral Higgs states mix to yield three neutral
Higgs mass-eigenstates of indefinite CP.  The neutral Higgs sector is
parameterized by three physical masses and three mixing angles.
The masses are, of course, U(2)-invariant quantities.  We have
demonstrated how to define the three mixing angles such that two of the
three angles are invariant and one is a pseudo-invariant quantity.
We then identified the invariants that directly enter
the various Higgs couplings to bosons and fermions of the model.
The end result is a complete listing of all Higgs interactions in an
invariant basis-independent form.

Using the above results, we addressed the significance of the parameter
$\tan\beta$, which appears in many of the Higgs boson
Feynman rules in the generic-basis formulation of the 2HDM~\cite{hhg}.   
Since $\tan\beta$ is a basis-dependent quantity, the
appearance of $\tan\beta$ in the Higgs Feynman boson rules is an
illusion.  In fact, $\tan\beta$ is completely absent in the invariant
basis-independent form of the Feynman rules.  For example, we
demonstrated that in the one-generation model, the Higgs-fermion
Feynman rules depend on three separate $\tan\beta$-like parameters.
However, in contrast to $\tan\beta$, these three separate parameters are
U(2)-invariant quantities that depend on physical Higgs-fermion couplings.
If one imposes constraints on the Higgs Lagrangian such as a discrete
symmetry $\Phi_1\to\Phi_1$ and $\Phi_2\to -\Phi_2$ (in some basis), or
supersymmetry (which selects a preferred basis), then $\tan\beta$ is
promoted to a physical parameter.  The latter can then be explicitly
associated with an invariant quantity
of the basis-independent formalism.

With the basis-independent formalism now fully developed, it is now time
to begin to apply these ideas to the precision Higgs programs at 
future colliders.
Instead of studying how to make precision measurements of
$\tan\beta$ (which does not make sense in the general 2HDM context), one
should examine the potential for precision measurements of physical
[U(2)-invariant] parameters.  In precision studies of the
Higgs-fermion interactions, it ought to be possible to make 
measurements of the three $\tan\beta$-like parameters introduced
in \sect{sec:seven}.  Close to the decoupling 
limit~\cite{habernir,ghdecoupling}, the lightest neutral Higgs boson 
$h_1$ of the 2HDM is nearly 
indistinguishable from the Standard Model 
CP-even Higgs boson.  Consequently, the couplings of $h_1$ are
expected to be quite insensitive to $\tan\beta$.  To extract
experimental information on the $\tan\beta$-like parameters will
therefore require the observation of the heavier
Higgs bosons $h_2$, $h_3$ and $H^\pm$.

For example, if  
$h_2$, $h_3$ are kinematically accessible at the ILC,
then one can probe the values of $\tan\beta_f$ [$f=t$, $b$ and $\tau$]
by studying Higgs production (in both $e^+e^-$ and $\gamma\gamma$
collisions) and Higgs
decay processes that involve $b$-quarks, $t$-quarks and $\tau$-leptons.   
Studies of $b\bar b h_k$~\cite{ilctanb2,ilctanb3}, 
$t\bar t h_k$~\cite{ilctanb2}
and $\tau^+\tau^-$~\cite{plctanb2} production provide initial estimates for
the sensitivity to $\tan\beta$.   
The production of $t\bar bH^-$~\cite{ilctanb1,plctanb1},
exhibits dependence on both $\tan\beta_t$ and $\tan\beta_b$,
although the latter dependence dominates if $\tan\beta_b\gg 1$.  
The $\tan\beta$-like parameters can also be probed by precision
studies of the heavy Higgs boson decays to heavy fermion pairs.
In particular, observation of $H^\pm\to\tau^\pm\nu_\tau$ would provide
independent information on the value of $\tan\beta_\tau$.
Opportunities also exist to study the couplings of the
heavy Higgs bosons to fermion pairs at the LHC if
$\tan\beta$ is large~\cite{Carena:2002es}.  
In \Ref{lhctanb}, $gg\to b\bar b h_{2,3}$
followed by $h_{2,3}\to\tau^+\tau^-$ provides an excellent channel for
measuring $\tan\beta$.  A number of $\tan\beta$-enhanced
effects that govern various $b$-quark decay processes 
can also provide useful information on 
$\tan\beta_b$ and $\tan\beta_\ell$ ($\ell=\tau$ or $\mu$).
Perhaps the most sensitive process of this kind is 
the rare (one-loop induced) decay $B_s\to\mu^+\mu^-$,
whose rate is enhanced in the MSSM by a factor of 
$\tan^6\beta$~\cite{Dedes:2003kp}.  

It remains to be seen how effective the
processes outlined above are for
distinguishing among the various $\tan\beta$-like parameters.  
If the three $\tan\beta$-like parameters are found to be close in
value, this result would provide an important clue to possible constraints
underlying the theoretical structure of the 2HDM.  In particular,
small deviations among these parameters could be related to
new TeV-scale physics associated with the Higgs sector.  A particular
example of this in the context of the MSSM was given at the end of
\sect{sec:seven}.  The application of the methods
of this paper to the  study of precision measurements of the
$\tanb$-like parameters at future colliders will be
treated in more detail elsewhere.

The potential phenomenological implications of the flavor structure of
the full three-generation model has not yet been examined in a
comprehensive way.  Strictly speaking, in the three-generation model,
the three $\tan\beta$-like parameters mentioned above would have to be
replaced with a more complicated set of parameters that reflect the full
flavor structure of the Higgs-fermion interactions.  
Since the third-generation Higgs-fermion interactions are expected to
dominate, the one-generation results should provide a
reasonable first approximation.  However, the absence of large FCNC
phenomena imposes some significant restrictions on the Higgs-fermion
interactions of the most general 2HDM.  A basis-independent analysis of
these restrictions will be addressed in a separate publication.

In conclusion, the basis-independent formalism provides a powerful
approach for connecting physical observables that can be measured in the
laboratory with fundamental invariant parameters of the 2HDM.
This will permit the development of two-Higgs doublet model-independent
analyses of data in Higgs studies at the LHC, ILC and beyond.
Ultimately, if 2HDM phenomena are discovered at future colliders, such
analyses will provide the most general setting for identifying the
fundamental nature of the 2HDM dynamics.

\acknowledgments

This work was supported in part by the U.S. Department of Energy.
We are grateful for many illuminating discussions
with Sacha Davidson and Jack Gunion.
We have also benefited from conversations with
Jan Kalinowski, Maria Krawczyk and John Mason.  
Finally, HEH is grateful to Sabine
Kraml for encouraging us to pursue this work as part of the
CP and non-standard Higgs (CPNSH) study group activities at CERN.

\appendix

\section{The 2HDM scalar potential in a generic
basis} \label{app:one}

Let $\Phi_1$ and
$\Phi_2$ denote two complex hypercharge-one, SU(2)$\ls{\rm L}$ doublets 
of scalar fields.
The most general gauge-invariant scalar potential is given
by
\beqa  \label{pot}
\mathcal{V}&=& m_{11}^2\Phi_1^\dagger\Phi_1+m_{22}^2\Phi_2^\dagger\Phi_2
-[m_{12}^2\Phi_1^\dagger\Phi_2+{\rm h.c.}]\nonumber\\[6pt]
&&\quad +\half\lambda_1(\Phi_1^\dagger\Phi_1)^2
+\half\lambda_2(\Phi_2^\dagger\Phi_2)^2
+\lambda_3(\Phi_1^\dagger\Phi_1)(\Phi_2^\dagger\Phi_2)
+\lambda_4(\Phi_1^\dagger\Phi_2)(\Phi_2^\dagger\Phi_1)
\nonumber\\[6pt]
&&\quad +\left\{\half\lambda_5(\Phi_1^\dagger\Phi_2)^2
+\big[\lambda_6(\Phi_1^\dagger\Phi_1)
+\lambda_7(\Phi_2^\dagger\Phi_2)\big]
\Phi_1^\dagger\Phi_2+{\rm h.c.}\right\}\,,
\eeqa
where $m_{11}^2$, $m_{22}^2$, and $\lam_1,\cdots,\lam_4$ are real parameters.
In general, $m_{12}^2$, $\lambda_5$,
$\lambda_6$ and $\lambda_7$ are complex.  The form of \eq{pot} holds
for any generic choice of $\Phi_1$--$\Phi_2$ basis, whereas the coefficients
$m_{ij}^2$ and $\lambda_i$ are basis-dependent quantities.  Matching
\eq{pot} to the U(2)-covariant form of \eq{genericpot}, we identify:
\beq \label{ynum}
Y_{11}=m_{11}^2\,,\qquad\qquad
Y_{12}=Y_{21}^\ast=-m_{12}^2 \,,\qquad\qquad
Y_{22}=m_{22}^2\,,
\eeq
and
\beqa \label{znum}
&& Z_{1111}=\lam_1\,,\qquad\qquad \,\,\phantom{Z_{2222}=}
Z_{2222}=\lam_2\,,\nonumber\\
&& Z_{1122}=Z_{2211}=\lam_3\,,\qquad\qquad
Z_{1221}=Z_{2112}=\lam_4\,,\nonumber \\
&& Z_{1212}=\lam_5\,,\qquad\qquad \,\,\phantom{Z_{2222}=}
Z_{2121}=\lam_5^\ast\,,\nonumber\\
&& Z_{1112}=Z_{1211}=\lam_6\,,\qquad\qquad
Z_{1121}=Z_{2111}=\lam_6^\ast\,,\nonumber \\
&& Z_{2212}=Z_{1222}=\lam_7\,,\qquad\qquad
Z_{2221}=Z_{2122}=\lam_7^\ast\,.
\eeqa
Explicit formulae for the coefficients of the Higgs basis scalar potential
in terms of the corresponding coefficients of \eq{pot} in a generic
basis can be found in \Ref{davidson}.

\section{The neutral Higgs boson squared-mass matrix in a generic
basis} \label{app:two}

Starting from \eq{genericpot}, one can obtain the neutral Higgs
squared-mass matrix from the quadratic part of the scalar potential:
\beq \label{vpmp}
\mathcal{V}_{\rm mass}=\frac{1}{2}\left(\begin{array}{cc}
\Phi^0_a & \quad \Phi^{0\,\dagger}_{\bbar}\end{array}\right) \mathscr{M}^2
\left(\begin{array}{c} \Phi^{0\,\dagger}_{\cbar} \\[4pt]
\!\!\!\!\Phi^0_d\end{array}\right)\,.
\eeq
Thus, $\mathscr{M}^2$ is given by the following matrix of second
derivatives:
\beq \label{scriptm2}
\mathscr{M}^2={\left(\begin{array}{cc}  \displaystyle
\frac{\partial^2 \mathcal{V}}{\partial\Phi^0_a
\partial\Phi^{0\,\dagger}_{\cbar}} &
\qquad\displaystyle
\frac{\partial^2 \mathcal{V}}{\partial\Phi^0_a\partial\Phi^0_d} \\ \\
\displaystyle
\frac{\partial^2 \mathcal{V}}{\partial\Phi^{0\,\dagger}_{\bbar}
{\partial\Phi^{0\,\dagger}_{\cbar}}} &
\qquad\displaystyle
\frac{\partial^2 \mathcal{V}}
{\partial\Phi^{0\,\dagger}_{\bbar}\partial\Phi^0_d}
\end{array}\right)}_{\Phi^0_a=v_a}\,,
\eeq
where $v_a\equiv v\widehat v_a/\sqrt{2}$ and
$\widehat v_{\abar}^*\widehat v_a=1$.  With $\mathcal{V}$ given by
\eq{genericpot}, one finds:
\beq \label{genericm2}
\mathscr{M}^2=\left(\begin{array}{cc} (Y_{a\cbar})^*
+\half v^2\bigl[(Z_{a\cbar f\ebar}
+Z_{f\cbar a\ebar})\,\widehat v_e\widehat v^*_{\fbar}\bigr]^* &\qquad
\quarter v^2(Z_{e\abar f\dbar}+Z_{e\dbar f \abar})
\widehat v^*_{\ebar}\widehat v^*_{\fbar} \\ \\
\quarter v^2(Z_{b\ebar c\fbar}+Z_{c\ebar b \fbar})
\widehat v_e\widehat v_f & \qquad
Y_{b\dbar}+\half v^2(Z_{e\fbar b\dbar}+Z_{e\dbar b\fbar})
\widehat v^*_{\ebar}\widehat v_f \end{array}\right)\,.
\eeq
In deriving this result, we used the hermiticity properties of
$Y$ and $Z$ to rewrite the upper left hand block so that the indices
appear in the standard order for matrix multiplication in \eq{vpmp}.  
In addition, we employed:
\beq
\frac{\partial\Phi^0_e}{\partial\Phi^0_a}=\delta_{e\abar}\,,\qquad\qquad
\frac{\partial\Phi^{0\,\dagger}_{\fbar}}
{\partial\Phi^{0\,\dagger}_{\bbar}}=\delta_{b\fbar}\,.
\eeq

It is convenient to express the squared-mass matrix in terms of
(pseudo)-invariants.  To do this, we note that we can expand an
hermitian second-ranked tensor [which satisfies $A_{a\bbar}=(A_{b\abar})^*$]
in terms of the eigenvectors of
$V_{a\bbar}\equiv \widehat v_a \widehat v^*_{\bbar}$:
\beq \label{aab}
A_{a\bbar}=
\Tr(VA)V_{a\bbar}+\Tr(WA)W_{a\bbar}+\bigl[(\widehat v^*_{\cbar}
\widehat w_d A_{c\dbar})\widehat v_a \widehat w^*_{\bbar}+
(\widehat w^*_{\cbar}\widehat v_d
 A_{c\dbar})\widehat w_a \widehat v^*_{\bbar}\bigr]\,,
\eeq
where $W_{a\bbar}\equiv \widehat w_a\widehat
w^*_{\bbar}=\delta_{a\bbar}-V_{a\bbar}$.
Likewise, we can expand a second-ranked symmetric tensor
with two unbarred (or two barred indices), \textit{e.g.}, 
\beq
A_{ab}= (\widehat v^*_{\cbar}\widehat v^*_{\dbar}
A_{cd})\widehat v_a \widehat v_b +
(\widehat w^*_{\cbar}\widehat w^*_{\dbar}A_{cd})\widehat w_a \widehat w_b +
(\widehat v^*_{\cbar}\widehat w^*_{\dbar}A_{cd})
(\widehat v_a \widehat w_b + \widehat w_a \widehat v_b)\,.
\eeq
We can therefore rewrite the upper and lower right hand
$2\times 2$ blocks of the squared-mass matrix [\eq{genericm2}]
respectively as:
\beqa
[\mathscr{M}^2]_{\abar \dbar}&=&\half v^2\left[Z_1 v^*_{\abar} v^*_{\dbar}
+Z_5 w^*_{\abar} w^*_{\dbar} +Z_6(\widehat v^*_{\abar} \widehat w^*_{\dbar}
+ \widehat w^*_{\abar} \widehat v^*_{\dbar})\right]\,, \label{upright}\\[5pt]
[\mathscr{M}^2]_{b\dbar}
&=& (Y_1+Z_1 v^2)V_{b\dbar}
+[Y_2+\half (Z_3+Z_4)v^2]W_{b\dbar}+[(Y_3+Z_6 v^2)\widehat v_b
\widehat w^*_{\dbar} +(Y^*_3+Z^*_6 v^2)\widehat w_b
\widehat v^*_{\dbar} ]\,.\nonumber \\
&&\phantom{line}\label{downright}
\eeqa
The upper and lower left hand blocks are
then given by the hermitian adjoints of the
lower and upper right hand blocks, respectively.
Note that \eq{downright} can be
simplified further by eliminating $Y_1$ and
$Y_3$ using the scalar potential minimum conditions [\eq{hbasismincond}].

Let us apply this result to the Higgs bases, where $\widehat
v=(1\,,\,0)$ and $\widehat w=(0\,,\,1)$.  After imposing the
scalar potential minimum conditions,
\beq \label{hbasism2}
\mathscr{M}^2=\frac{1}{2} v^2\left(\begin{array}{cccc} Z_1 &\quad
    Z_6^* &\quad Z_1 &\quad Z_6 \\
  Z_6  &\quad Z_3+Z_4+2Y_2/v^2 &\quad  Z_6  &\quad
 Z_5  \\
 Z_1 &\quad    Z_6^* &\quad Z_1 &\quad Z_6\\
 Z_6^* &\quad Z_5^* &\quad Z_6^*
 &\quad Z_3+Z_4+2Y_2/v^2 \end{array}\right)\,.
\eeq
The massless Goldstone boson eigenvector
\beq
G^0=\frac{-i}{\sqrt{2}}\left(\begin{array}{c} 1\\\ \!0\\ \!\!-1\\ \,\,0
\end{array}
\right)\,,
\eeq
can be determined by inspection [the normalization factor is chosen
for consistency with \eq{hbasisfields}].
Thus, we can perform a (unitary) similarity
transformation on $\mathscr{M}^2$ to remove the Goldstone boson from
the neutral Higgs squared-mass matrix.  Explicitly, with the unitary matrix
\beq
V=\frac{1}{\sqrt{2}}\left(\begin{array}{cccc} 1 &\quad 0 &\quad \,\,0
& \quad -i\\  0 &\quad 1 &\quad -i & \quad \,\,\,0\\
 1 &\quad 0 &\quad \,\,0 & \quad \!\phm i\\
0 &\quad 1 &\quad \!\phm i & \quad
\,\,\,0\end{array}\right)\,,
\eeq
it follows from \eq{hbasism2} that:
\beq
V^\dagger \mathscr{M}^2 V=\left(\begin{array}{cc}\mathcal{M} &\quad 0
    \\
0 &\quad 0 \end{array}\right)\,,
\eeq
where $\mathcal{M}$ is the $3\times 3$ neutral Higgs squared-mass
matrix in the $\varphi_1^0$--$\varphi_2^0$--$a^0$ basis
obtained in \eq{matrix33}.

We diagonalize $\mathcal{M}$ as described in \sect{sec:four}.
The corresponding diagonalization of $\mathscr{M}^2$ is given by:
\beq \label{rvmvr}
\mathcal{D} \mathscr{M}^2\mathcal{D}^\dagger\equiv
\left(\begin{array}{cc} R &\quad 0\\ 0 &\quad 1\end{array}\right)
V^\dagger \mathscr{M}^2 V
\left(\begin{array}{cc} R^T &\quad 0\\ 0 &\quad 1\end{array}\right)
= \left(\begin{array}{cc}\mathcal{M}_D &\quad 0\\
0 &\quad 0 \end{array}\right)\,,
\eeq
where $\mathcal{M}_D= {\rm diag}~(m_1^2\,,\,m_2^2\,,\,m_3^2)$
and $m_k$ is the mass of the neutral Higgs mass-eigenstate $h_k$.
The diagonalizing matrix $\mathcal{D}$ is given by:
\beq
\mathcal{D}\equiv
\left(\begin{array}{cc} R &\quad 0\\ 0 &\quad 1\end{array}\right) V^\dagger
=\frac{1}{\sqrt{2}}\left(\begin{array}{cccc} d_{11} &\quad d_{12}
&\quad d^*_{11} &\quad d^*_{12} \\
 d_{21} &\quad d_{22}
&\quad d^*_{21} &\quad d^*_{22} \\ d_{31} &\quad d_{32}
&\quad d^*_{31} &\quad d^*_{32} \\ d_{41} &\quad d_{42}
&\quad d^*_{41} &\quad d^*_{42}\end{array}\right)\,,
\eeq
where
\beqa
\!\!\!\!\! d_{11}&=&c_{13}c_{12}\,,\qquad\qquad\,
d_{21}=c_{13}s_{12}\,,\qquad\qquad\,
d_{31}=s_{13}\,,\qquad\qquad\,\,\,\,
d_{41}=i\,,\nonumber \\
\!\!\!\!\! d_{12}&=&-s_{123}e^{-i\theta_{23}}\,,\qquad\!
d_{22}=c_{123}e^{-i\theta_{23}}\,,\qquad\quad\!
d_{32}= ic_{13}e^{-i\theta_{23}}\,,\qquad
d_{42}=0\,,
\eeqa
with the $c_{ij}$ and $s_{ij}$ defined in \eq{rmatrix} and
\beq
c_{123}\equiv c_{12}-is_{12}s_{13}\,,\qquad\quad s_{123}\equiv
s_{12}+ic_{12}s_{13}\,.
\eeq
Note that $\mathcal{D}$ is a unitary matrix and $\det\mathcal{D}=1$.
Unitarity implies that:
\beqa
&&\qquad\qquad\qquad\quad
\Re\left(d_{k1} d_{\ell 1}^* + d_{k2}d_{\ell 2}^* \right)=\delta_{k\ell}\,,
\label{dunitarity1} \\[6pt]
&&\half\sum_{k=1}^4\,|d_{k1}|^2=\half\sum_{k=1}^4\,|d_{k2}|^2=1\,,\qquad\qquad
\sum_{k=1}^4\,d_{k2}^{\,2}=\sum_{k=1}^4\,d_{k1}d_{k2}=0\,. \label{dunitarity2}
\eeqa
Noting that $d_{41}=i$ and $d_{42}=0$ [and using \eq{dtoq}], 
these equations reduce to
\eqs{unitarity1}{unitarity2} given in \sect{sec:four}.  In addition, 
${\rm det}~\mathcal{D}=-i\det RW =1$,
where $RW$ is given in \eq{RW}.  This yields an additional constraint
on the $d_{k\ell}$ [\textit{c.f.} \eq{detrw}].

The matrix $\mathcal{D}$ converts the neutral
Higgs basis fields into the neutral Higgs mass-eigenstates:
\beq \label{conversion}
\left(\begin{array}{c} h_1\\ h_2 \\ h_3 \\ G^0\end{array}\right)
=\mathcal{D}
\left(\begin{array}{c} \overline{H}\lsup{0\,\dagger}_1\\ H^{0\,\dagger}_2 \\
\overline{H}_1\lsup{0} \\ H^0_2\end{array}\right)\,,
\eeq
where $\overline{H}_1\lsup{0}\equiv H_1^0-v/\sqrt{2}$.

The mass-eigenstate fields do not depend on the choice of basis.
Using the fact that $H_1$ is invariant and $H_2$ is pseudo-invariant
with respect to flavor-U(2) transformations,
\eq{conversion} implies that the $d_{k1}$ are invariants whereas the
$d_{k2}$ are pseudo-invariants with the same transformation law as $H_2$
[\eq{h2pseudo}].  One can also check this directly from \eq{rvmvr},
using the fact that the physical Higgs masses must be basis-independent.
These results then imply that $\theta_{12}$ and $\theta_{13}$ are
invariant whereas $e^{i\theta_{23}}$ is a pseudo-invariant, \textit{i.e.},
$e^{i\theta_{23}}\to (\det U)^{-1} e^{i\theta_{23}}$
under an arbitrary flavor-U(2) transformation $U$.

Finally, using the results of this appendix, we can eliminate the
Higgs basis fields entirely and obtain the diagonalizing matrix that
converts the neutral Higgs fields in the generic basis into the
neutral Higgs mass-eigenstates:
\beq \label{gentomass}
\left(\begin{array}{c} h_1\\ h_2 \\ h_3 \\ G^0\end{array}\right)
=\mathcal{D}\left(\begin{array}{c}
\overline\Phi_{\bbar}\lsup{0\,\dagger}\widehat U\lsup{\dagger}_{b\abar} \\
\widehat U_{a\bbar}\overline\Phi_b\lsup{0}
\end{array}\right)\,,
\eeq
where $\overline\Phi_a\lsup{0}\equiv \Phi_a^0-v\widehat v_a/\sqrt{2}$ and
$\widehat U$ is the matrix
that converts the generic basis fields into the Higgs basis fields
[see \eq{ugenhiggs}].
\Eq{gentomass} then yields:
\beq
h_k=\frac{1}{\sqrt{2}}\left[\overline\Phi_{\abar}\lsup{0\,\dagger}
(d_{k1}\widehat v_a+d_{k2}\widehat w_a)
+(d^*_{k1}\widehat v^*_{\abar}+d^*_{k2}\widehat w^*_{\abar})
\overline\Phi_a\lsup{0}\right]\,,
\eeq
where $h_4\equiv G^0$.
Note that the U(2)-invariance of the $h_k$ imply
that the $d_{k1}$ are invariants and the $d_{k2}$ are pseudo-invariants
that transform oppositely to $\widehat w$ as $d_{k2}\to (\det
U)d_{k2}$ in agreement with the previous results above.  Indeed, it
is useful to define:
\beq \label{dtoq}
d_{k1}\equiv q_{k1}\,,\qquad {\rm and}\qquad
d_{k2}\equiv q_{k2} e^{-i\theta_{23}}\,,
\eeq
where \textit{all} the $q_{k\ell}$ are U(2)-invariant [see \eq{qtrans}].
In particular, $\widehat w_a e^{-i\theta_{23}}$ is a proper vector
with respect to flavor-U(2) transformations.
Hence,
\beq \label{apphk}
h_k=\frac{1}{\sqrt{2}}\left[\overline\Phi_{\abar}\lsup{0\,\dagger}
(q_{k1} \widehat v_a+q_{k2}\widehat w_a e^{-i\theta_{23}})
+(q^*_{k1}\widehat v^*_{\abar}+q^*_{k2}\widehat w^*_{\abar}e^{i\theta_{23}})
\overline\Phi_a\lsup{0}\right]
\eeq
provides an invariant expression for the neutral
Higgs mass-eigenstates.

\section{Explicit formulae for the neutral Higgs masses and mixing angles}
\label{app:three}

To obtain expressions for the neutral Higgs masses and mixing angles,
we insert \eq{master} into \eq{genericpot}, and expand out the
resulting expression, keeping only terms that are linear and quadratic
in the fields.  Using \eqs{invariants}{pseudoinvariants}, one can
express the resulting expression in terms of the invariants
($Y_1$, $Y_2$ and $Z_{1,2,3,4}$) and pseudo-invariants ($Y_3$, $Z_{5,6,7}$).
The terms linear in the fields vanish if the potential
minimum conditions [\eq{hbasismincond}] are satisfied.  We then
eliminate $Y_1$ and $Y_3$ from the expressions of the quadratic terms.
The result is:
\beqa \label{V2}
\mathcal{V}_2 &=& H^+H^-(Y_2+\half v^2 Z_3)+\half v^2 h_j h_k \biggl\{
Z_1 \Re(q_{j1})\Re(q_{k1})
+[\half(Z_3+Z_4)+Y_2/v^2]\Re(q_{j2}q^*_{k2}) \nonumber \\
&&\!\!\!\qquad\quad +\half\Re(Z_5 q_{j2}q_{k2}\,e^{-2i\theta_{23}})
+\Re(q_{j1})\Re(Z_6 q_{k2}\, e^{-i\theta_{23}})
+\Re(q_{k1})\Re(Z_6 q_{j2}\, e^{-i\theta_{23}})
\biggr\}\nonumber \\[6pt]
&=&
m_{H^\pm}^2 H^+H^- +\half \sum_k m_k^2 (h_k)^2 +\half v^2\sum_{j\neq k}
C_{jk}h_j h_k\,,
\eeqa
where there is an implicit sum over $j,k=1,\ldots,4$ (with $h_4\equiv G^0$).
However, since $q_{41}=i$ and $q_{42}=0$, it is clear that there are
no terms in \eq{V2} involving $G^0$.  Hence, we may restrict the sum
to run over $j,k=1,2,3$.  The charged Higgs mass obtained above
confirms the result quoted in \eq{hplus}.  The neutral Higgs boson
masses are given by:
\beq \label{hmassesinv}
m_k^2 = |q_{k2}|^2 A^2+ v^2\left[q_{k1}^2 Z_1
+\Re(q_{k2})\,\Re(q_{k2}Z_5\,e^{-2i\theta_{23}})
+2q_{k1}\Re(q_{k2}Z_6\, e^{-i\theta_{23}})\right]\,,
\eeq
where $A^2$ is defined in \eq{madef}.  It is often convenient to
assume that $m_1\leq m_2\leq m_3$.

Note that the right-hand side of \eq{hmassesinv} is manifestly
U(2)-invariant.  Moreover, by using \eqs{unitarity1}{unitarity2}, 
one finds that
the sum of the three neutral Higgs boson squared-masses is given by
\beq \label{tracesum}
{\rm Tr}~\mathcal{M}=\sum_k\,m_k^2=2Y_2+(Z_1+Z_3+Z_4)v^2\,,
\eeq
as expected. A more explicit form for the neutral Higgs squared-masses
than the one obtained in \eq{hmassesinv}
would require the solution of the cubic characteristic equation
[\eq{charpoly}].  Although an analytic solution can be found, it is too
complicated to be of much use (a numerical evaluation is more practical).

The coefficients $C_{jk}$ of \eq{V2} must vanish.  Since $C_{jk}$ is
symmetric under the interchange of its indices, the conditions $C_{jk}=0$
yield three independent equations that determine the two
mixing angles $\theta_{12}$ and $\theta_{13}$ and 
an invariant combination of $\theta_{23}$ and the phase of $Z_6$ (or
$Z_5$).  These three invariant angles are defined
modulo $\pi$ once a definite convention is established for the signs
of neutral Higgs mass-eigenstate fields (as discussed at the end
of \sect{sec:four}).  Unique solutions for the invariant angles within
this domain are obtained after a mass ordering for the three neutral 
Higgs bosons is specified (except at certain singular points of the 2HDM
parameter space as noted in footnote \ref{fnmass}). 

To determine explicit formulae for the invariant angles, we shall initially
assume that $Z_6\equiv |Z_6|e^{i\theta_6}\neq 0$ and define the
invariant angles $\phi$ and $\theta_{56}$:
\beq\label{app:invang}
\begin{array}{c}
\phm\phi\equiv\theta_6-\theta_{23}\,, \\[6pt]
\theta_{56}\equiv \theta_5-\theta_6\,,\end{array}
\quad\qquad {\rm where} \qquad \begin{cases}
\,\theta_6\equiv\arg Z_6\,, &\\
\,\theta_{5}\equiv\half\arg Z_5\,. & \end{cases}
\eeq
The factor of $1/2$ in the definition of $\theta_5$ has been inserted for
convenience.  As discussed in \sect{sec:four}, we can fix the
conventions for the overall signs of the $h_k$ fields by restricting 
the domain of $\theta_{12}$, $\theta_{13}$ and $\phi$ to the region:
\beq \label{app:domains}
-\pi/2\leq\theta_{12}\,,\,\theta_{13}<\pi/2\,,\qquad\quad
0\leq \phi<\pi\,.
\eeq

One can obtain more tractable equations for $\theta_{13}$ and $\phi$
by taking appropriate linear combinations of the $C_{jk}$:
\beqa
C_{23} c_{12}-C_{13} s_{12} &=& s_{13}\,\Re(Z_6\,e^{-i\theta_{23}})-
\half c_{13}\,\Im(Z_5\,e^{-2i\theta_{23}}) \label{c23c12a}\,,\\
C_{23} s_{12}+C_{13} c_{12} &=& \half(Z_1 - A^2/v^2) \sin 2\theta_{13}
-\cos 2\theta_{13}\,\Im(Z_6\,e^{-i\theta_{23}})\,.
\label{c23c12b}
\eeqa
Setting $C_{13}=C_{23}=0$ yields:
\beqa
\tan\theta_{13}&=&\frac{\Im(Z_5\,e^{-2i\theta_{23}})}
{2\,\Re(Z_6\,e^{-i\theta_{23}})}
\,, \label{tan13}\\[8pt]
\tan 2\theta_{13}&=&\frac{2\,\Im(Z_6\,e^{-i\theta_{23}})}
{Z_1-A^2/v^2}\,. \label{tan213}
\eeqa
Using the well known identity
$\tan 2\theta_{13}=2\tan\theta_{13}/(1-\tan^2\theta_{13})$,
one can use \eqs{tan13}{tan213} to eliminate $\theta_{13}$ and 
obtain an equation for $\phi$.%
\footnote{Recall that the
quantity $A^2$ [\eq{madef}] depends on $\phi$ via
$\Re(Z_5\,e^{-2i\theta_{23}})=|Z_5|\cos 2(\theta_{56}+\phi)$.}
The resulting equation for $\phi$ has more than one solution.
Plugging a given solution for $\phi$ back into \eq{tan13} yields
a corresponding solution for $\theta_{13}$. 
Note that if $(\theta_{13}\,,\,\phi)$ is a solution to
\eqs{tan13}{tan213}, then so is $(-\theta_{13}\,,\,\phi\pm\pi)$,
in agreement with \eqs{signflip13}{signflip23}.  
By restricting to the domain of $\theta_{13}$ and $\phi$ specified
by \eq{app:domains}, only one of these two solutions survives.
However, multiple solutions to \eqs{tan13}{tan213} still exist within the
allowed domain, which
correspond to different choices for the mass
ordering of the three neutral Higgs fields.  
By imposing a particular mass ordering, a unique
solution is selected [see \eqs{s13}{c13s12}].
  
Finally, having obtained $\phi$ and
$\tan\theta_{13}$, we use $C_{12}=0$ to compute $\theta_{12}$.  The
result is:
\beq \label{tan12}
\tan 2\theta_{12}=\frac{s_{13}\,\Im(Z_5\,e^{-2i\theta_{23}})+2c_{13}\,
\Re(Z_6\,e^{-i\theta_{23}})}{c_{13}^2\left(A^2/v^2-Z_1\right)
+\Re(Z_5\,e^{-2i\theta_{23}})-2s_{13} c_{13}\,
\Im(Z_6\,e^{-i\theta_{23}})}\,.
\eeq
We can simplify the above result by using \eqs{tan13}{tan213}
to solve for $\Im(Z_5\,e^{-2i\theta_{23}})$ and
$\Im(Z_6\,e^{-i\theta_{23}})$ and eliminate these factors from \eq{tan12}.
The end result is:
\beq \label{tan12f}
\tan 2\theta_{12}=\frac{2\cos 2\theta_{13}\,\Re(Z_6\,e^{-i\theta_{23}})}
{c_{13}\left[c_{13}^2(A^2/v^2-Z_1)+\cos 2\theta_{13}\,
\Re(Z_5\,e^{-2i\theta_{23}})\right]}\,.
\eeq
Note that if $\theta_{12}$ is a solution to \eq{tan12f}, then
$\theta_{12}\pm\pi/2$ is also a solution.  That is, \eq{tan12f}
yields two solutions for $\theta_{12}$ 
in the allowed domain [\eq{app:domains}], which
correspond to the two possible mass orderings of $h_1$ and $h_2$
as shown below \eq{m2m1}.

The neutral Higgs boson masses were given in \eq{hmassesinv}.
With the help of \eqthree{tan13}{tan213}{tan12f}, one can 
can express these masses in terms of 
$Z_1$, $Z_6$ and the invariant angles:
\beqa
m_1^2 &=&
\left[Z_1-\frac{s_{12}}{c_{12}c_{13}}\Re(Z_6\,e^{-i\theta_{23}})
+\frac{s_{13}}{c_{13}}\Im(Z_6\,e^{-i\theta_{23}})\right]v^2\,,
\label{m12}\\[5pt]
m_2^2 &=&
\left[Z_1+\frac{c_{12}}{s_{12}c_{13}}\Re(Z_6\,e^{-i\theta_{23}})
+\frac{s_{13}}{c_{13}}\Im(Z_6\,e^{-i\theta_{23}})\right]v^2\,,
\label{m22}\\[5pt]
m_3^2 &=&
\left[Z_1-\frac{c_{13}}{s_{13}}\Im(Z_6\,e^{-i\theta_{23}})\right]v^2\,.
\label{m32}
\eeqa

For the subsequent analysis, it is useful to invert \eqst{m12}{m32}
and solve for $Z_1$, $\Re(Z_6\,e^{-i\theta_{23}})$ and 
$\Im(Z_6\,e^{-i\theta_{23}})$:
\beq
Z_1 v^2=m_1^2 c_{12}^2 c_{13}^2+m_2^2 s_{12}^2 c_{13}^2 + m_3^2
s_{13}^2\,,\label{z1v} 
\eeq
\beqa
\Re(Z_6\,e^{-i\theta_{23}})\,v^2 &=& c_{13}s_{12}c_{12}(m_2^2-m_1^2)\,,
\label{z6rv} \\[5pt]
\Im(Z_6\,e^{-i\theta_{23}})\,v^2 &=& s_{13}c_{13}(c_{12}^2 m_1^2+s_{12}^2
m_2^2-m_3^2) \,. \label{z6iv}
\eeqa
In addition, \eqs{tan213}{tan12f} can be used to express 
$\Re(Z_5\,e^{-i\theta_{23}})$ in terms of~$Z_6$:
\beq \label{z56}
\Re(Z_5\,e^{-2i\theta_{23}})=\frac{c_{13}}{s_{13}}
\Im(Z_6\,e^{-i\theta_{23}})
+\frac{c_{12}^2-s_{12}^2}{c_{13}s_{12}c_{12}}\,
\Re(Z_6\,e^{-i\theta_{23}})\,.
\eeq
Inserting \eqs{z6rv}{z6iv} into \eqs{tan13}{z56} then yields
expressions for $\Im(Z_5\,e^{-i\theta_{23}})$
and $\Re(Z_5\,e^{-i\theta_{23}})$ in terms of the invariant angles and
the neutral Higgs masses.

The above results can be used to derive an expression for
\beqa \label{imz56}
\Im(Z_5^* Z_6^2)&=& 2\,\Re(Z_5 e^{-2i\theta_{23}})\,
\Re(Z_6\,e^{-i\theta_{23}})\,\Im(Z_6\,e^{-i\theta_{23}}) \nonumber \\
&& \qquad -\,\Im(Z_5 e^{-2i\theta_{23}})\left\{
[\Re(Z_6\,e^{-i\theta_{23}})]^2-[\Im(Z_6\,e^{-i\theta_{23}})]^2\right\}\,.
\eeqa
Using \eq{tan13} and \eqst{z6rv}{z56}, 
one can simplify the right hand side of \eq{imz56} to obtain:
\beq \label{imz56f}
\Im(Z_5^* Z_6^2)\,v^6=2s_{13}c_{13}^2 s_{12}c_{12}\,(m_2^2-m_1^2)(m_3^2-m_1^2)
(m_3^2-m_2^2)\,.
\eeq
\Eq{imz56f} was first derived in \Ref{cpx}; it is 
equivalent to a result initially obtained in \Ref{pomarol}.
In particular, if any two of the 
neutral Higgs masses are degenerate, then $\Im(Z_5^* Z_6^2)=0$,
in which case one can always find a basis in which 
the pseudo-invariants $Z_5$ and $Z_6$ are simultaneously real. 
The neutral scalar squared-mass matrix [\eq{matrix33}] then 
breaks up into a block diagonal form consisting of a
$2\times 2$ block and a $1\times 1$ block.  The 
diagonalization of the $2\times 2$ block has a simple analytic form,
and the neutral scalar mixing
can be treated more simply by introducing one invariant
mixing angle instead of the three needed in the general case.
Note that  $\Im(Z_5^* Z_6^2)=0$ is a necessary (although not
sufficient) requirement for a CP-conserving Higgs sector, as discussed
in \App{app:four}.  For the remainder of this Appendix, 
we shall assume that the
neutral Higgs boson masses are non-degenerate.

In order to facilitate the discussion of the CP-conserving limit and
the decoupling limit of the 2HDM (which are treated in \App{app:four}),
it is useful to derive a number of additional relations for the
invariant angles.  First, we employ \eqst{m12}{m32} to
eliminate $\theta_{12}$ and $\phi$ and obtain a single equation
for $\theta_{13}$:
\beq \label{s13}
s_{13}^2=\frac{(Z_1 v^2-m_1^2)(Z_1 v^2-m_2^2)+|Z_6|^2 v^4}
{(m_3^2-m_1^2)(m_3^2-m_2^2)}\,.
\eeq
\Eq{s13} determines $c_{13}$ (in the convention where 
$c_{13}\geq 0$).
The sign of $s_{13}$ is determined
from \eq{m32}, which can be rewritten as: 
\beq \label{phieq}
\sin\phi=\frac{(Z_1v^2-m_3^2)\tan\theta_{13}}{|Z_6|v^2}\,.
\eeq
Since $\sin\phi\geq 0$ in the angular domain specified by \eq{app:domains}, it
follows that the sign of $s_{13}$ is equal to the sign of 
the quantity $Z_1 v^2-m_3^2$.  In particular, if $m^2_3$ is the largest 
eigenvalue of $\widetilde{\mathcal{M}}$ [\eq{matrix33}], then it must
be greater than the largest diagonal element of
$\widetilde{\mathcal{M}}$.  That is, $Z_1 v^2-m_3^2< 0$ if 
$m_3> m_{1,2}$, in which case $s_{13}\leq 0$.

However,
\eq{phieq} does not fix the sign of $\cos\phi$.  To determine this sign, we
can use \eq{tan13} to eliminate $\theta_{13}$ from
\eq{phieq}.  Consequently, one obtains a single equation for $\phi$:
\beq \label{tan2phi}
\tan 2\phi=\frac{\Im(Z_5^* Z_6^2)}{\Re(Z_5^* Z_6^2)+
\displaystyle{\frac{|Z_6|^4 v^2}{m_3^2-Z_1 v^2}}}\,.
\eeq
Given $\sin\phi\geq 0$ and $\tan 2\phi$ in the region
$0\leq\phi<\pi$, one can uniquely determine the value of $\phi$ 
(and hence the sign of $\cos\phi$).
Thus, for a fixed ordering of the neutral Higgs masses, 
\eqst{s13}{tan2phi} provide a unique solution for $(\theta_{13},\phi)$
in the domain $-\pi/2\leq\theta_{13}<\pi/2$ and $0\leq\phi<\pi$.

Next, we note that \eq{z6rv} can be rewritten as:
\beq \label{m2m1}
\sin 2\theta_{12}=\frac{2\,|Z_6|\,v^2\cos\phi}
{c_{13}(m_2^2-m_1^2)}\,.
\eeq
As advertised below \eq{tan12f}, the mass ordering of $m_1$ and $m_2$ 
fixes the sign of $\sin 2\theta_{12}$.
In particular, in the angular domain 
of \eq{app:domains}, $m_2>m_1$ implies that $s_{12}\cos\phi\geq 0$.
The sign of $s_{12}$ is then fixed after using \eq{imz56f} to infer that
$\sin 2\theta_{56}\cos\phi\geq 0$ for $m_3>m_2>m_1$.

An alternative expression for $\theta_{12}$ can be obtained by
combining \eqs{z1v}{s13}: 
which yields:
\beq \label{c13s12}
c_{13}^2 s_{12}^2=\frac{(Z_1 v^2-m_1^2)(m_3^2-Z_1 v^2)-|Z_6|^2 v^4}
{(m_2^2-m_1^2)(m_3^2-m_2^2)}\,.
\eeq 
Note the similarity of the expressions given by \eqs{s13}{c13s12};
both these results play an important role in determining the 
conditions that govern the decoupling limit.
 
A simpler form for $\tan^2\theta_{13}$ can also be obtained by combining 
\eqs{tan213}{phieq}:
\beq \label{t213}
\tan^2\theta_{13}=\frac{m_3^2-A^2}{m_3^2-Z_1 v^2}\,.
\eeq
Finally, one can derive an
expression for $m_2^2-m_3^2$, after eliminating $\Im(Z_6
e^{-i\theta_{23}})$ in favor of $\Re(Z_5 e^{-2i\theta_{23}})$ using \eq{z56}:
\beq \label{m2m3}
m_2^2-m_3^2=\frac{v^2}{c_{13}^2}\left[\Re(Z_5 e^{-2i\theta_{23}})
+\frac{c_{13}(s_{12}^2-c_{12}^2 s_{13}^2)}{s_{12}c_{12}}
\Re(Z_6 e^{-i\theta_{23}})\right]\,.
\eeq
The expressions for the differences of squared-masses
[\eqs{m2m1}{m2m3}] 
take on rather simple forms in the CP-conserving limit.

For completeness, we end this Appendix with a treatment of the case
where $Z_6=0$.  Since all three neutral Higgs
squared-masses are assumed to be non-degenerate, we require that
$Z_5\equiv |Z_5|e^{2i\theta_5}\neq 0$ in what follows,\footnote{If
$Z_5=Z_6=0$, then the neutral Higgs squared-mass matrix is diagonal in
the Higgs basis, with two degenerate Higgs boson mass-eigenstates.}
and define the invariant angle
$\phi_5\equiv \theta_5-\theta_{23}$.  Once the sign conventions of
the neutral Higgs fields are fixed, the invariant angles 
$\theta_{12}$, $\theta_{13}$ and $\phi_5$ are defined modulo $\pi$.
We first note that \eqthree{c23c12a}{c23c12b}{tan12}
are valid when $Z_6=0$.  Thus, setting \eq{c23c12b} to zero
implies that $\sin 2\theta_{13}=0$,\footnote{If $Z_6=0$ and 
$A^2=Z_1 v^2$, then Eq. (C7) is automatically
equal to zero.  In this case, we set Eq. (C6) to zero and
conclude that either $c_{13}=0$ or $\Im(Z_5
e^{-2i\theta_{23}})=0$.  If the latter holds true,
then the squared-mass matrix $\widetilde{\mathcal{M}}$ 
[see \eq{mtilmatrix}] is diagonal with degenerate eigenvalues.} 
which yields two possible cases:
(\textit{i})~$s_{13}=0$ or (\textit{ii})~$c_{13}=0$. 

If (\textit{i}) $s_{13}=0$, then \eq{c23c12a} yields $\Im(Z_5
e^{-2i\theta_{23}})=0$, \textit{i.e.}, $\sin 2\phi_5=0$, and
\eq{tan12} implies that $\sin 2\theta_{12}=0$.  Thus, we obtain four
possible solutions for the invariant angles modulo~$\pi$, which
correspond to four of the possible six mass orderings of the three neutral
Higgs states.  If (\textit{ii})~$c_{13}=0$ [and
$s_{13}=-1$ in the convention of \eq{app:domains}],
then \eq{tan12} implies that $\tan 2\theta_{12}= -\tan2\phi_5$, or
equivalently $\sin 2(\theta_{12}+\phi_5)=
\sin 2(\theta_{12}-\theta_{23}+\theta_5)=0$.  
In particular, in a basis in which $Z_5$ is real, the rotation
matrix $R$ [\eq{rmatrix}] depends only on the combination 
$\theta_{12}-\theta_{23}$ when $s_{13}=-1$ (so that 
$\theta_{12}+\theta_{23}$ is indeterminate).  
Note that
$q_{31}=-1$, $q_{41}=i$, $q_{22}=-iq_{12}=e^{i\theta_{12}}$
and $q_{11}=q_{21}=q_{32}=q_{42}=0$.  Indeed, only the combination
$\theta_{12}-\theta_{23}$ enters the Higgs couplings in a real $Z_5$
basis.  Consequently, the condition $\sin 2(\theta_{12}+\phi_5)=0$
yields two solutions modulo $\pi$, which correspond to the final two
possible mass orderings of the neutral Higgs states.

\section{The CP-conserving and the Decoupling Limits of the 2HDM}
\label{app:four}

To make contact with the 2HDM literature, we consider two limiting
cases of the most general 2HDM---the 
CP-conserving limit and the decoupling limit.

\subsection{The CP-conserving limit of the 2HDM}

In the CP-conserving limit, we impose CP-invariance on 
all bosonic couplings and fermionic couplings of the Higgs bosons.
The requirement of a CP-conserving bosonic sector is
equivalent to the requirement that the scalar potential is explicitly
CP-conserving and that the Higgs vacuum is CP-invariant
(\textit{i.e.}, there is no spontaneous CP-violation).  
Basis-independent conditions for a CP-conserving bosonic sector have been
given in refs.~\cite{cpx,cpx2,davidson,cpbasis}.  In \Ref{davidson}, these
conditions were recast into the following form.  The bosonic sector is
CP-conserving if and only if:\footnote{Since the scalar potential
minimum conditions imply that $Y_3=-\half Z_6 v^2$, no separate
condition involving $Y_3$ is required.}  
\beq \label{cpoddinv}
\Im[Z_6 Z_7^\ast]=\Im[Z_5^*Z_6^2]=\Im[Z_5^*(Z_6+Z_7)^2]=0\,.  
\eeq 
\Eq{cpoddinv} is equivalent to the requirement that
\beq \label{cpang}
\sin 2(\theta_5-\theta_6)=
\sin 2(\theta_5-\theta_7)=\sin(\theta_6-\theta_7)=0\,,
\eeq
where $\theta_5$ and $\theta_6$ are defined in \eq{app:invang} 
and $\theta_7\equiv\arg Z_7$ 
(note that $\theta_5$ is defined modulo
$\pi$ and $\theta_6$ and $\theta_7$ are defined modulo $2\pi$).

Additional constraints arise when the Higgs-fermion couplings are
included.  Consider the most general coupling of the two Higgs
doublets to three generations of quarks and leptons, as described in
\sect{sec:six}.  As we shall demonstrate below \eq{antihermitian}, if
\beq \label{zrho}
Z_5[\rho^Q]^2\,,\,\,\,Z_6\rho^Q\,,\,\,\,{\rm and}\,\,\, Z_7\rho^Q\quad
\mbox{\text{are hermitian matrices}}\qquad (Q=U,D,E)\,,
\eeq
then the couplings of the neutral Higgs bosons to fermion pairs 
are CP-invariant.  Thus, if \eqs{cpoddinv}{zrho} are satisfied, then
the neutral Higgs bosons are eigenstates of CP, and the
only possible source of CP-violation in the 2HDM is the 
unremovable phase in the CKM matrix $K$
that enters via the charged current interactions mediated by either
$W^\pm$ or $H^\pm$ exchange\footnote{One can also formulate a
basis-independent condition 
(that is invariant with respect to
separate redefinitions of the Higgs doublet fields and the quark
fields) for the absence of CP-violation in the charged current
interactions.  This condition involves the Jarlskog 
invariant~\cite{jarlskog}, and can also be written 
as~\cite{cpinvariants,branco}: 
$\Tr\ls{\rm f}\,\bigl[H^{U,0},\,H^{D,0}\bigr]\lasup{\,3}=0$
(summed over three quark generations), where
$H^{Q,0}\equiv M^{Q,0} M^{Q,0\,\dagger}$
and the $M^{Q,0}$ are defined below \eq{diagdmass}.
Since CP-violating phenomena in the charged current interactions are observed
and well described by the CKM matrix, we shall not impose 
this latter condition here.}
[see \eq{hffu2}].

One can explore the consequences of CP-invariance by studying the
pattern of Higgs couplings and the structure of the neutral Higgs
boson squared-mass matrix [\eq{matrix33}].  
The tree-level couplings of $\go$ are
CP-conserving, even in the general CP-violating 2HDM.  In particular,
the couplings $\go\go\go$, $\go G^+G^-$, $\go H^+ H^-$
and $ZZ\go$ are absent.  Moreover, \eq{YG} implies that $\go$
possesses purely pseudoscalar couplings to the fermions.  Hence,
$\go$ is a CP-odd scalar, independently of the 
structure of the scalar potential.
We can therefore use the couplings of $\go$ to the neutral Higgs bosons as a
probe of the CP-quantum numbers of these states.  
The analysis of the neutral
Higgs boson squared-mass matrix (which does not depend on $Z_7$)
simplifies significantly when
$\Im[Z_5^* Z_6^2]=0$.  One can then choose a basis
where $Z_5$ and $Z_6$ are simultaneously real, in which case
the scalar squared-mass matrix decomposes into 
diagonal block form.  The upper 
$2\times 2$ block can be diagonalized analytically and yields the
mass-eigenstates $\hl$ and $\hh$ (with $\mhl\leq\mhh$).  The lower $1\times 1$
block yields the mass-eigenstate $\ha$.  If all the conditions of
\eqs{cpoddinv}{zrho} are satisfied, then the neutral Higgs boson
mass-eigenstates are also states of definite CP quantum number.
We shall demonstrate below that 
$\hl$ and $\hh$ are CP-even scalars and $\ha$ is a CP-odd scalar.

We first consider the special case of $Z_6=0$.  
In this case,  the
scalar squared-mass matrix is diagonal in a basis
where $Z_5$ is real.  
It is convenient
to choose $c_{12}=c_{13}=c_{23}=1$ (corresponding to one of the
possible Higgs mass orderings, as discussed at the end of
\App{app:three}).  For this choice of Higgs mixing angles, 
$q_{11}=q_{22}=1$, $q_{41}=q_{32}=i$, and all other $q_{k\ell}$
vanish.  These results then fix all the Higgs couplings.
The existence of non-zero $Z\go h_1$ and
$Z h_2 h_3$ couplings imply that $h_1$ is CP-even, and $h_2$ and
$h_3$ are states of opposite CP quantum number.  If
\eqs{cpoddinv}{zrho} hold with $Z_7\neq 0$ and/or
$\rho^Q\neq 0$, 
then one can use the non-zero $H^+ H^- h_2$ and/or the $\overline QQh_2$
couplings to conclude that $h_2$ is CP-even and $h_3$ is CP-odd.  
Thus, we can identify\footnote{If $Z_6=Z_7=\rho^Q=0$, then
only the relative CP quantum numbers of $h_2$ and $h_3$ are
well-defined.}
$h_3=\ha$ and $h_{1,2}=\hl\,,\,\hh$ (the mass ordering of the latter two
states is not yet determined).  
The scalar squared masses [\eq{diagtil}] are given by: 
\beq \label{z6case}
m_1^2=Z_1 v^2\,,\qquad\qquad
m_{2,3}^2=Y_2+\half\left[Z_3+Z_4\pm\varepsilon|Z_5|\right]v^2\,,
\eeq
where 
$\varepsilon\equiv\cos2(\theta_5-\theta_{23})=\pm 1$.
In general, these scalar squared-masses will \text{not} satisfy
$m_1\leq m_2\leq m_3$.
In a real $Z_5$ basis with $\theta_{23}=0$, it follows that
$\varepsilon|Z_5|=Z_5$.
For example, if $m_1\leq m_2$ (in which case, $h_1=\hl$ and
$h_2=\hh$), then
$\mhh^2-\mha^2=Z_5 v^2$.
A different choice for the $\theta_{ij}$ 
can be made to reorder the
neutral Higgs states in ascending mass order. 

If $\Im(Z_5^* Z_6^2)=0$ and $Z_6\neq 0$, one possible choice is
$c_{13}=c_{23}=1$ (in a basis where $Z_5$ and
$Z_6$ are real), in which case $\theta_{12}$ is
determined by diagonalizing the upper left $2\times 2$ block of
\eq{matrix33}.  However, it is instructive to consider the
implications of the more general results of \App{app:three}.
Since $Z_6\neq 0$ by assumption, \eq{tan2phi} yields
$\sin\phi\cos\phi=0$, and
\eq{imz56f} implies that either some of the neutral
Higgs boson masses are degenerate
or $s_{13}s_{12}c_{12}=0$.\footnote{Since
$Z_6\neq 0$, one can use \eqs{tan13}{tan213} to show
that $c_{13}\neq 0$.}  
In the case of degenerate masses, some of 
the invariant angles are not well defined, since any linear
combination of the degenerate states is also a mass-eigenstate.
Hence, the degenerate case must be treated separately.  In what
follows, we shall assume that all three neutral Higgs boson masses are
non-degenerate.  Note that if
$\sin\phi=0$, then \eq{phieq} yields
$s_{13}=0$, whereas if $\cos\phi=0$, then \eq{m2m1} yields
$\sin 2\theta_{12}=0$.%
\footnote{The same constraints are obtained by imposing the
requirement of CP-conserving Higgs couplings.  In particular,
the existence of a $\go h_k h_k$ coupling would imply that
$h_k$ is a state of mixed CP-even and CP-odd components.  All such
couplings must therefore be absent in the CP-conserving limit.
Using the results of \eqs{G3}{cpang} one can easily check that
at least one of these CP-violating couplings is present unless
$s_{13}=\sin\phi=0$ or $\cos\phi=\sin 2\theta_{12}=0$.}
Thus, we shall consider separately the two cases:
(I)~$s_{13}=\sin\phi=0$ and (II)~$\cos\phi=\sin 2\theta_{12}=0$.

Consider Case~I where $s_{13}=\sin\phi=0$.  With the angles
restricted as specified in \eq{app:domains},
\beq \label{case1}
\theta_{13}=0\,,\qquad\qquad \phi=0\,,\qquad \quad
[\rm{Case~I}]\,.
\eeq
Next, we examine Case~II where $\cos\phi=\sin 2\theta_{12}=0$.  
It is convenient to consider separately two subcases:
(IIa)~$s_{12}=\cos\phi=0$ and
(IIb)~$c_{12}=\cos\phi=0$.  More explicitly, 
\beqa 
\theta_{12}&=&0\,,\phm\qquad\,\,\quad 
\phi=\half\pi\,,
\qquad [\rm{Case~IIa}]\,\,,\label{case2a}\\
\theta_{12}&=&-\half\pi\,,\phm\qquad 
\phi=\half\pi\,,\qquad [\rm{Case~IIb}]\,. \label{case2b}
\eeqa
The values of the $q_{k\ell}$ corresponding to cases I, IIa and IIb
are given in Tables~\ref{tab2}---\ref{tab4}.

\begin{table}[ht!]
\centering
\caption{The U(2)-invariant quantities $q_{k\ell}$ in the
CP-conserving limit.  Case I: $s_{13}=0$ and $\sin(\theta_6-\theta_{23})=0$.  
In the standard notation of the
CP-conserving 2HDM (with real scalar potential parameters), 
$e^{-i\theta_{23}}={\rm sgn}~Z_6\equiv\varepsilon_6$, 
which implies that 
the neutral Higgs fields are $h_1=\hl$, $h_2=-\varepsilon_6\hh$, 
$h_3=\varepsilon_6\ha$ and $h_4=\go$, and the angular factors are
$c_{12}=\sbma$ and $s_{12}=-\varepsilon_6\cbma$.} \vskip 0.08in
\label{tab2}
\begin{tabular}{|c||c|c|}\hline
$\phaa k\phaa $ &\phaa $q_{k1}\phaa $ & \phaa $\phm q_{k2} \phaa $ \\ \hline
$1$ & $c_{12}$ & $-s_{12}$ \\
$2$ & $s_{12}$ & $\phm c_{12}$ \\
$3$ & $0$ & $\phm i$ \\
$4$ & $i$ & $\phm 0$ \\ \hline 
\end{tabular}
\end{table}
\begin{table}[ht!]
\centering
\caption{The U(2)-invariant quantities $q_{k\ell}$ in the
CP-conserving limit. Case IIa: $s_{12}=0$ and $\cos(\theta_6-\theta_{23})=0$. 
In the standard notation of the
CP-conserving 2HDM (with real scalar potential parameters),  
$e^{-i\theta_{23}}=i\,{\rm sgn}~Z_6\equiv i\varepsilon_6$, 
which implies that
the neutral Higgs fields are $h_1=\hl$,
$h_2=\varepsilon_6\ha$, $h_3=\varepsilon_6\hh$ 
and $h_4=\go$, and the angular factors are
$c_{13}=\sbma$ and $s_{13}=\varepsilon_6\cbma$.}
\vskip 0.08in
\label{tab3}
\begin{tabular}{|c||c|c|}\hline
$\phaa k\phaa $ &\phaa $q_{k1}\phaa $ & \phaa $\phm q_{k2} \phaa $ \\ \hline
$1$ & $c_{13}$ & $ -is_{13} $\\
$2$ & $0$ & $\phm 1$ \\
$3$ & $s_{13}$ & $\phm i c_{13}$ \\
$4$ & $i$ & $\phm 0$ \\ \hline 
\end{tabular}
\end{table}
\begin{table}[ht!]
\centering
\caption{The U(2)-invariant quantities $q_{k\ell}$ in the
CP-conserving limit. Case IIb: $c_{12}=0$ and $\cos(\theta_6-\theta_{23})=0$.
In the standard notation of the
CP-conserving 2HDM (with real scalar potential parameters),  
$e^{-i\theta_{23}}=i\,{\rm sgn}~Z_6\equiv i\varepsilon_6$,
which implies that
the neutral Higgs fields are $h_1=\varepsilon_6\ha$,
$h_2=-\hl$, $h_3=\varepsilon_6\hh$ and $h_4=\go$, and the angular factors are
$c_{13}=\sbma$ and $s_{13}=\varepsilon_6\cbma$.} 
\vskip 0.08in
\label{tab4}
\begin{tabular}{|c||c|c|}\hline
$\phaa k\phaa $ &\phaa $\phm q_{k1}\phaa $ & \phaa $q_{k2} \phaa $ \\ \hline
$1$ & $\phm 0$ & $1$ \\
$2$ & $-c_{13}$ & $is_{13}$ \\
$3$ & $\phm s_{13}$ & $i c_{13}$ \\
$4$ & $\phm i$ & $0$ \\ \hline
\end{tabular}
\end{table}

Using these values of the $q_{k\ell}$ [along with the values of
$\phi$ given in \eqst{case1}{case2b}],
we can employ the bosonic couplings of the
Higgs bosons to determine the CP-quantum numbers of the three neutral
states, $h_k$.
Using the same techniques as in
the $Z_6=0$ case treated previously, we again conclude that one
of the three neutral
Higgs states is CP-odd and the other two are CP-even.  The existence
of three cases [I, IIa and IIb above] corresponds to the three possible 
neutral Higgs 
fields that can be identified as the CP-odd scalar.  The three cases
can also be understood from the structure of the matrix
$\widetilde{\mathcal{M}}$ given in \eq{mtilmatrix}.  In particular,
if $\Im(Z_5^*Z_6^2)=0$ and $Z_6\neq 0$, then two possible cases exist:
\beqa
{\rm Case~I:}\phantom{I}\quad \sin\phi=0   &\Longrightarrow& 
\Im(Z_5 e^{-2i\theta_{23}})=\Im(Z_6 e^{-i\theta_{23}})=0\,,\label{sp0}\\
{\rm Case~II:}\quad \cos\phi=0   &\Longrightarrow& 
\Im(Z_5 e^{-2i\theta_{23}})=\Re(Z_6 e^{-i\theta_{23}})=0\,.\label{cp0}
\eeqa

In both Case I and Case II, $\widetilde{\mathcal{M}}$ assumes a block
diagonal form consisting of a $2\times 2$ block (corresponding to the
the CP-even Higgs bosons) and a $1\times 1$ block (corresponding to the
CP-odd Higgs boson).  In Case I, the $1\times 1$ block associated with
$\ha$ is the $33$ element of $\widetilde{\mathcal{M}}$.  Thus,
$s_{13}=0$ and we identify $h_3$ as the CP-odd Higgs boson, with mass:  
\beq \label{mhaI}
m^2_{\ha}=  A^2 = 
Y_2+\half v^2 \left[Z_3+Z_4-\Re(Z_5\,e^{-2i\theta_{23}})\right]
\qquad\qquad\,\, [\rm{Case~I}]\,.
\eeq
In Case IIa, the $1\times 1$ block associated with 
$\ha$ is the $22$ element of  $\widetilde{\mathcal{M}}$.  Thus,
$\theta_{12}=0$, and we identify $h_2$ as the CP-odd Higgs boson, with mass: 
\beq 
m^2_{\ha}=  A^2+v^2\,\Re(Z_5\,e^{-2i\theta_{23}}) =
Y_2+\half v^2 \left[Z_3+Z_4+\Re(Z_5\,e^{-2i\theta_{23}})\right]
\quad [\rm{Case~II}]\,.\label{twoma}
\eeq
If we choose $\theta_{12}=-\pi/2$ instead of of $\theta_{12}=0$ above, 
then this corresponds to an additional orthogonal transformation 
$\mathcal{\widetilde{M}}\to R_{12}\widetilde{\mathcal{M}}R_{12}^T$,
where $R_{12}$ is defined in \eq{rmatrix}.  One can easily
check that
$R_{12}\widetilde{\mathcal{M}}R_{12}^T$ is also in block diagonal
matrix form, consisting of a $2\times 2$ block and a $1\times 1$ block.
The latter, associated with the CP-odd scalar state, 
is the $11$ element of $R_{12}\widetilde{\mathcal{M}}R_{12}^T$.  This is Case
IIb, and we identify $h_1$ as the CP-odd Higgs boson, 
with mass given by \eq{twoma}. 
Note that \eqst{case1}{case2b} imply that
the value of $Z_5\,e^{-2i\theta_{23}}$ in Case II has the opposite
sign from the corresponding result in Case I.  Thus, \eqs{mhaI}{twoma}
yield the same result for the
mass of the CP-odd scalar in terms of the model parameters.

In the standard notation of the CP-conserving 2HDM,
one considers only real basis choices, in which the Higgs Lagrangian
parameters and the scalar vacuum expectation values are 
real.  We can therefore restrict basis changes to O(2) 
transformations~\cite{davidson}.\footnote{If $Z_6=Z_7=\rho^Q=0$, then
the possible transformations among real bases are elements of
O(2)$\times\mathbb{Z}_2$.  In particular, the sign of $Z_5$ changes when
when the Higgs basis field
$H_2\to iH_2$. In this case, $Z_5$ is an O(2)-invariant but it is
a pseudo-invariant with respect to $\mathbb{Z}_2$.}
In this context, pseudo-invariants are SO(2)-invariant quantities that
change sign under an O(2) transformation with determinant
equal to $-1$.  Note that
$Z_5$ is now an invariant with respect to O(2) transformations, but
$Z_6$, $Z_7$ and $e^{-i\theta_{23}}$ are pseudo-invariants.  
In particular, for $Z_6\neq 0$ in the convention where $0\leq\phi<\pi$,
\beq \label{vareps}
e^{-i\theta_{23}}=e^{i\phi}e^{-i\theta_6}=\begin{cases}
\,\,\varepsilon_6 \quad & \text{[Case I]}\,, \\ \,\,i\varepsilon_6 \quad
& \text{[Case II]}\,,\end{cases}
\eeq
where $Z_6\equiv\varepsilon_6|Z_6|$ in the real basis.  That is,
$\varepsilon_6$ is a pseudo-invariant quantity (in contrast, the sign of $Z_5$
is invariant) with respect to O(2) transformations.  Using \eq{vareps}
in either \eq{mhaI} or \eq{twoma} yields
$\mha$ in terms of the real-basis parameters:
\beq
m_{\ha}^2=Y_2+\half v^2\left(Z_3+Z_4-Z_5\right)\,.
\eeq

The generic real basis fields can be expressed in
terms of 
the two neutral CP-even scalar mass-eigenstates $\hl$, $\hh$
(with $\mhl\leq\mhh$) and the CP-odd scalar mass-eigenstate $\ha$,
$\go$ as follows~\cite{ghhiggs,hhg}:
\beqa
\Phi_1^0&=&\frac{1}{\sqrt{2}}\left[v\widehat v_1-\hl s_\alpha
+\hh c_\alpha+i(\go\cb-\ha\sb)\right]\,,\\
\Phi_2^0&=&\frac{1}{\sqrt{2}}\left[v\widehat v_2+\hl c_\alpha
+\hh s_\alpha+i(\go\sb+\ha\cb)\right]\,,
\eeqa
with $\mhl\leq\mhh$,
where $\widehat v_a=(\cb\,,\,\sb)$,
$s_\alpha\equiv\sin\alpha$, $c_\alpha\equiv\cos\alpha$, and
$\alpha$ is the CP-even neutral Higgs boson mixing angle.  These
equations can be written more compactly as:
\beq \label{cpchdm}
\Phi^0_a=\frac{1}{\sqrt{2}}\left[(v+\hl\sbma+\hh\cbma+i\go)\widehat v_a
+(\hl\cbma-\hh\sbma+i\ha)\widehat w_a\right]\,,
\eeq
where $\sbma\equiv\sin(\beta-\alpha)$ and
$\cbma\equiv\cos(\beta-\alpha)$.

Using the results of Tables~\ref{tab2}---\ref{tab4}
and comparing \eq{cpchdm} to \eq{master} [with
$e^{-i\theta_{23}}$ determined from \eq{vareps}],
one can identify the neutral Higgs fields $h_k$ with the eigenstates
of definite CP quantum numbers, $\hl$, $\hh$ and $\ha$, and relate
the angular factor $\beta-\alpha$ with the
appropriate invariant angle:\footnote{The extra minus signs 
in the identification of $h_2=-\varepsilon_6\hh$ 
in Case I and $h_2=-\hl$ in Case IIb arise due to
the fact that the standard conventions of the CP-conserving 2HDM
correspond to $\det R=-1$ (whereas $\det R=+1$ in Case~IIa).}
\beqa \label{cases12}
&&\hspace{-0.5in}{\rm Case~I:}\phm\quad h_1=\hl\,,\,\,
h_2=-\varepsilon_6\hh\,,\,\, {\rm and} \,\, h_3=\varepsilon_6\ha\,,\quad\,\,
c_{12}=\sbma\,\,\, {\rm and}\,\, s_{12}=-\varepsilon_6\cbma\,,\nonumber \\
&&\hspace{-0.5in}{\rm Case~IIa:}\quad h_1=\hl\,,\,\,
h_2=\varepsilon_6\ha\,,\,\, {\rm and} \,\, 
h_3=\varepsilon_6\hh\,,\qquad\,\,\,\!\!\!
c_{13}=\sbma\,\,\, {\rm and}\,\, s_{13}=\varepsilon_6\cbma\,,\nonumber \\
&&\hspace{-0.5in}{\rm Case~IIb:}\quad h_1=\varepsilon_6\ha\,,\,\,
h_2=-\hl\,,\,\, {\rm and} \,\, h_3=\varepsilon_6\hh\,,\quad\,\,
c_{13}=\sbma\,\,\, {\rm and}\,\, s_{13}=\varepsilon_6\cbma\,.
\eeqa
In the convention for the angular domain given by \eq{app:domains},
$c_{12}$ and $c_{13}$ are non-negative and therefore
$\sbma\geq 0$. 
The appearance of the pseudo-invariant quantity $\varepsilon_6$
in \eq{cases12}
implies that $\hh$, $\ha$ (and $H^\pm$) are pseudo-invariant fields,
and $\cbma$ is a pseudo-invariant with respect to O(2) 
transformations.\footnote{Note that $\sbma$ is invariant with respect to O(2)
transformations, which is consistent with our convention that
$\sbma\geq 0$.   The analogous results have
been obtained in \Ref{davidson} in a convention where $\cbma\geq 0$.}
In contrast, $\hl$ is an invariant field.

At this stage, we have not imposed any mass ordering of the three
neutral scalar states.  Since one can distinguish between the CP-odd
and the CP-even neutral scalars, it is sufficient to require that
$\mhl\leq\mhh$.  (If one does not care about the mass ordering
of $\ha$ relative to the CP-even states, then
Cases IIa and IIb can be discarded without loss of generality.)  
We can compute the masses of the CP-even scalars and
the angle $\beta-\alpha$~\cite{ghdecoupling} in any of the three cases:
\beqa
\mhl^2 &=& \mha^2\,\cbma^2+v^2\left[Z_1\sbma^2+Z_5\cbma^2+2\sbma\cbma
  Z_6\right]\,,\label{mhl2}\\
\mhh^2 &=& \mha^2\,\sbma^2+v^2\left[Z_1\cbma^2+Z_5\sbma^2-2\sbma\cbma
  Z_6\right]\,,\label{mhh2}
\eeqa
and
\beq \label{bma}
\tan[2(\beta-\alpha)]=\frac{2Z_6 v^2}{m_{\ha}^2+(Z_5-Z_1) v^2}\,,\qquad
\sin [2(\beta-\alpha)]=\frac{-2Z_6 v^2}{\mhh^2-\mhl^2}\,.
\eeq
Note that
\eqst{mhl2}{bma} are covariant with respect to O(2) transformations,
since $Z_6$ and $\cbma$ are both pseudo-invariant quantities.

Additional constraints arise by requiring that the  
neutral Higgs-fermion couplings are CP-invariant, as previously noted
in \eq{zrho}.  To derive this latter result, we employ the 
possible values of the $q_{k\ell}$ given in
Tables~\ref{tab2}---\ref{tab4} in 
the Higgs-fermion interactions given by \eq{hffu2}.
In particular, we demand that the couplings of $\hl$ and $\hh$
to fermions are scalar interactions, whereas the couplings of $\ha$ 
to fermions are pseudoscalar interactions.
These requirements produce the following basis-independent conditions:
\beq \label{antihermitian}
e^{i\theta_{23}} \rho^Q\,\,{\rm is}\quad \begin{cases}
\mbox{\text{hermitian{\phantom{anti}}}} &
\qquad\text{in~Case~I}\,,\\
\mbox{\text{anti-hermitian}} &
\qquad\text{in~Cases~IIa~and~IIb}\,.\end{cases}
\eeq
In both Cases I and II, 
the results of \eqthree{sp0}{cp0}{antihermitian} imply that the matrix
$Z_6 \rho^Q=(Z_6 e^{-i\theta_{23}})(e^{i\theta_{23}}\rho^Q)$ is hermitian.
Combining this result with \eq{cpoddinv} then yields \eq{zrho}.

Invariant techniques for describing the constraints on the Higgs-fermion
interaction due to CP-invariance have also been
considered in \Refs{cpx2}{branco}.  In these works, the authors
construct invariant expressions that are both U(2)-invariant and
invariant with respect to the redefinition of the quark fields.  For
example, the invariants denoted by $J_a$ and $J_b$ in \Ref{cpx2}
are given by $J_a\equiv \Im\, J^D$ and $J_b\equiv \Im\, J^U$ where
\beq \label{jqdef}
J^Q= \Tr(VYT^Q)\,,\qquad T^Q_{a\bbar}\equiv\Tr\ls{\rm f}(\eta^{Q,0}_a
\eta^{Q,0\,\dagger}_{\bbar}) =\Tr\ls{\rm f}(\eta^{Q}_a
\eta^{Q\,\dagger}_{\bbar})\,,
\eeq
and the trace $\Tr\ls{\rm f}$ sums over the diagonal quark
generation indices.  Note that the trace over generation indices
ensures that the resulting expression is invariant with respect to
unitary redefinitions of the quark fields [\eq{biunitary}].
Using \eq{aab} [with $A=Y$], it is straightforward to re-express
\eq{jqdef} as:
\beq
J^Q=Y_1\Tr\ls{\rm f}[(\kappa^Q)^2]+Y_3\Tr\ls{\rm f}[\kappa^Q\rho^Q]\,,
\eeq
after using \eqthree{invariants}{pseudoinvariants}{kapparho}.
Indeed, $J^Q$ is invariant with respect to U(2) transformations
since the product of pseudo-invariants $Y_3\,\rho^Q$ is a U(2)-invariant
quantity.  Moreover, taking the trace over the quark generation 
indices ensures that $J^Q$ is invariant with respect to 
unitary redefinitions of the quark fields.
In \Ref{cpx2}, a proof is given 
that $\Im\, J^Q=0$ is one of the invariant conditions 
for CP-invariance of the Higgs-fermion interactions. 
In our formalism, this result is easily verified. Using the 
scalar potential minimum conditions [\eq{hbasismincond}], we obtain:
\beq
\Im\, J^Q=-\frac{v}{\sqrt{2}}\, \Im\bigl[Z_6
\Tr\ls{\rm f}(M_Q \rho^Q)\bigr]\,.
\eeq
But, CP-invariance requires [by \eq{zrho}] that
$Z_6\rho ^Q$ is hermitian.  Since $M_Q$ is a real diagonal
matrix, it then immediately follows that
$\Im\,J^Q=0$.

We end this subsection with a very brief outline of the tree-level MSSM
Higgs sector.  Since this model is CP-conserving, it is conventional
to choose the phase conventions of the Higgs fields that yield
a real basis.  In the natural
supersymmetric basis, the $\lambda_i$ of \eq{pot}
are given by:
\beq
\lambda_1=\lambda_2=\quarter(g^2+g^{\prime\,2})\,,\quad
\lambda_3=\quarter(g^2-g^{\prime\,2})\,,\quad
\lambda_4=-\half g^2\,,\quad
\lambda_5=\lambda_6=\lambda_7=0\,,
\eeq
where $g$ and $g'$ are the usual electroweak couplings [with
$m_Z^2=\quarter(g^2+g^{\prime\,2})v^2$].
From these results, one can compute the (pseudo-)invariants:
\beqa \label{mssmz}
\hspace{-0.25in} Z_1=Z_2 &=& \quarter (g^2+g^{\prime\,2})\cos^2
2\beta\,,
\quad\qquad
\! Z_3= Z_5+\quarter(g^2-g^{\prime\,2})\,,\quad\qquad
Z_4= Z_5-\half g^2\,,\nonumber \\ \hspace{-0.5in}
Z_5&=& \quarter (g^2+g^{\prime\,2})\sin^2 2\beta\,,\qquad\quad
Z_6=-Z_7= -\quarter (g^2+g^{\prime\,2})\sin 2\beta\cos2\beta\,.
\eeqa
The standard MSSM 
tree level Higgs sector formulae~\cite{ghhiggs} for the Higgs masses and 
$\beta-\alpha$ are easily reproduced 
using \eq{mssmz} and the results of this Appendix.

\subsection{The decoupling limit of the 2HDM}
\label{app:decoupling}

The decoupling limit corresponds
to the limiting case in which one of the two Higgs doublets of the 2HDM
receives a very large mass and is therefore decoupled from the
theory~\cite{habernir,ghdecoupling}.  This can be achieved by
assuming that $Y_2\gg v^2$ and $|Z_i|\lsim\mathcal{O}(1)$ [for all~$i$].  
The effective low energy theory is a one-Higgs-doublet
model that corresponds to the Higgs sector of the Standard Model.
We shall order the neutral scalar masses according to $m_1< m_{2,3}$
and define the invariant Higgs mixing angles accordingly.
Thus, we expect one light CP-even Higgs boson, $h_1$, with couplings
identical (up to small corrections) to those of the 
Standard Model (SM) Higgs boson.
Using the fact that $m^2_1$, $|Z_i| v^2 \ll m^2_2$, $m^2_3$, $\mhpm^2$ in the
decoupling limit, \eqs{m22}{m32} yield:\footnote{We 
assumed that $Z_6\neq 0$ in the derivation of 
\eqs{decoupconds}{z5decoup}.  In the case of $Z_6=0$,
we may use \eqthree{s13}{c13s12}{z6case} to conclude that 
$s_{12}=s_{13}=0$ are exactly
satisfied as long as $m_1<m_{2,3}$.  Setting \eq{c23c12a} to zero, it then
follows that $\Im(Z_5\,e^{-2i\theta_{23}})=0$.}
\beq \label{decoupconds}
|s_{12}|\lsim\mathcal{O}\left(\frac{v^2}{m_2^2}\right)\ll 1\,,
\qquad \qquad |s_{13}|\lsim\mathcal{O}\left(\frac{v^2}{m_3^2}\right)\ll 1\,,
\eeq
and \eq{tan2phi} imples that $\tan 2\phi+\tan 2\theta_{56}\ll 1$,
where $\theta_{56}\equiv \half\arg Z_5-\arg Z_6$.  This latter
inequality is equivalent to:
\beq \label{z5decoup}
\Im(Z_5\,e^{-2i\theta_{23}})\lsim\mathcal{O}
\left(\frac{v^2}{m_3^2}\right)\ll 1\,.
\eeq
Note that \eq{z5decoup} is also satisfied if $\theta_{23}\to\theta_{23}+\pi/2$.
These two respective solutions (modulo~$\pi$) correspond to the two
possible mass orderings of $h_2$ and $h_3$.

One can explicitly verify the assumed mass hierarchy of the  
Higgs bosons in the decoupling limit.  
Using \eqs{m12}{decoupconds}, it follows 
that $m_1^2=Z_1 v^2$, with corrections 
$\lsim\mathcal{O}(v^4/m^2_{2,3})$.
\Eq{t213} yields $m_3^2=A^2$, with corrections 
$\lsim\mathcal{O}(v^2)$, and \eq{m2m3} yields
$m_3^2-m_2^2\lsim\mathcal{O}(v^2)$.  Finally, \eqs{hplus}{madef} 
imply that $\mhpm^2-m_3^2\lsim\mathcal{O}(v^2)$.  That is,
$m_1\ll m_2\simeq m_3\simeq \mhpm$.

The values of the $q_{k\ell}$ in the exact decoupling 
limit, where $s_{12}=s_{13}=\Im(Z_5\,e^{-2i\theta_{23}})=0$,
are tabulated in Table~\ref{tab5}.
\begin{table}[ht!]
\centering
\caption{The U(2)-invariant quantities $q_{k\ell}$ in the exact
decoupling limit.}\vskip 0.15in
\label{tab5}
\begin{tabular}{|c||c|c|}\hline
$\phaa k\phaa $ &\phaa $q_{k1}\phaa $ & \phaa $q_{k2} \phaa $ \\ \hline
$1$ & $1$ & $0$ \\
$2$ & $0$ & $1$ \\
$3$ & $0$ & $i$ \\
$4$ & $i$ & $0$ \\ \hline
\end{tabular}
\end{table}

\noindent
It is a simple exercise to insert the values of the
$q_{k\ell}$  in the exact decoupling limit
into the Higgs couplings of \sects{sec:five}{sec:six}.
The couplings of $h_1\equiv h$ are then given by:
\beqa
&&\hspace{-0.2in} \mathscr{L}_{\rm h}= \half(\partial_\mu h)^2-\half
Z_1v^2 h^2
-\half vZ_1 h^3 -\eighth vZ_1 h^4
+\left(gm_W W_\mu^+W^{\mu\,-}+\frac{g}{2c_W}
m_Z Z_\mu Z^\mu\right) h \nonumber \\
&& \qquad
+\left[\quarter g^2  W_\mu^+W^{\mu\,-}
+\frac{g^2}{8c_W^2}Z_\mu Z^\mu\right] h^2
+\biggl\{\left(\half eg A^\mu W_\mu^+ -\frac{g^2s_W^2}{2c_W}Z^\mu
  W_\mu^+\right)G^-h+{\rm h.c.}\biggr\} \nonumber \\
&& \qquad
-\half ig\left[W_\mu^+G^-\ddel\lsup{\,\mu} h+{\rm h.c.}\right]
+\frac{g}{2c_W}  Z^\mu G^0\ddel_\mu h
+\frac{1}{v}\overline D M_D D h+\frac{1}{v}\overline U M_U U h\,.
\eeqa
This is precisely the SM Higgs Lagrangian.  Even in the
most general CP-violating 2HDM, the interactions of the $h$ in the
decoupling limit are CP-conserving and diagonal in quark flavor space.
CP-violating and flavor non-diagonal effects in the Higgs interactions
are suppressed by factors of $\mathcal{O}(v^2/m^2_{2,3})$ as shown in detail in
\Ref{ghdecoupling2}.
In contrast to the SM-like Higgs boson $h$, 
the interactions of the heavy neutral Higgs bosons ($h_2$
and $h_3$) and the charged Higgs bosons ($\hpm$) 
exhibit both CP-violating and quark flavor non-diagonal couplings
(proportional to the $\rho^Q$) in the decoupling limit.
In particular, whereas \eq{z5decoup} implies that
$\sin 2(\theta_5-\theta_{23})\ll 1$, the CP-violating invariant
quantities
$\sin(\theta_6-\theta_{23})$ and $\sin(\theta_7-\theta_{23})$ 
[\textit{c.f.} \eq{cpang}] need
not be small in the most general 2HDM.

One can understand the origin of 
the decoupling conditions [\eqs{decoupconds}{z5decoup}] as follows.
First, using \eq{hmassesinv}, we see that we can decouple $h_2$ and $h_3$
(and $H^\pm$) by
taking $A^2\gg v^2$ while sending $q_{12}\to 0$.  Thus, in the
convention in which the mass ordering of
the three neutral Higgs states is $m_1\leq m_2\leq m_3$, it follows
that the exact decoupling limit is formally achieved when
$A^2\to\infty$ and
$|q_{12}|^2=s_{12}^2+c_{12}^2 s_{13}^2=0$, which implies that
$s_{12}=s_{13}=0$.  Inserting these results into
\eq{rtil} yields $\widetilde R=I$, where $I$ is the
$3\times 3$ identity matrix.  Consequently, 
$\wtil{\mathcal{M}}$
[see \eqst{mtilmatrix}{diagtil}] must be diagonal up to
corrections of $\mathcal{O}(v^2/A^2)$.  However, because 
\eq{mtilmatrix} is dominated in the decoupling limit
by its $22$ and $33$ elements (which are approximately degenerate), 
it follows that the $23$ element must vanish exactly in leading order.
Thus, in the exact decoupling limit, 
$\Im(Z_5\,e^{-2i\theta_{23}})=0$.  Note that this latter
constraint is consistent with \eq{tan13}, as $\theta_{13}=0$ in
the decoupling limit.

For further details and a more comprehensive treatment of the decoupling
limit, see \Refs{ghdecoupling}{ghdecoupling2}.


\begin{thebibliography}{99}

\bibitem{Lee:1973iz}
T.D.~Lee,
Phys.\ Rev.\ {\bf D8} 1226 (1973);
Phys.\ Rep. {\bf 9}, 143 (1974).

\bibitem{Peccei:1977hh}
R.D.~Peccei and H.R.~Quinn,
Phys.\ Rev.\ Lett.\  {\bf 38}, 1440 (1977).

\bibitem{susyhiggs}
P.~Fayet, Nucl.\ Phys.\ {\bf B78}, 14 (1974);
{\bf B90}, 104 (1975);
K.~Inoue, A.~Kakuto, H.~Komatsu and S.~Takeshita,
Prog.\ Theor.\ Phys.\  {\bf 67}, 1889 (1982);
R.A.~Flores and M.~Sher,
Annals Phys. (NY)\ {\bf 148}, 95 (1983).

\bibitem{ghhiggs}
J.F.~Gunion and H.E.~Haber,
Nucl.\ Phys.\ {\bf B272}, 1 (1986);
{\bf B278}, 449 (1986)
[Erratum: {\bf B402}, 567 (1993)].

\bibitem{hhg}
J.F.~Gunion, H.E.~Haber, G.~Kane and S.~Dawson,
{\it The Higgs Hunter's Guide} (Perseus Publishing,
Cambridge, MA, 1990).

\bibitem{Carena:2002es}
M.~Carena and H.E.~Haber,
Prog.\ Part.\ Nucl.\ Phys.\ {\bf 50}, 63 (2003)
[arXiv:hep-ph/0208209].

\bibitem{branco}
G.C.~Branco, L.~Lavoura and J.P.~Silva, {\it CP Violation}
(Oxford University Press, Oxford, England, 1999),
chapters 22 and 23.

\bibitem{susyreviews}
P.~Fayet and S.~Ferrara,
Phys.\ Rept.\  {\bf 32}, 249 (1977);
H.P.~Nilles,
Phys.\ Rep.\  {\bf 110}, 1 (1984);
H.E.~Haber and G.~L.~Kane,
Phys.\ Rep.\  {\bf 117}, 75 (1985).

\bibitem{desh}
N.G.~Deshpande and E.~Ma,
Phys.\ Rev.\ {\bf D18}, 2574 (1978).

\bibitem{type1}
H.E.~Haber, G.L.~Kane and T.~Sterling,
Nucl.\ Phys.\ {\bf B161}, 493 (1979).


\bibitem{type2}
J.F.~Donoghue and L.F.~Li,
Phys.\ Rev.\ {\bf D19}, 945 (1979).

\bibitem{hallwise}
L.J.~Hall and M.B.~Wise,
Nucl.\ Phys.\ {\bf B187}, 397 (1981).

\bibitem{lavoura2}
L.~Lavoura,
Phys.\ Rev.\ {\bf D50}, 7089 (1994)
[arXiv:hep-ph/9405307].

\bibitem{Weinberg}
S.L.~Glashow and S.~Weinberg,
Phys.\ Rev.\ {\bf D15}, 1958 (1977);
E.A.~Paschos,
Phys.\ Rev.\ {\bf D15}, 1966 (1977).

\bibitem{Georgi}
H.~Georgi and D.V.~Nanopoulos,
Phys.\ Lett. {\bf 82B}, 95 (1979).

\bibitem{ilcbook}
J.F.~Gunion, H.E.~Haber and Rick Van Kooten, in \textit{Linear
Collider Physics 
in the New Millennium}, edited by K.~Fujii, D.J.~Miller and A.~Soni,
(World Scientific, Singapore, 2005) pp.~41--133 [arXiv:hep-ph/0301023].

\bibitem{davidson}
S.~Davidson and H.E.~Haber, Phys.\ Rev.\ {\bf D72}, 035004 (2005)
[Erratum: {\bf D72}, 099902 (2005)] [arXiv:hep-ph/0504050].


\bibitem{cpbasis}
J.F.~Gunion and H.E.~Haber,
Phys.\ Rev.\ {\bf D72}, 095002 (2005)
[arXiv:hep-ph/0506227].

\bibitem{Branco:2005em}
G.C.~Branco, M.N.~Rebelo and J.I.~Silva-Marcos,
Phys.\ Lett.\ {\bf B614}, 187 (2005)
[arXiv:hep-ph/0502118].

\bibitem{ivanov}
I.P.~Ivanov,
Phys.\ Lett.\ {\bf B632}, 360 (2006)
[arXiv:hep-ph/0507132];
C.C.~Nishi, arXiv:hep-ph/0605153.
 

\bibitem{cpx}
L.~Lavoura and J.P.~Silva,
Phys.\ Rev.\ {\bf D50}, 4619 (1994)
[arXiv:hep-ph/9404276].

\bibitem{cpx2}
F.J.~Botella and J.P.~Silva,
Phys.\ Rev.\ {\bf D51}, 3870 (1995)
[arXiv:hep-ph/9411288].



\bibitem{Ginzburg:2004vp}
I.F.~Ginzburg and M.~Krawczyk,
Phys.\ Rev.\ {\bf D72}, 115013 (2005)
[arXiv:hep-ph/0408011].

\bibitem{murnaghan}
F.D.~Murnaghan, \textit{The Unitary and Rotation Groups} (Spartan
Books, Washington, DC, 1962).

\bibitem{synge}
J.L.~Synge and A.~Schild, \textit{Tensor Calculus} (Dover publications, Inc.,
New York, NY, 1978).

\bibitem{matrixref}
D.S.~Bernstein, \textit{Matrix Mathematics} (Princeton University
Press, Princeton, NJ, 2005).

\bibitem{cpcarlos}
A.~Pilaftsis and C.E.M.~Wagner,
Nucl.\ Phys.\ {\bf B553}, 3 (1999)
[arXiv:hep-ph/9902371].

\bibitem{Gunion:1997aq}
J.F.~Gunion, B.~Grzadkowski, H.E.~Haber and J.~Kalinowski,
Phys.\ Rev.\ Lett.\  {\bf 79}, 982 (1997)
[arXiv:hep-ph/9704410].

\bibitem{Grzadkowski:1999ye}
B.~Grzadkowski, J.F.~Gunion and J.~Kalinowski,
Phys.\ Rev.\ {\bf D60}, 075011 (1999)
[arXiv:hep-ph/9902308].

\bibitem{typeiii}
W.S.~Hou, Phys.\ Lett.\ {\bf B296}, 179 (1992);
D.~Chang, W.S.~Hou and W.Y.~Keung,
Phys.\ Rev.\ {\bf D48}, 217 (1993)
[arXiv:hep-ph/9302267];
D.~Atwood, L.~Reina and A.~Soni,
Phys.\ Rev.\  {\bf D55}, 3156 (1997)
[arXiv:hep-ph/9609279].

\bibitem{horn}
R.A.~Horn and C.R.~Johnson, {\it Matrix Analysis} (Cambridge
Univ. Press, Cambridge UK, 1990).

\bibitem{ilctanb1}
J.L.~Feng and T.~Moroi,
Phys.\ Rev.\ {\bf D56}, 5962 (1997)
[arXiv:hep-ph/9612333].

\bibitem{tanbprecision1}
J.~Kamoshita,
Prog.\ Theor.\ Phys.\  {\bf 100}, 773 (1998)
[arXiv:hep-ph/9809353].

\bibitem{ilctanb2}
V.D.~Barger, T.~Han and J.~Jiang,
Phys.\ Rev.\ {\bf D63}, 075002 (2001)
[arXiv:hep-ph/0006223].

\bibitem{ilctanb3}
J.F.~Gunion, T.~Han, J.~Jiang and A.~Sopczak,
Phys.\ Lett.\ {\bf B565}, 42 (2003)
[arXiv:hep-ph/0212151].

\bibitem{boosetal}
E.~Boos, H.U.~Martyn, G.~Moortgat-Pick, M.~Sachwitz, A.~Sherstnev 
and P.M.~Zerwas,
Eur.\ Phys.\ J.\ {\bf C30}, 395 (2003)
[arXiv:hep-ph/0303110].

\bibitem{plctanb1}
M.A.~Doncheski, S.~Godfrey and S.~Zhu,
Phys.\ Rev.\ {\bf D68}, 053001 (2003)
[arXiv:hep-ph/0306126].

\bibitem{plctanb2}
S.Y.~Choi, J.~Kalinowski, J.S.~Lee, M.M.~Muhlleitner, M.~Spira and 
P.M.~Zerwas,
Phys.\ Lett.\ {\bf B606}, 164 (2005)
[arXiv:hep-ph/0404119].

\bibitem{lhctanb}
R.~Kinnunen, S.~Lehti, F.~Moortgat, A.~Nikitenko and M.~Spira,
Eur.\ Phys.\ J.\  {\bf C40}, Supplement 1, 23 (2005)
[arXiv:hep-ph/0503075].



\bibitem{deltab}
R.~Hempfling,
Phys.\ Rev.\ {\bf D49}, 6168 (1994);
L.J.~Hall, R.~Rattazzi and U.~Sarid,
Phys.\ Rev.\ {\bf D50}, 7048 (1994)
[arXiv:hep-ph/9306309];
M.~Carena, M.~Olechowski, S.~Pokorski and C.E.M.~Wagner,
Nucl.\ Phys.\ {\bf B426}, 269 (1994)
[arXiv:hep-ph/9402253];
J.A.~Coarasa, R.A.~Jimenez and J.~Sola,
Phys.\ Lett.\ {\bf B389}, 312 (1996)
[arXiv:hep-ph/9511402];
D.M.~Pierce, J.A.~Bagger, K.T.~Matchev and R.J.~Zhang,
Nucl.\ Phys.\ {\bf B491}, 3 (1997)
[arXiv:hep-ph/9606211].

\bibitem{Carena:2001uj}
M.~Carena, D.~Garcia, U.~Nierste and C.E.M.~Wagner,
Phys.\ Lett.\ {\bf B499}, 141 (2001) [arXiv:hep-ph/0010003].

\bibitem{habernir}
H.E.~Haber and Y.~Nir,
Nucl.\ Phys.\ {\bf B335}, 363 (1990).

\bibitem{ghdecoupling}
J.F.~Gunion and H.E.~Haber,
Phys.\ Rev.\ {\bf D67}, 075019 (2003)
[arXiv:hep-ph/0207010].

\bibitem{Dedes:2003kp}
For a review and guide to the literature, see A.~Dedes,
Mod.\ Phys.\ Lett.\ {\bf A18}, 2627 (2003)
[arXiv:hep-ph/0309233].

\bibitem{pomarol}
A.~Mendez and A.~Pomarol,
Phys.\ Lett.\  {\bf B272}, 313 (1991).

\bibitem{jarlskog}
C.~Jarlskog,
Phys.\ Rev.\ Lett.\  {\bf 55}, 1039 (1985);
Z.\ Phys.\ {\bf C29}, 491 (1985).

\bibitem{cpinvariants}
J.~Bernabeu, G.C.~Branco and M.~Gronau,
Phys.\ Lett.\ {\bf B169} (1986) 243.

\bibitem{ghdecoupling2}
J.F.~Gunion, H.E.~Haber, and J.~Kalinowski, in preparation.

\end{thebibliography}
\end{document}